\documentclass[preprintnumbers,amsmath,amssymb,twocolumn,superscriptaddress]{revtex4}

\usepackage{graphicx}
\usepackage{subfigure}
\usepackage{dcolumn}
\usepackage{bm}
\usepackage{booktabs}
\usepackage{threeparttable}
\usepackage{multirow}
\usepackage{array}
\usepackage{makecell}
\usepackage{enumerate}
\usepackage[colorlinks,
            citecolor=blue,
            anchorcolor=red,
            menucolor=red,
            linkcolor=red,
            filecolor=red,
            runcolor=red,
            urlcolor=blue,
            frenchlinks=red]{hyperref}

\newcommand{\PreserveBackslash}[1]{\let\temp=\\#1\let\\=\temp}
\newcolumntype{C}[1]{>{\PreserveBackslash\centering}p{#1}}
\newcolumntype{R}[1]{>{\PreserveBackslash\raggedleft}p{#1}}
\newcolumntype{L}[1]{>{\PreserveBackslash\raggedright}p{#1}}

\begin{document}

\title{Elastic and inelastic $J/\psi$ photoproduction in $p$-$p$ collisions at LHC energies: the feature of Weizs\"{a}cker-Williams approximation}

\author{Zhi-Lei Ma}
\affiliation {Department of Physics, Yunnan University, Kunming 650091, China}
\affiliation {Department of Astronomy, Key Laboratory of Astroparticle Physics of Yunnan Province, Yunnan University, Kunming 650091, China}

\author{Zhun Lu}
\email{zhunlu@seu.edu.cn}
\affiliation {School of Physics, Southeast University, Nanjing 211189, China}

\author{Jia-Qing Zhu}
\email{zhujiaqing@ynu.edu.cn}
\affiliation {Department of Physics, Yunnan University, Kunming 650091, China}

\author{Li Zhang}
\email{lizhang@ynu.edu.cn}
\affiliation {Department of Astronomy, Key Laboratory of Astroparticle Physics of Yunnan Province, Yunnan University, Kunming 650091, China}

\date{\today}

\begin{abstract}
The $J/\psi$ production originating from elastic and inelastic photoproduction processes in p-p collisions at LHC energies is investigated, where the fragmentation processes are involved.
An exact treatment is performed, which adopts the Martin-Ryskin method to weight the contribution from different channels, and can return to Weizs\"{a}cker-Williams approximation (WWA) when $Q^{2}\rightarrow0$.
The relevant kinematical relations are also achieved.
We present a comprehensive analysis for the feature of WWA by comparing with the exact treatment.
The results are expressed in $Q^{2}$, $y$, $z$, $p_{T}$, and $y_{r}$ (rapidity) distributions, and the total cross sections are also estimated.
The numerical results indicate that, the incoherent-photon emission can provide the meaningful contribution to elastic photoproduction, and starts to play a very important role in the inelastic processes.
The photoproduction and fragmentation processes can improve the contribution of $J/\psi$ production in $p$-$p$ collisions at LHC energies.
Moreover, the WWA is only effective in very restricted domains,
and the exact treatment is needed to deal accurately with the $J/\psi$ photoproduction, which can naturally avoid double counting and WWA errors.
\end{abstract}


\maketitle

\section{INTRODUCTION}
\label{Introduction}

\begin{figure}
\setlength{\abovecaptionskip}{1mm}
\centering
\includegraphics[width=0.42\columnwidth]{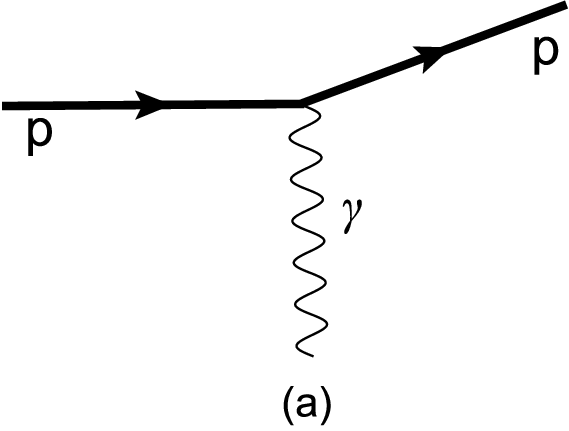}\hspace{5mm}
\includegraphics[width=0.5\columnwidth]{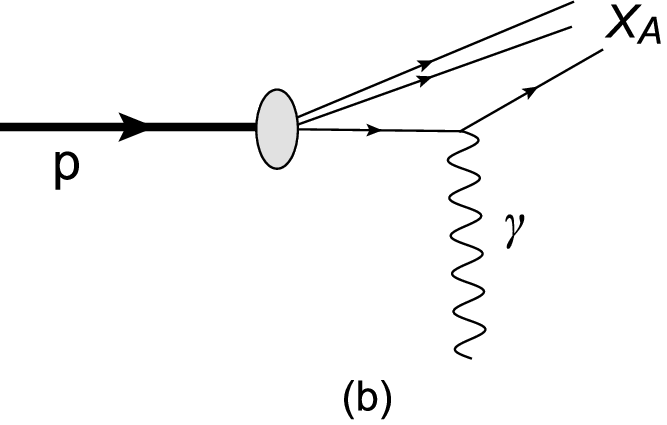}
\caption{(a): Coherent-photon emission, virtual photon is radiated coherently by the whole proton which remains intact after scattering.
(b): Incoherent-photon emission, virtual photon is radiated incoherently by the quarks inside proton which is allowed to break up after scattering.}
\label{fig:photoemi.}
\end{figure}

During the last years the study of photon - induced interactions at hadronic colliders has been strongly motivated by the possibility of constraining the dynamics of the strong interactions at large energies.
This mechanism is the dominant channel in ultra-peripheral collisions~\cite{Baltz:2007kq, Nucl.Rev.Part.Sci_55_271, Djuvsland:2010qs}, and plays a fundamental role in $\emph{ep}$ deep inelastic scattering at HERA~\cite{Butterworth:1996zw}, and is also an important part of current experimental efforts at LHC~\cite{Acharya:2019vlb}.
In the case of heavy vector meson photoproduction, it sheds light on the low-$x$ physics and helps to constraint the nuclear parton distributions.
On the side of projectile, there are two kinds of photon emission mechanisms [Fig.~\ref{fig:photoemi.}] need to be distinguished~\cite{Baur:1998ay}: coherent-photon emission (coh.) and incoherent-photon emission (incoh.).
In the first type, the virtual photons are radiated coherently by the whole nucleus which remains intact after photons emitted.
In the second type, the virtual photons are emitted incoherently by the individual constituents inside nucleus, and as a weakly bound system nucleus will dissociate after photons emitted.
On the target side, there are two types of photoproduction: elastic process in which the target nucleus remains intact after scattering with photons; inelastic process in which the target nucleus is allowed to break up.
When these different channels are considered simultaneously, we have to weight the different contributions for avoiding double counting.
But in fact, this serious trouble is encountered in most works and caused the large fictitious contributions~\cite{Zhu:2015qoz, Fu:2011zzm, Fu:2011zzf, Chin.Phys.C_36_721, Yu:2015kva, Yu:2017rfi, Yu:2017pot}.

There are a lot of studies for these processes, and the incoherent-photon emission mechanism has been adopted in the two-photon processes~\cite{Drees:1994zx, Ohnemus:1993qw}.
However, the application of this mechanism from the individual quarks, to our knowledge, is insufficient in $J/\psi$ photoproduction.
In Refs.~\cite{Klein:2019avl, Baltz:2002pp, Klein:1999qj, Klein:2003vd}, Klein and Nystrand studied the coherent and elastic vector meson production via photon-Pomeron and photon-meson interactions, where $J/\psi$ photoproduction in $p$-$p$ collisions is also analyzed by adopting the photon spectrum associated with a proton.
These authors also presented a Monte Carlo simulation program, STARTlight, that calculated the cross sections for a variety of UPC final states~\cite{Klein:2016yzr}.
In Refs.~\cite{Yang:2019lls, Wu:2020ujf}, authors studied the inclusive diffractive heavy quarkonium photoproduction using the resolved pomeron model, where the direct and resolved contributions are included.
Gon\c{c}alves et al. presented a comprehensive analysis of elastic vector mesons photoproduction in hadronic collisions in Refs.~\cite{Goncalves:2019txs, Goncalves:2018blz, Goncalves:2018htp, Xie:2021seh, Xie:2022sjm, daSilveira:2021bzs, Goncalves:2015dia}, where the predictions for transverse momentum and rapidity distributions considering different QCD dynamics were estimated.
They also presented their results for the inelastic quarkonium photoproduction by using the semiclassical photon spectrum~\cite{Goncalves:2013ixa}.
Machado et al. studied the inclusive and exclusive $J/\psi$, $\psi(2S)$, and $\Upsilon$ photoproductions in the proton-nucleus and nucleus-nucleus collisions at the LHC within the color dipole formalism~\cite{Jenkovszky:2021sis, Kopp:2018xvu, Goncalves:2017wgg, SampaiodosSantos:2014puz}.
There are also a lot of other works for the photoproduction of heavy quarkonium.
However, the photon emission types in all of theses above works are coherent, the incoherent contribution is neglected.

On the other hand, it is well known that the $J/\psi$ photoproduction can be theoretically studied by using the Weizs\"{a}cker-Williams approximation (WWA), which can be traced back to early works by Fermi \cite{Fermi:1924tc}, Weizs\"{a}cker and Williams \cite{vonWeizsacker:1934nji}, and Landau and Lifshitz~\cite{Phys.Rev._45_729, Sov.Phys._6_244}.
The central idea of WWA is that the moving electromagnetic field of charged particles can be treated as a flux of photons distributed with some density $n(\omega)$ on a frequency spectrum~\cite{Phys.Rev._51_1037, Phys.Rev._105_1598, Nucl.Phys._23_1295}.
Thus, the cross section is given by the convolution between the photon flux and the relevant real photoproduction cross section.
In the calculations, an important function is the photon flux function or equivalent photon spectrum, which has different forms for different charged sources.
Although the great success has been achieved, the discussion about the properties of WWA in $J/\psi$ photoproduction is still insufficient.
WWA is usually employed to the processes which are actually not applicable, and a number of imprecise statements and some widely used equivalent photon spectra are achieved beyond the WWA validity range~\cite{Drees:1989vq, Drees:1988pp, Frixione:1993yw, Zhu:2015via, Zhu:2015qoz, Fu:2011zzm, Fu:2011zzf, Chin.Phys.C_36_721, Yu:2015kva, Yu:2017rfi, Yu:2017pot, Nystrand:2004vn, Kniehl:2001tk, Kniehl:1990iv, Fu:2012xm, sp}.
Especially at LHC energies, the WWA significantly decides the accuracy of photoproduction processes, since photon flux function $f_{\gamma}\propto\ln\sqrt{s}/m$, the collision energy $\sqrt{s}$ is very large and thus $f_{\gamma}$ becomes important.
For these reasons, we consider that it is necessary to systematically study the properties of WWA, and to discuss important inaccuracies encountered in its application.

According to the purposes discussed above, in this work, we consider the $J/\psi$ photoproduction in $p$-$p$ collisions at LHC energies.
We present the comparison between WWA and the exact treatment which reduces to the WWA in the limit $Q^{2}\rightarrow0$ and can be considered as the generalization of Leptoproduction~\cite{Graudenz:1993tg,Fleming:1997fq}.
The kinematical relations matched with the exact treatment are also achieved.
And then we discuss the production rate of photoproduction and fragmentation processes by comparing with leading order (LO) contribution.

This paper is organized as follows.
Section~\ref{Exac} presents the formalism of exact treatment, where the direct, resolved and fragmentation contributions are included.
Based on Martin-Ryskin method, the coherent and incoherent-photon emissions, and the elastic and inelastic processes are considered simultaneously.
In Section~\ref{WWA}, the WWA is discussed by switching the exact formulae into the WWA ones in the limit $Q^{2}\rightarrow0$.
Section~\ref{Numerical results} provides the numerical results, the distributions of $Q^{2}$, $y$, $z$, $p_{T}$, and $y_{r}$, and the total cross sections are presented.
The summary and conclusions are given in Section~\ref{Summary and conclusions}.

\begin{figure*}
\setlength{\abovecaptionskip}{1mm}
\centering
\includegraphics[width=0.36\textwidth]{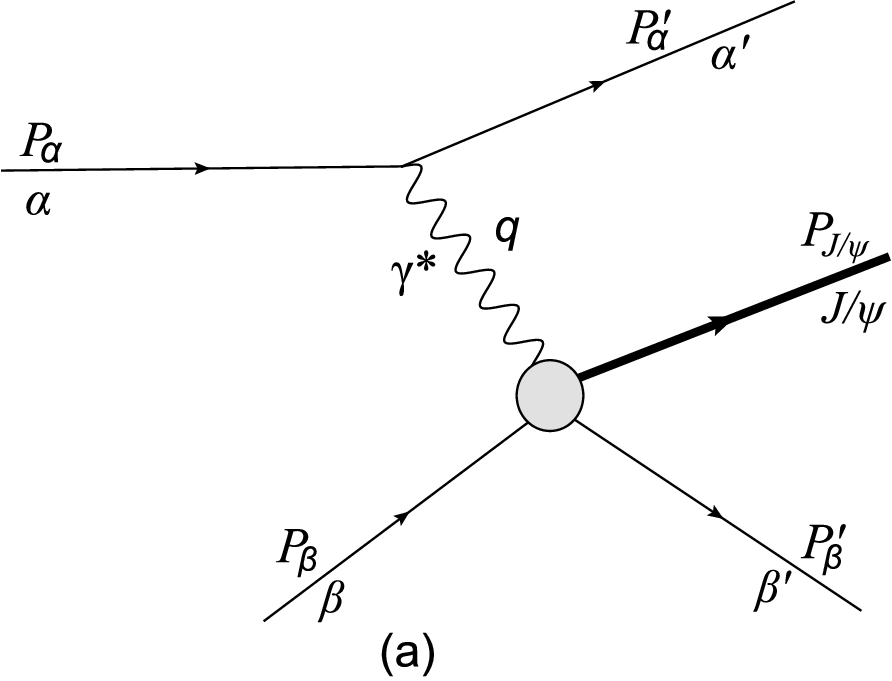}\hspace{25mm}
\includegraphics[width=0.25\textwidth]{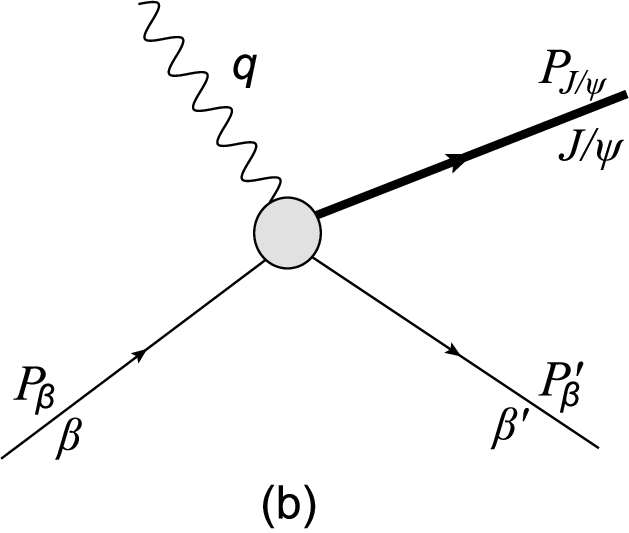}
\caption{(a) The general $J/\psi$ photoproduction processes.
The virtual photon emitted from the projectile $\alpha$ interacts with target $\beta$, $\alpha$ can be the proton or its charged parton (quarks).
$\beta$ can be the proton or its parton for elastic and inelastic processes, respectively.
$\alpha'$ and $\beta'$ are the scattered $\alpha$ and $\beta$.
(b): photo-absorption.}
\label{fig:feyn_ab}
\end{figure*}

\section{Exact treatment}
\label{Exac}

Heavy quarkonium production involves both perturbative and nonperturbative aspects of QCD.
The factorization formalism of nonrelativistic QCD (NRQCD) provides a rigorous theoretical framework for the description of heavy-quarkonium production and decay~\cite{Bodwin:1994jh}.
This formalism implies a separation of short-distance coefficients, which can be calculated perturbatively as expansions in the strong-coupling constant $\alpha_{s}$, from long-distance matrix elements (LDMEs), which must be extracted from experiment.
The LDMEs are process independent, and can be classified in terms of their scaling in $\nu$, the relative velocity of the heavy quarks in the bound state.
A crucial feature of this formalism is that it takes into account the $Q\bar{Q}$ Fock space, which is spanned by the states $n=^{2S+1}L_{J}^{(c)}$ with definite spin $S$, orbital angular momentum $L$, total angular momentum $J$, and color multiplicity $c$.
In particular, this formalism predicts the existence of color-octet processes by the nonperturbative emission of soft gluons in nature.

In this section, we employ the NRQCD factorization formalism to describe the inelastic $J/\psi$ photoproduction.
For elastic process, we adopt a phenomenological saturation model developed by Marquet-Peschanski-Soyez (MPS)~\cite{Marquet:2007qa}, which is based on the solutions of the BK equation in the momentum space, and is characterized by a saturation scale that is proportional to the squared momentum transfer $t$ and by the factorization of the $t$ - dependence associated to the proton vertex.
This model is recently updated by Xie and Gon\c{c}alves in Ref.~\cite{Xie:2022sjm}, through the fitting to the latest high precision HERA data for the reduced and vector meson cross sections.
In their work they performed a comprehensive study of the exclusive $\rho$, $J/\psi$, and $\gamma$ production in $ep$ collisions at the kinematical range that will be probed by the EIC and LHeC.
They compared the MPS model with BSAT and BCGC phenomenological saturation models in Refs.~\cite{Kowalski:2006hc,Watt:2007nr} based on distinct assumptions for the treatment of the dipole scattering amplitude, and indicated that a future experimental analysis of exclusive processes, considering events characterized by large values of the squared transferred momentum, has the potentiality of constraining the description of the QCD dynamics.

\subsection{General formalism}
\label{formalism}

As a generalization of Leptoproduction framework, the exact treatment for the $J/\psi$ photoproduction in p-p collisions consists of two important parts.
Firstly, the photon radiated from the projectile is off the mass shell and no longer transversely polarized, thus we can expand the density of this virtual photon using the linear combinations.
The formalism is analogous to that in Ref.~\cite{Kniehl:2001tk}.
Secondly, the square of the electric form factor $F^{2}_{1}(Q^{2})$ is applied as the probability or weighting factor (WF) to distinguish the contributions from the different photon emissions and from the different interaction types, thus we can avoid the trouble of double counting.

Actually, a consistent analysis of the terms neglected in going from the accurate expression of diagram of Fig.~\ref{fig:feyn_ab}(a) to the WWA one, allows us in a natural manner to estimate the properties of WWA in $J/\psi$ photoproduction.
This can be performed in a general form for every reaction.
In our case, for achieving the first part of exact treatment we derive the general form of cross section in Fig.~\ref{fig:feyn_ab}(a).
Denoting the scattering amplitude as $\mathcal{M}_{\alpha\beta}$, we obtain
\begin{align}\label{dab.Gen1}
&d\sigma(\alpha+\beta\rightarrow \alpha+J/\psi+\beta)\displaybreak[0]\nonumber\\
&=\frac{|\mathcal{M}_{\alpha\beta}|^{2}}{4\left[(p_{\alpha}\cdot p_{\beta})^{2}-m_{\alpha}^{2}m_{\beta}^{2}\right]^{1/2}}
d\mathrm{PS}_{3}(p_{\alpha}+p_{\beta};p_{\alpha}',p_{J/\psi},p_{\beta}'),\displaybreak[0]
\end{align}
where we employ the short-hand notation
\begin{align}\label{dPSn}
d\mathrm{PS}_{n}(P;p_{1},...,p_{n})=(2\pi)^{4}\delta^{4}(P-\sum_{i=1}^{n}p_{i})\prod_{i=1}^{n}\frac{d^{3}p_{i}}{(2\pi)^{3}2E_{i}},\displaybreak[0]
\end{align}
for the Lorentz invariant N-particle phase-space element.
By decomposing the squared scattering amplitude as $|\mathcal{M}_{\alpha \beta}|^{2}=4\pi\alpha_{\mathrm{em}}e_{\alpha}^{2}\rho^{\mu\nu}M^{\mu}M^{*\nu}/Q^{2}$ and rearranging $d\mathrm{PS}_{n}$, the cross section can be further expressed as
\begin{align}\label{dab.Gen2}
&d\sigma(\alpha+\beta\rightarrow \alpha+J/\psi+\beta)\displaybreak[0]\nonumber\\
=&4\pi e_{\alpha}^{2}\alpha_{\mathrm{em}}\frac{\rho_{\mu\nu}M^{\mu}M^{*\nu}}{Q^2}\left[\frac{(q\cdot p_{\beta})^{2}-q^{2}m_{\beta}^{2}}{(p_{\alpha}\cdot p_{\beta})^{2}-m_{\alpha}^{2}m_{\beta}^{2}}\right]^{1/2}\displaybreak[0]\nonumber\\
&\times\frac{d^{3}p'_{\alpha}}{(2\pi)^{3}2E'_{\alpha}}\frac{d\mathrm{PS}_{2}(q+p_{\beta};p_{J/\psi},p_{\beta}')}{4\hat{p}_{\mathrm{CM}}\sqrt{\hat{s}}},\displaybreak[0]
\end{align}
where $e_{\alpha}$ is the charge of the projectile, $\alpha_{\mathrm{em}}=1/137$ is the electromagnetic coupling constant, $E_{\alpha'}$ is the energy of $\alpha'$, $M^{\mu}$ is the virtual photo-absorption amplitude, and
\begin{align}\label{lambda}
\left[\frac{(q\cdot p_{\beta})^{2}-q^{2}m_{\beta}^{2}}{(p_{\alpha}\cdot p_{\beta})^{2}-m_{\alpha}^{2}m_{\beta}^{2}}\right]^{1/2}
=\frac{\hat{p}_{\mathrm{CM}}\sqrt{\hat{s}}}{p_{\mathrm{CM}}\sqrt{s_{\alpha \beta}}},\displaybreak[0]
\end{align}
$s_{\alpha \beta}=(p_{\alpha}+p_{\beta})^{2}$ and $\hat{s}=(q+p_{\beta})^{2}$ are the energy square in the $\alpha \beta$ and $\gamma^{*}\beta$ CM frames, respectively.
$p_{\mathrm{CM}}$ and $\hat{p}_{\mathrm{CM}}$ are the momenta in the corresponding CM frames.
The details can be found in Appendix~\ref{FKR}.
The quantity $\rho^{\mu\nu}$ is the density matrix of the virtual photon produced by $\alpha$,
\begin{align}\label{Rou.Gen}
\rho^{\mu\nu}=&(-g^{\mu\nu}+\frac{q^{\mu}q^{\nu}}{q^{2}})F_{2}(Q^{2})\displaybreak[0]\nonumber\\
-&\frac{(2P_{\alpha}-q)^{\mu}(2P_{\alpha}-q)^{\nu}}{q^{2}}F_{1}(Q^{2}),\displaybreak[0]
\end{align}
$F_{1}(Q^{2})$ and $F_{2}(Q^{2})$ are the general notations of form factors.

After integrating over the phase space volume of the produced system of particles, the following quantity can be included in Eq.~(\ref{dab.Gen2})
\begin{align}\label{W}
W^{\mu\nu}=\frac{1}{2}\int M^{\mu}M^{*\nu}d\mathrm{PS}_{2}(q+p_{\beta};p_{c},p_{\beta}').\displaybreak[0]
\end{align}
Here, $W^{\mu\nu}$ is the absorptive part of the $\gamma\beta$ amplitude [Fig.~\ref{fig:feyn_ab}(b)], connected with the cross section in the usual way.
The tensors according to which $W^{\mu\nu}$ is expanded, can be constructed only from the $q$, $p_{\beta}$ and $g^{\mu\nu}$ tensor.
In order to consider explicitly gauge invariance: $q^{\mu}W^{\mu\nu}=q^{\nu}W^{\mu\nu}=0$, it is convenient to use the following linear combinations~\cite{Budnev:1974de}
\begin{align}\label{Line.QR}
Q^{\mu}=&\left[\frac{-q^{2}}{(q\cdot p_{\beta})^{2}-q^{2}m_{\beta}^{2}}\right]^{\frac{1}{2}}\left(p_{\beta}-q\frac{q\cdot p_{\beta}}{q^{2}}\right)^{\mu},\displaybreak[0]\nonumber\\
R^{\mu\nu}=&-g^{\mu\nu}+\frac{(q\cdot p_{\beta})(q^{\mu}p_{\beta}^{\nu}+q^{\nu}p_{\beta}^{\mu})-q^{2}p_{\beta}^{\mu}p_{\beta}^{\nu}
-m_{\beta}^{2}q^{\mu}q^{\nu}}{(q\cdot p_{\beta})^{2}-q^{2}m_{\beta}^{2}},\displaybreak[0]\nonumber\\
\end{align}
which satisfy the relations: $q_{\mu}Q^{\mu}=q_{\mu}R^{\mu\nu}=Q_{\mu}R^{\mu\nu}=0$, $Q^{\mu}Q_{\mu}=1$.
After expending $W^{\mu\nu}$ in these tensors, we have
\begin{align}\label{WEp}
W^{\mu\nu}=R^{\mu\nu}W_{\mathrm{T}}(q^{2},q\cdot p_{\beta})+Q^{\mu}Q^{\nu}W_{\mathrm{L}}(q^{2},q\cdot p_{\beta}).\displaybreak[0]
\end{align}
The dimensionless invariant functions $W_{\mathrm{T}}$ and $W_{\mathrm{L}}$ are simply connected with the cross section of transverse or longitudinal photon absorption $\sigma_{\mathrm{T}}$ and $\sigma_{\mathrm{L}}$, respectively:
\begin{align}\label{W.TL}
W_{\mathrm{T}}=&2\hat{p}_{\mathrm{CM}}\sqrt{\hat{s}}\sigma_{\mathrm{T}}(\gamma^{*}+\beta\rightarrow J/\psi+\beta),\displaybreak[0]\nonumber\\
W_{\mathrm{L}}=&2\hat{p}_{\mathrm{CM}}\sqrt{\hat{s}}\sigma_{\mathrm{L}}(\gamma^{*}+\beta\rightarrow J/\psi+\beta).\displaybreak[0]
\end{align}
Substituting Eqs.~(\ref{WEp}), (\ref{W.TL}) into (\ref{dab.Gen2}), we finally achieve the differential cross section as follows:
\begin{align}\label{dabTL}
&d\sigma(\alpha+\beta\rightarrow \alpha+J/\psi+\beta)\displaybreak[0]\nonumber\\
=&\frac{e_{\alpha}^{2}\alpha_{\mathrm{em}}}{2\pi^{2}Q^{2}}\frac{\rho^{\mu\nu}}{4\hat{p}_{\mathrm{CM}}\sqrt{\hat{s}}}
\left[R^{\mu\nu}W_{T}+Q^{\mu}Q^{\nu}W_{L}\right]\frac{\hat{p}_{\mathrm{CM}}\sqrt{\hat{s}}}{p_{\mathrm{CM}}\sqrt{s_{\alpha \beta}}}\frac{d^{3}p'_{\alpha}}{E'_{\alpha}}\displaybreak[0]\nonumber\\
=&\frac{e_{\alpha}^{2}\alpha_{\mathrm{em}}}{2\pi^{2}y}\left[\frac{y\rho^{++}}{Q^{2}}\sigma_{T}(\gamma^{*}+\beta\rightarrow J/\psi+\beta)+\frac{y\rho^{00}}{Q^{2}}\right.\displaybreak[0]\nonumber\\
&\left.\times \frac{\sigma_{L}}{2}(\gamma^{*}+\beta\rightarrow J/\psi+\beta)\right]\frac{\hat{p}_{\mathrm{CM}}\sqrt{\hat{s}}}{p_{\mathrm{CM}}\sqrt{s_{\alpha \beta}}}\frac{d^{3}p'_{\alpha}}{E'_{\alpha}}.\displaybreak[0]
\end{align}
Here, the coefficients $\rho^{ab}$ are the elements of the density matrix Eq.~(\ref{Rou.Gen}) in the $\gamma\beta$-helicity basis:
\begin{align}\label{Rouzz00}
\rho^{++}=&\frac{1}{2}\left[\frac{(2-y)^{2}}{y^{2}+Q^{2}m_{\beta}^{2}/(p_{\alpha}\cdot p_{\beta})^{2}}-\frac{4m_{\alpha}^{2}+Q^{2}}{Q^{2}}\right]F_{1}(Q^{2})\displaybreak[0]\nonumber\\
&+F_{2}(Q^{2}),\displaybreak[0]\nonumber\\
\rho^{00}=&\frac{(2-y)^{2}}{y^{2}+Q^{2}m_{\beta}^{2}/(p_{\alpha}\cdot p_{\beta})^{2}}F_{1}(Q^{2})-F_{2}(Q^{2}),\displaybreak[0]
\end{align}
where the relations $2\rho^{++}=\rho^{\mu\nu}R^{\mu\nu}$ and $\rho^{00}=\rho^{\mu\nu}Q^{\mu}Q^{\nu}$ are used.

We are now in a position to achieve the second part of exact treatment, we have to derive the detailed expressions of the form factors for each photon emission mechanism and for each interaction type.
We adopt the central idea of the Martin-Ryskin method~\cite{Martin:2014nqa} in the calculations, in which the coherent probability or weighting factor (WF) is described by the square of the electric proton form factor $G_{\mathrm{E}}^{2}(Q^{2})$, and the effect of the magnetic form factor is neglected.
$G_{\mathrm{E}}(Q^{2})$ can be parameterized by the dipole form: $G_{\mathrm{E}}(Q^{2})=1/(1+Q^{2}/0.71~\mathrm{GeV^{2}})^{2}$.
In the cases of coherent-photon emission or elastic process, the WF is: $w_{\mathrm{coh}}=w_{\mathrm{el}}=G_{\mathrm{E}}^{2}(Q^{2})$.
In the cases of incoherent-photon emission or inelastic process, the remaining probability, $1-w=1-G_{\mathrm{E}}^{2}(Q^{2})$, has to be considered to avoid double counting.

Therefore, in elastic photoproduction processes, the general notations $F_{1}(Q^{2})$ and $F_{2}(Q^{2})$ in Eq.~(\ref{Rouzz00}) can be written as:
\begin{align}\label{F12el}
F_{1~\mathrm{coh}}^{elastic}(Q^{2})&=F_{2~\mathrm{coh}}^{elastic}(Q^{2})=w_{\mathrm{el}}w_{\mathrm{coh}}=G_{\mathrm{E}}^{4}(Q^{2}),\displaybreak[0]\nonumber\\
F_{1~\mathrm{incoh}}^{elastic}(Q^{2})&=F_{2~\mathrm{incoh}}^{elastic}(Q^{2})=w_{\mathrm{el}}(1-w_{\mathrm{coh}})\displaybreak[0]\nonumber\\
&=G_{\mathrm{E}}^{2}(Q^{2})[1-G_{\mathrm{E}}^{2}(Q^{2})],\displaybreak[0]
\end{align}
and those for inelastic processes are changed accordingly:
\begin{align}\label{F12inel}
F_{1~\mathrm{coh}}^{inelastic}(Q^{2})&=F_{2~\mathrm{coh}}^{inelastic}(Q^{2})=(1-w_{\mathrm{el}})w_{\mathrm{coh}}\displaybreak[0]\nonumber\\
&=[1-G_{\mathrm{E}}^{2}(Q^{2})]G_{\mathrm{E}}^{2}(Q^{2}),\displaybreak[0]\nonumber\\
F_{1~\mathrm{incoh}}^{inelastic}(Q^{2})&=F_{2~\mathrm{incoh}}^{inelastic}(Q^{2})=(1-w_{\mathrm{el}})(1-w_{\mathrm{coh}})\displaybreak[0]\nonumber\\
&=[1-G_{\mathrm{E}}^{2}(Q^{2})]^{2}.\displaybreak[0]
\end{align}

\subsection{The $Q^{2}$ and $y$ distributions of $J/\psi$ production}
\label{subsec:Q2_y_dis.}

In the present section, we employ the accurate expression Eq.~(\ref{dabTL}) to give the $Q^{2}$ and $y$ dependent differential cross sections.
It is convenient to do the calculations in the rest frame of $\alpha$, where $|\mathbf{q}|=|\mathbf{p}_{\alpha'}|=r$, the photon virtuality $Q^{2}=-q^{2}=(p_{\alpha}-p_{\alpha'})^{2}=2m_{\alpha}(\sqrt{r^{2}+m_{\alpha}^{2}}-m_{\alpha})$, $d^{3}p_{\alpha}'=r^{2}drd\cos\theta d\varphi$, and $y=(q\cdot p_{\beta})/(p_{\alpha}\cdot p_{\beta})=(q_{0}-|\textbf{p}_{\beta}|r\cos\theta/E_{\beta})/m_{\alpha}$.
By doing the Jacobian transformation,
\begin{align}\label{Jac.Q2}
dydQ^{2}=\mathcal{J}d\cos\theta dr=\left|\frac{D(Q^{2},y)}{D(\cos\theta,r)}\right|d\cos\theta dr,\displaybreak[0]
\end{align}
the details are given in Appendix~\ref{FKR},
the differential cross section in Eq.~(\ref{dabTL}) can be cast into
\begin{align}\label{dabTL.Q2}
&\frac{d\sigma}{dydQ^{2}}(\alpha+\beta\rightarrow \alpha+J/\psi+\beta)\displaybreak[0]\nonumber\\
=&\frac{e_{\alpha}^{2}\alpha_{\mathrm{em}}}{2\pi}\left[\frac{y\rho^{++}}{Q^{2}}\sigma_{T}(\gamma^{*}+\beta\rightarrow J/\psi+\beta)+\frac{y\rho^{00}}{Q^{2}}\right.\displaybreak[0]\nonumber\\
&\left.\times \frac{\sigma_{L}}{2}(\gamma^{*}+\beta\rightarrow J/\psi+\beta)\right]f(s_{\alpha\beta},p_{\mathrm{CM}},\hat{s},\hat{p}_{\mathrm{CM}}),\displaybreak[0]
\end{align}
and
\begin{align}\label{f.kine}
&f(s_{\alpha\beta},p_{\mathrm{CM}},\hat{s},\hat{p}_{\mathrm{CM}})\displaybreak[0]\nonumber\\
=&\frac{\hat{p}_{\mathrm{CM}}\sqrt{\hat{s}}}{yp_{\mathrm{CM}}\sqrt{s_{\alpha\beta}}}
\frac{s_{\alpha\beta}-m_{\alpha}^{2}-m_{\beta}^{2}}{\sqrt{(s_{\alpha\beta}-m_{\alpha}^{2}-m_{\beta}^{2})^{2}-4m_{\alpha}^{2}m_{\beta}^{2}}}.\displaybreak[0]
\end{align}

Now we switch the general expression Eq.~(\ref{dabTL.Q2}) to each specific channel.
First of all, we discuss the case of elastic photoproduction processes.
The density matrix $\rho^{\mu\nu}$ in this case is the same as Eq.~(\ref{Rouzz00}), but $F_{1}(Q^{2})$ and $F_{2}(Q^{2})$ should be substituted by Eq.~(\ref{F12el}).
We adopt the Marquet-Peschanski-Soyez (MPS) model which predicts that the saturation scale becomes proportional to the squared momentum transfer $t$ and that nonperturbative contributions can be factorized when the scattering amplitude is described in the momentum transfer representation.
This model is recently updated by Xie and Gon\c{c}alves in Ref.~\cite{Xie:2022sjm}, through the fitting to the latest high precision HERA data.
In the color dipole formalism, the scattering amplitude for the $\gamma^{*}p\rightarrow Ep$ process can be expressed in the momentum representation as follows~\cite{Marquet:2007qa}:
\begin{align}\label{AmpTL}
&\mathcal{A}_{T,L}^{\gamma^{*}p\rightarrow Ep}(x,Q^{2},t=-\mathbf{q}^{2})\displaybreak[0]\nonumber\\
&=i\int d^{2}\mathbf{r}\int\frac{dz}{4\pi}(\psi_{E}^{*}\psi_{\gamma})_{T,L}(z,\mathbf{r},Q^{2})e^{-iz\mathbf{q}\cdot \mathbf{r}}\mathcal{T}(x,\mathbf{r},\mathbf{q}),\displaybreak[0]
\end{align}
where $\mathbf{q}$ is the momentum transfer, $x = (Q^2 + M^2)/(\hat{s}+Q^2)$, $r$ is the size of the $q\bar{q}$ dipole,  and $z$ and $(1-z)$ are the momentum fractions of the incoming photon momentum carried by the quark and anti-quark, respectively.
The overlap  functions $(\psi_E^*\psi_{\gamma})_{T,L}$  describe the fluctuation of a photon with transverse or longitudinal polarization into a color dipole and the subsequent formation of the final state $E$.
The modelling of the overlap function for the vector meson production is still a theme of debate, we will follow Ref.~\cite{Xie:2022sjm} to assume that the vector meson is predominantly a quark-antiquark state and that its spin and polarization structures are the same of the photon~\cite{Goncalves:2004bp,Kowalski:2003hm,Kowalski:2006hc}.
Such assumptions imply that the overlap functions are given by
\begin{align}\label{Ol}
(\Psi_V^*\Psi_{\gamma})_T =&\hat{e}_fe\frac{N_c}{\pi z(1-z)}\lbrace  m_f^2K_0(\epsilon r)\phi_T(r,z)\displaybreak[0]\nonumber\\
&-(z^2+(1-z)^2)\epsilon K_1(\epsilon r)\partial_r\phi_T(r,z)\rbrace,\notag\displaybreak[0]\\
(\Psi_V^*\Psi_{\gamma})_L =&\hat{e}_fe\frac{N_c}{\pi}2Qz(1-z)K_0(\epsilon r)\Bigg[M_V\phi_L(r, z)\displaybreak[0]\nonumber\\
&+\delta \frac{m_f^2-\nabla_r^2}{M_Vz(1-z)}\phi_L(r, z)\Bigg],\displaybreak[0]
\end{align}
where $\hat{e}_f$ is the effective charge of the quarks,  $N_c=3$ and $\phi_{T,L}(r,z)$ define the scalar part of the vector meson wave functions.
There are two popular models employed in the literature -- the Boosted Gaussian (BG)  and Light - Cone Gauss (LCG) -- which differ in the assumptions: $\delta$ and $\phi_{T,L}(r,z)$.
In the BG model, $\delta = 1$ and
\begin{align}\label{BG}
\phi_{T,L}(z,r)=&\mathcal{N}_{T,L}z(1-z)\exp\Big(-\frac{m_f^2\mathcal{R}^2}{8z(1-z)}\displaybreak[0]\nonumber\\
&-\frac{2z(1-z)r^2}{\mathcal{R}^2}+\frac{m_f^2\mathcal{R}^2}{2}\Big),\displaybreak[0]
\end{align}
where $\mathcal{N}_{T,L}$ and $\mathcal{R}$ are free parameters to be determined by the normalization condition of the wave function and by the decay width.
In the LCG model, $\delta = 0$ and
\begin{align}\label{LCG}
\phi_T(z,r) &=N_T [z(1-z)]^2 \exp\left(-\frac{r^2}{2R_T^2}\right),\displaybreak[0]\\
\phi_L(z,r) &=N_Lz(1-z)  \exp\left(-\frac{r^2}{2R_L^2}\right).\displaybreak[0]
\end{align}
In our work, we also follow the authors in Ref.~\cite{Xie:2022sjm} to employ the LCG model in the calculations, which has smaller $\chi^{2}$ for $J/\psi$ production.
The relevant parameters can be found in Table~I of Ref.~\cite{Xie:2022sjm}.
In MPS model, the scattering amplitude can be expressed as~\cite{Marquet:2007qa}:
\begin{align}\label{TMPS}
&\mathcal{T}_{\mathrm{MPS}}(x,\mathbf{r},\mathbf{b})=2\pi R_p^2 f(\mathbf{b})\displaybreak[0]\nonumber\\
\times&\!\!\begin{cases}
\mathcal{N}_0(\frac{rQ_s(x,\mathbf{b})}{2})^{2[\gamma_s+(1/\kappa\lambda Y)\ln(2/rQ_s(x,\mathbf{b}))]},~rQ_s(x,\mathbf{b})\leqslant 2\displaybreak[0]\\
1-\exp\big(-a\ln^2(b rQ_s(x,\mathbf{b}))\big),\quad\qquad rQ_s(x,\mathbf{b})>2,\displaybreak[0]
\end{cases}
\end{align}
where $\mathcal{N}_0=0.7$, $\kappa=9.9$, $Y=\ln(1/x)$, and the constants $a$ and $b$ are given by
\begin{align}\label{pab}
\begin{split}
&a=-\frac{\mathcal{N}^2_0\gamma_s^2}{(1-\mathcal{N}_0)^2\ln(1-\mathcal{N}_0)},\\
&b=\frac{1}{2}(1-\mathcal{N}_0)^{-(1-\mathcal{N}_0)/(2\mathcal{N}_0\gamma_s)}.
\end{split}
\end{align}
The MPS model is characterized by a $\mathbf{q}$-dependent saturation scale: $Q_s(x,\mathbf{b})=(x_0/x)^{\lambda/2}(1+c\mathbf{q}^2)$, and by the factorization of the form factor: $f(\mathbf{b}) = e^{-B\mathbf{q}^2}$, which describes the $\mathbf{q}$-dependence of the proton vertex.
The parameters $c$, $B$, $R_p$, $\gamma_s$, $\lambda$ and $x_0$ have been obtained by fitting the latest HERA data in Table~II of Ref.~\cite{Xie:2022sjm}.

Finally, the transverse and longitudinal cross sections for elastic $J/\psi$ photoproduction in Eq.~(\ref{dabTL.Q2}) can be calculated including the corrections associated to the real part of the amplitude and the skewedness factor, which is related to the fact that the gluons emitted from the quark and antiquark into the dipole can carry different momentum fractions \cite{Shuvaev:1999ce}.
As a consequence,
\begin{align}\label{dcsTL}
\frac{d\sigma_{T,L}^{\gamma^* p\to J/\psi p}}{dt}
=\frac{R_g^2(1+\beta^2)}{16\pi}|\mathcal{A}_{T,L}(x,Q^2,\mathbf{b})|^2,
\end{align}
where $\beta$ is the ratio of real and imaginary parts of the amplitude~\cite{Kowalski:2006hc}
\begin{align}\label{betarat}
\beta=\tan(\frac{\pi}{2}\lambda),~~\mbox{with}~~\lambda=\frac{\partial \ln (\mathrm{Im}\mathcal{A}_{T,L})}{\partial\ln1/x}.
\end{align}
Moreover, the skewedness factor $R_g^2$ is~\cite{Kowalski:2006hc}
\begin{align}\label{Rg}
R_g=\frac{2^{2\lambda+3}}{\sqrt{\pi}}\frac{\Gamma(\lambda+5/2)}{\Gamma(\lambda+4)},
\end{align}

Considering the different photon emission mechanisms, the elastic process has to further divide into the elastic-coherent and elastic-incoherent processes.
For elastic-coherent process, virtual photon is radiated coherently by the whole incident proton, and then interacts with the another incident proton which remains intact after scattering.
The corresponding cross section can be derived directly from Eqs. (\ref{dabTL.Q2}) and (\ref{dcsTL}), where both the projectile and target are the proton: $m_{\alpha}=m_{\beta}=m_{p}$.

For elastic-incoherent process, the virtual photon emitter is quarks inside incoming proton, the corresponding differential cross section is
\begin{align}\label{CS.elincoh}
&\frac{d\sigma^{elastic}_{\mathrm{incoh.}}}{dydQ^{2}}(p+p\rightarrow X_{A}+J/\psi+p)\displaybreak[0]\nonumber\\
=&2\sum_{a}\int dx_{a}f_{a/p}(x_{a},\mu^{2})\frac{d\sigma}{dydQ^{2}}(a+p\rightarrow a+J/\psi+p),\displaybreak[0]\nonumber\\
\end{align}
where $x_{a}=p_{a}/p_{A}$ is the parton's momentum fraction, $f_{a/p}(x_{a},\mu^{2})$ is the parton distribution function of proton A, and the factorized scale is chosen as $\mu=\sqrt{m_{J/\psi}^{2}+Q^{2}}$.
The factor of two in Eq.~(\ref{CS.elincoh}) arises because both protons emit photons and thus serve as targets.
The partonic cross section can also be obtained from Eqs. (\ref{dabTL.Q2}) and (\ref{dcsTL}) with $m_{\alpha}=m_{a}=0$, $m_{\beta}=m_{p}$.

Here we discuss the inelastic $J/\psi$ photoproduction.
In the initial state, the inelastic photoproduction process may be direct or resolved~\cite{Ma:2018zzq}.
In the inelastic direct photoproduction process, the high-energy photon, emitted from the projectile $\alpha$, interacts with the partons of target proton directly.
In the inelastic resolved photoproduction process, the uncertainty principle allows the high-energy hadron-like photon to fluctuate into a color singlet state with multiple $q\bar{q}$ pairs and gluons.
Due to this fluctuation, the photon interacts with the partons inside the target proton like a hadron, and the subprocesses are almost purely strong interaction processes.
We must keep in mind that the distinction between these two types of contributions does not really exist, only the sum of them has a physical meaning.
Actually, as always with photons, the situation is quite complex.
Together with the two different photon emissions mentioned earlier, we have four classes of inelastic processes: inelastic coherent-direct (coh.dir.), inelastic coherent-resolved (coh.res.), inelastic incoherent-direct (incoh.dir.), and inelastic incoherent-resolved (incoh.res.) processes.
These abbreviations will appear in many places of the remaining content, and we do not explain its meaning again.

In the case of inelastic coherent-direct process, the virtual photon emitted from the whole incident proton $A$ interacts with parton $b$ of target proton $B$, and the proton $A$ remains intact after photon emitted.
According to the NRQCD factorization formalism, the corresponding differential cross section can be derived with $m_{\alpha}=m_{p}$ and $m_{\beta}=m_{b}=0$,
\begin{align}\label{CSyQ2.cohdir}
&\frac{d\sigma^{inelastic}_{\mathrm{coh.dir.}}}{dydQ^{2}}(p+p\rightarrow p+J/\psi+X)\displaybreak[0]\nonumber\\
=&2\sum_{b}\int dx_{b}d\hat{t}f_{b/p}(x_{b},\mu_{b}^{2})\sum_{n}\langle\mathcal{O}^{J/\psi}[n]\rangle\displaybreak[0]\nonumber\\
&\times\frac{d\sigma}{dydQ^{2}d\hat{t}}(p+b\rightarrow p+c\bar{c}[n]+b),\displaybreak[0]
\end{align}
where $\langle\mathcal{O}^{J/\psi}[n]\rangle$ is the LDMEs of NRQCD.
$x_{b}=p_{b}/p_{B}$ is the momentum fraction of the proton, $f_{b/p}(x_{b},\mu^{2})$ is the parton distribution function of massless parton $b$ in proton $B$.
Based on Eq.~(\ref{dabTL.Q2}), the cross section of the subprocess $\alpha+b\rightarrow \alpha+c\bar{c}[n]+b$ should be rewritten as
\begin{align}\label{abTLinel}
&\frac{d\sigma}{dydQ^{2}d\hat{t}}(\alpha+b\rightarrow \alpha+c\bar{c}[n]+b)\displaybreak[0]\nonumber\\
=&\frac{e_{\alpha}^{2}\alpha_{\mathrm{em}}}{2\pi}F_{b}[n]\left[\frac{y\rho^{++}}{Q^{2}}T_{b}[n]-y\rho^{00}
L_{b}[n]\right]f(s,p_{\mathrm{CM}},\hat{s},\hat{p}_{\mathrm{CM}}),\displaybreak[0]\nonumber\\
\end{align}
where $F_{b}[n]$, $T_{b}[n]$ and $L_{b}[n]$ can be found in Ref.~\cite{Kniehl:2001tk}.

In the case of inelastic incoherent-direct process, the virtual photon emitted from the quarks $a$ inside proton $A$ interacts with parton $b$ of proton $B$, and $A$ is allowed to break up after photon emitted.
Similarly, the corresponding differential cross section has the form of
\begin{align}\label{CSyQ2.incohdir}
&\frac{d\sigma^{inelastic}_{\mathrm{incoh.dir.}}}{dydQ^{2}}(p+p\rightarrow X_{A}+J/\psi+X)\displaybreak[0]\nonumber\\
=&2\sum_{a,b}\int dx_{a}dx_{b}d\hat{t}f_{a/p}(x_{a},\mu_{a}^{2})f_{b/p}(x_{b},\mu_{b}^{2})\displaybreak[0]\nonumber\\
&\times\sum_{n}\langle\mathcal{O}^{J/\psi}[n]\rangle\frac{d\sigma}{dydQ^{2}d\hat{t}}(a+b\rightarrow a+c\bar{c}[n]+b),\displaybreak[0]
\end{align}
the differential cross section $d\sigma/dydQ^{2}(a+b\rightarrow a+c\bar{c}[n]+b)$ can be derived from Eq.~(\ref{abTLinel}) with $m_{\alpha}=m_{a}=0$ and $e_{\alpha}=e_{a}$, where $e_{a}$ is the charge of massless quark $a$.

In the inelastic coherent-resolved process, the parton $a'$ of hadron-like photon which emitted from proton $A$, interacts with the parton $b$ of proton $B$ via the interactions of quark-antiquark annihilation, quark-gluon Compton scattering and gluon-gluon fusion.
The relevant differential cross section reads
\begin{align}\label{CSyQ2.cohres}
&\frac{d\sigma^{inelastic}_{\mathrm{coh.res.}}}{dydQ^{2}}(p+p\rightarrow p+J/\psi+X)\displaybreak[0]\nonumber\\
=&2\sum_{b}\sum_{a'}\int dx_{b}dz_{a'}d\hat{t}f_{b/p}(x_{b},\mu_{b}^{2})f_{a'/\gamma}(z_{a'},\mu_{\gamma}^{2})\displaybreak[0]\nonumber\\
&\times\frac{\alpha_{\mathrm{em}}}{2\pi}\frac{y\rho^{++}}{Q^{2}}\sum_{n}\langle\mathcal{O}^{J/\psi}[n]\rangle\frac{d\sigma_{a'b\rightarrow c\bar{c}[n]d}}{d\hat{t}},\displaybreak[0]
\end{align}
where $f_{\gamma}(z_{a'},\mu^{2})$ is the parton distribution function of the resolved photon \cite{Gluck:1999ub}, $z_{a}'=p_{a'}/q$ denotes the parton's momentum fraction of the resolved photon radiated from the proton $A$.
The complete list for the partonic cross sections of $a'+b\rightarrow c\bar{c}[n]+b$ can be found in Refs. \cite{Klasen:2003zn}.

In the inelastic incoherent-resolved process, the quarks inside proton $A$ emit a hadron-like virtual photon, then the parton $a^\prime$ of this resolved photon interacts with parton $b$ inside proton $B$, and $A$ is break up after photon emitted.
The relevant differential cross section is
\begin{align}\label{CSyQ2.incohres}
&\frac{d\sigma^{inelastic}_{\mathrm{incoh.res.}}}{dydQ^{2}}(p+p\rightarrow X_{A}+J/\psi+X)\displaybreak[0]\nonumber\\
=&2\sum_{a,b}\sum_{a'}\int dx_{a}dx_{b}dz_{a'}d\hat{t}f_{a/p}(x_{a},\mu_{a}^{2})f_{b/p}(x_{b},\mu_{b}^{2})\displaybreak[0]\nonumber\\
&\times f_{a'/\gamma}(z_{a'},\mu_{\gamma}^{2})\frac{e_{a}^{2}\alpha_{\mathrm{em}}}{2\pi}\frac{y\rho^{++}}{Q^{2}}
\sum_{n}\langle\mathcal{O}^{J/\psi}[n]\rangle\frac{d\sigma_{a'b\rightarrow c\bar{c}[n]d}}{d\hat{t}}.\displaybreak[0]\nonumber\\
\end{align}

\subsection{The $p_{T}$, $y_{r}$, and $z$ distributions of $J/\psi$ production}
\label{pT.yr.z.distributions}

The $p_{T}$, rapidity $y_{r}$, and inelastic variable $z=(p_{J/\psi}\cdot p_{\beta})/(q\cdot p_{\beta})$ distributions can be obtained by using the Jacobian transformation.
It is needed to emphasize that one should add a term with the exchange $(y_r\rightarrow -y_r)$ in the formulae of $y_{r}$ distribution, which reflects the fact that each colliding proton can serve as a photon emitter and as a target.

Actually in the final state, the $J/\psi$ photoproduction can be divided into two categories: direct $J/\psi$ produced from the $\gamma$-$g$ fusion, annihilation, and Compton scattering of partons; fragmentation $J/\psi$ produced by the fragmentation process from the final state partons.
In the following we will take into account all of these aspects.

\subsubsection{Direct $J/\psi$ production}
\label{sec:D_Jpsi_PT_yr_z}

For deriving the $p_{T}$ and $y_{r}$ distributions, we have to reordering and redefining the integration variables, and the Mandelstam variables in $\gamma^{*}\beta$ CM frame should be written as
\begin{align}\label{Mant.yr}
\hat{s}=&2\cosh y_{r}m_{T}\sqrt{\cosh^{2}y_{r}m_{T}^{2}+m_{\beta}^{2}-M_{J/\psi}^{2}}\displaybreak[0]\nonumber\\
&+2\cosh^{2}y_{r}m_{T}^{2}+m_{\beta}^{2}-M_{J/\psi}^{2},\displaybreak[0]\nonumber\\
\hat{t}=&M_{J/\psi}^{2}-Q^{2}-2m_{T}(\hat{E}_{\gamma}\cosh y_{r}-\hat{p}_{\mathrm{CM}}\sinh y_{r}),\displaybreak[0]\nonumber\\
\hat{u}=&M_{J/\psi}^{2}+m_{\beta}^{2}-2m_{T}(\hat{E}_{\beta}\cosh y_{r}+\hat{p}_{\mathrm{CM}}\sinh y_{r}),\displaybreak[0]
\end{align}
where $y_{r}=(1/2)\ln(E+p_{z})/(E-p_{z})$, $m_{T}=\sqrt{M_{J/\psi}^{2}+p_{T}^{2}}$ is the $J/\psi$
transverse mass, $\hat{E}_{\gamma}$, $\hat{E}_{\beta}$ and $\hat{p}_{\mathrm{CM}}$ are the energies and momentum in $\gamma^{*}\beta$ CM frame, respectively.
The details are given in Appendix~\ref{FKR}.

In the case of direct photoproduction processes, the variables $x_{b}$ and $\hat{t}$ should be chosen to do the following transformation,
\begin{align}\label{Jac.dir.}
d\hat{t}dx_{b}=\mathcal{J}dy_{r}dp_{T}^{2}=\left|\frac{D(x_{b},\hat{t})}{D(y_{r},p_{T}^{2})}\right|dy_{r}dp_{T}^{2},\displaybreak[0]
\end{align}
and then the relevant cross sections for direct $J/\psi$ production can be written as,
\begin{align}
&\frac{d\sigma^{inelastic}_{\mathrm{coh.dir.}}}{dy_{r}dp_{T}}(p+p\rightarrow p+J/\psi+X)\displaybreak[0]\nonumber\\
=&2\sum_{b}\int 2p_{T}dQ^{2}dyf_{b/p}(x_{b},\mu_{b}^{2})\mathcal{J}\sum_{n}\langle\mathcal{O}^{J/\psi}[n]\rangle\nonumber\\
&\times\frac{d\sigma}{dQ^{2}dyd\hat{t}}(p+b\rightarrow p+c\bar{c}[n]+b),\label{dPT.coh.dir.}\displaybreak[0]\\
&\displaybreak[0]\nonumber\\
&\frac{d\sigma^{inelastic}_{\mathrm{incoh.dir.}}}{dy_{r}dp_{T}}(p+p\rightarrow X_{A}+J/\psi+X)\displaybreak[0]\nonumber\\
=&2\sum_{a,b}\int 2p_{T}dQ^{2}dydx_{a}f_{a/p}(x_{a},\mu_{a}^{2})f_{b/p}(x_{b},\mu_{b}^{2})\mathcal{J}\nonumber\\
&\times\sum_{n}\langle\mathcal{O}^{J/\psi}[n]\rangle\frac{d\sigma}{dQ^{2}dyd\hat{t}}(a+b\rightarrow a+c\bar{c}[n]+b).
\label{dPT.incoh.dir.}\displaybreak[0]
\end{align}

In the case of resolved contributions, we should choose the variables $\hat{t}_{\gamma}$ and $z_{a'}$ to do the similar transformation,
\begin{align}\label{Jac.res.}
d\hat{t}_{\gamma}dz_{a'}=\mathcal{J}dy_{r}dp_{T}^{2}=
\left|\frac{D(z_{a'},\hat{t}_{\gamma})}{D(y_{r},p_{T}^{2})}\right|dy_{r}dp_{T}^{2},\displaybreak[0]
\end{align}
the corresponding differential cross sections are
\begin{align}
&\frac{d\sigma^{inelastic}_{\mathrm{coh.res.}}}{dy_{r}dp_{T}}(p+p\rightarrow p+J/\psi+X)\displaybreak[0]\nonumber\\
=&2\sum_{b}\sum_{a'}\int 2p_{T}dQ^{2}dydx_{b}f_{b/p}(x_{b},\mu_{b}^{2})f_{\gamma}(z_{a'},\mu_{\gamma}^{2})\mathcal{J}\displaybreak[0]\nonumber\\
&\times \frac{\alpha_{\mathrm{em}}}{2\pi}\frac{y\rho^{++}}{Q^{2}}
\sum_{n}\langle\mathcal{O}^{J/\psi}[n]\rangle\frac{d\sigma_{a'b\rightarrow c\bar{c}[n]d}}{d\hat{t}},\label{dPT.coh.res.}\displaybreak[0]\\
&\displaybreak[0]\nonumber\\
&\frac{d\sigma^{inelastic}_{\mathrm{incoh.res.}}}{dy_{r}dp_{T}}(p+p\rightarrow X_{A}+J/\psi+X)\displaybreak[0]\nonumber\\
=&2\sum_{a,b}\sum_{a'}\int 2p_{T}dQ^{2}dydx_{a}dx_{b}f_{a/p}(x_{a},\mu_{a}^{2})f_{b/p}(x_{b},\mu_{b}^{2})\displaybreak[0]\nonumber\\
&\times f_{\gamma}(z_{a'},\mu_{\gamma}^{2})\mathcal{J}e_{a}^{2}\frac{\alpha_{\mathrm{em}}}{2\pi}\frac{y\rho^{++}}{Q^{2}}
\sum_{n}\langle\mathcal{O}^{J/\psi}[n]\rangle\frac{d\sigma_{a'b\rightarrow c\bar{c}[n]d}}{d\hat{t}},\label{dPT.incoh.res.}\displaybreak[0]
\end{align}
where the Mandelstam variables of resolved photoproduction are the same as Eq.~(\ref{Mant.yr}), but with $Q^{2}=0$.

For the $z$ distribution, we can do the similar transformation and the Mandelstam variables in Eq.~(\ref{Mant.yr}) should be rewritten as:
\begin{align}\label{Mant.z}
&\hat{s}=\frac{M_{J/\psi}^{2}}{z}+\frac{p_{T}^{2}}{z(1-z)},\displaybreak[0]\nonumber\\
&\hat{t}=-(1-z)(\hat{s}+Q^{2}-m_{\beta}^{2}),\displaybreak[0]\nonumber\\
&\hat{u}=M_{J/\psi}^{2}+m_{\beta}^{2}-z(\hat{s}+Q^{2}-m_{\beta}^{2}).\displaybreak[0]
\end{align}
We do not list here the relevant cross sections again.

\subsubsection{Fragmentation $J/\psi$ production}
\label{sec:F_Jpsi_PT_yr_z}

The fragmentation $J/\psi$ production is also an important channel which involves a nonperturbative part described by the $J/\psi$ fragmentation function,
$D_{c\rightarrow c\bar{c}[n]}(z_{c},Q^{2})$, its detailed expression was calculated in Refs.~\cite{Ma:1995vi, Ma:2013yla}.
$z_{c}=2p_{T}\cosh y_{r}/\sqrt{\hat{s}}$ is the momentum fraction of the final state $J/\psi$.
First of all, we should rewrite the Mandelstam variables as follows
\begin{align}\label{Mant.coh.dir.frag.}
&\hat{s}=yx_{b}(s-2m_{p}^{2})-Q^{2},\displaybreak[0]\nonumber\\
&\hat{t}=-Q^{2}-\frac{\hat{s}}{2\cosh y_{r}}e^{-y_{r}}+\frac{Q^{2}}{2\cosh y_{r}}e^{y_{r}},\displaybreak[0]\nonumber\\
&\hat{u}=-(\hat{s}+Q^{2})\frac{e^{y_{r}}}{2\cosh y_{r}}.\displaybreak[0]
\end{align}
Then the variables $z_{c}$ and $\hat{t}$ should be chosen to do the following transformation
\begin{align}\label{Jac.frag}
d\hat{t}dz_{c}=\mathcal{J}dy_{r}dp_{T}^{2}=\left|\frac{D(z_{c},\hat{t})}{D(y_{r},p_{T}^{2})}\right|dy_{r}dp_{T}^{2}.\displaybreak[0]
\end{align}

For direct photoproduction processes, the corresponding cross sections of fragmentation $J/\psi$ production are
\begin{align}
&\frac{d\sigma^{inelastic}_{\mathrm{coh.dir.-frag.}}}{dy_{r}dp_{T}}(p+p\rightarrow p+J/\psi+X)\displaybreak[0]\nonumber\\
=&2\sum_{b,c}\int 2p_{T}dQ^{2}dydx_{b}f_{b/p}(x_{b},\mu_{b}^{2})\frac{\mathcal{J}}{z_{c}}\sum_{n}\langle\mathcal{O}^{J/\psi}[n]\rangle\displaybreak[0]\nonumber\\
&\times D_{c\rightarrow Q\bar{Q}[n]}(z_{c},Q^{2})\frac{d\sigma}{dQ^{2}dyd\hat{t}}(p+b\rightarrow p+c+d),\label{ddPT.coh.dir.frag}\displaybreak[0]\\
&\displaybreak[0]\nonumber\\
&\frac{d\sigma^{inelastic}_{\mathrm{incoh.dir.-frag.}}}{dy_{r}dp_{T}}(p+p\rightarrow X_{A}+J/\psi+X)\displaybreak[0]\nonumber\\
=&2\sum_{a,b,c}\int 2p_{T}dQ^{2}dydx_{a}dx_{b}f_{a/p}(x_{a},\mu_{a}^{2})\displaybreak[0]\nonumber\\
&\times f_{b/p}(x_{b},\mu_{b}^{2})\frac{\mathcal{J}}{z_{c}}\sum_{n}\langle\mathcal{O}^{J/\psi}[n]\rangle D_{c\rightarrow Q\bar{Q}[n]}(z_{c},Q^{2})\displaybreak[0]\nonumber\\
&\times \frac{d\sigma}{dQ^{2}dyd\hat{t}}(a+b\rightarrow a+c+d),\label{dPT.incoh.dir.frag.}\displaybreak[0]
\end{align}
where the partonic subprocesses involved here are $q\gamma^{*}\rightarrow q\gamma$, $q\gamma^{*}\rightarrow qg$ and $g\gamma^{*}\rightarrow q\bar{q}$~\cite{Ma:2019mwr}.

For resolved contributions, the differential cross sections are
\begin{align}
&\frac{d\sigma^{inelastic}_{\mathrm{coh.res.-frag.}}}{dy_{r}dp_{T}}(p+p\rightarrow p+J/\psi+X)\displaybreak[0]\nonumber\\
=&2\sum_{b}\sum_{a',c}\int 2p_{T}dQ^{2}dydx_{b}dz_{a'}f_{b/p}(x_{b},\mu_{b}^{2})f_{\gamma}(z_{a'},\mu_{\gamma}^{2})
\frac{\mathcal{J}}{z_{c}}\displaybreak[0]\nonumber\\
&\times\frac{\alpha_{\mathrm{em}}}{2\pi}\frac{y\rho^{++}}{Q^{2}}\sum_{n}\langle\mathcal{O}^{J/\psi}[n]\rangle D_{c\rightarrow Q\bar{Q}[n]}(z_{c},Q^{2})
\frac{d\sigma_{a'b\rightarrow cd}}{d\hat{t}},\label{ddPT.coh.res.frag}\displaybreak[0]\\
\displaybreak[0]\nonumber\\
&\frac{d\sigma^{inelastic}_{\mathrm{incoh.res.-frag.}}}{dy_{r}dp_{T}}(p+p\rightarrow X_{A}+J/\psi+X)\displaybreak[0]\nonumber\\
=&2\sum_{a,b}\sum_{a',c}\int 2p_{T}dQ^{2}dydx_{a}dx_{b}dz_{a'}f_{a/p}(x_{a},\mu_{a}^{2})\displaybreak[0]\nonumber\\
&\times f_{b/p}(x_{b},\mu_{b}^{2})f_{\gamma}(z_{a'},\mu_{\gamma}^{2})\frac{\mathcal{J}}{z_{c}}
\frac{e_{a}^{2}\alpha_{\mathrm{em}}}{2\pi}\frac{y\rho^{++}}{Q^{2}}\displaybreak[0]\nonumber\\
&\times \sum_{n}\langle\mathcal{O}^{J/\psi}[n]\rangle D_{c\rightarrow Q\bar{Q}[n]}(z_{c},Q^{2})\frac{d\sigma_{a'b\rightarrow cd}}{d\hat{t}},
\displaybreak[0]\label{ddPT.incoh.res.frag}\displaybreak[0]
\end{align}
where the involved subprocesses are $qq\rightarrow qq$, $qq'\rightarrow qq'$,
$q\bar{q}\rightarrow q\bar{q}$, $q\bar{q}\rightarrow q'\bar{q}'$, $q\bar{q}'\rightarrow q\bar{q}'$,
$qg\rightarrow q\gamma$, $qg\rightarrow qg$ and $gg\rightarrow q\bar{q}$~\cite{Owens:1986mp}.
The Mandelstam variables of resolved contributions are the same as Eq.~(\ref{Mant.coh.dir.frag.}), but with $Q^{2}=0$.

\section{Weizs\"{a}cker-Williams approximation}
\label{WWA}

The connection between the processes in Fig.~\ref{fig:feyn_ab}(a) and (b) can help us to understand the central idea of Weizs\"{a}cker-Williams approximation.
By replacing the electromagnetic fields from a fast-moving charged particle to an equivalent flux of photon, in which the number of photons with energy $\omega$, $n(\omega)$, is given by the Fourier transform of the time-dependent electromagnetic field.
The photoproduction process in Fig.~\ref{fig:feyn_ab}(a) can be factorized into the real photo-absorption cross section [Fig.~\ref{fig:feyn_ab}(b)] and the photon spectrum.
This idea was first raised by Fermi~\cite{Fermi:1924tc}, and was developed by Weizs\"{a}cker and Williams to describe the high energy interactions, and the method is known as the Weizs\"{a}cker-Williams approximation (WWA)~\cite{vonWeizsacker:1934nji}.
An essential advantage of WWA consists in the fact that, when using it, it is sufficient to obtain the photo-absorption cross section on the mass shell only.
Details of its off mass-shell behavior are not essential.
Thus, the WWA, as a useful technique, has been widely applied to obtain various cross sections for charged particle productions~\cite{Budnev:1974de}.
Although the great success has been achieved, the properties of WWA in $J/\psi$ photoproduction are rarely noticed, and a number of imprecise statements pertaining to the essence of WWA were given~\cite{Drees:1989vq, Drees:1988pp, Frixione:1993yw, Zhu:2015via, Zhu:2015qoz, Fu:2011zzm, Fu:2011zzf, Chin.Phys.C_36_721, Yu:2015kva, Yu:2017rfi, Yu:2017pot, Nystrand:2004vn, Kniehl:2001tk, Kniehl:1990iv, Fu:2012xm, sp}.
And also the serious double counting exists when different charged sources are considered simultaneously~\cite{Zhu:2015qoz, Fu:2011zzm, Fu:2011zzf, Chin.Phys.C_36_721, Yu:2015kva, Yu:2017rfi, Yu:2017pot}.

The exact treatment developed in Section~\ref{Exac} can reduce to the WWA when $Q^{2}\rightarrow 0$.
This provides us an overall approach to compare our results with the WWA ones.
In the present section we switch the accurate expression Eq.~(\ref{dabTL}) into the WWA form and discuss a number of widely employed photon spectra.
There are two simplifications should be performed: the longitudinal photon contribution $\sigma_{\mathrm{L}}$ should be neglected;
the term of $\sigma_{\mathrm{T}}$ is substituted by its on-shell value.
Taking $Q^{2}\rightarrow0$, Eq.~(\ref{dabTL}) switches to:
\begin{align}\label{dWWA.Gen.}
&\lim_{Q^{2}\rightarrow0}\frac{d\sigma}{d\hat{t}}(\alpha+\beta\rightarrow\alpha+J/\psi+\beta)\displaybreak[0]\nonumber\\
=&\left(e_{\alpha}^{2}\frac{\alpha_{\mathrm{em}}}{2\pi^{2}}\frac{y\rho^{++}}{Q^{2}}\frac{d^{3}p'_{\alpha}}{E'_{\alpha}}\right)
\frac{\hat{p}_{\mathrm{CM}}\sqrt{\hat{s}}}{yp_{\mathrm{CM}}\sqrt{s_{\alpha b}}}\frac{d\sigma_{T}}{d\hat{t}}\bigg|_{Q^{2}=0}\displaybreak[0]\nonumber\\
=&\left[\frac{e_{\alpha}^{2}\alpha_{\mathrm{em}}}{2\pi}(y\rho^{++})\frac{dydQ^{2}}{Q^{2}}\right]
\frac{d\sigma_{T}}{d\hat{t}}f(s_{\alpha\beta},p_{\mathrm{CM}},\hat{s},\hat{p}_{\mathrm{CM}})\bigg|_{Q^{2}=0}\displaybreak[0]\nonumber\\
=&dn_{\gamma}\frac{d\sigma_{T}}{d\hat{t}}f(s_{\alpha\beta},p_{\mathrm{CM}},\hat{s},\hat{p}_{\mathrm{CM}})\bigg|_{Q^{2}=0},\displaybreak[0]
\end{align}
we can see that the contribution of $d\sigma_{L}/d\hat{t}$ and the terms proportional to $Q^{2}$ are neglected.
Therefore, the general form of the photon spectrum $f_{\gamma}(y)$, which is associated with various particles, reads
\begin{align}\label{fgamma.Gen.}
&f_{\gamma}(y)=\frac{dn_{\gamma}}{dy}=\frac{e_{\alpha}^{2}\alpha_{\mathrm{em}}}{2\pi}\int\frac{dQ^{2}}{Q^{2}}y\rho^{++}\displaybreak[0]\nonumber\\
=&\frac{e_{\alpha}^{2}\alpha_{\mathrm{em}}}{2\pi}\int\frac{dQ^{2}}{Q^{2}}\left\{yF_{2}(Q^{2})+\left[\frac{2(1-y)}{y}
-\frac{2ym_{\alpha}^{2}}{Q^{2}}\right]F_{1}(Q^{2})\right\}.\displaybreak[0]\nonumber\\
\end{align}
Actually, the origin of various widely employed photon spectra is another plane wave form, which is given in Ref.~\cite{Budnev:1974de} and can be written as follows
\begin{align}\label{fgamma.Gen.V}
dn_{\gamma}=&\frac{e_{\alpha}^{2}\alpha_{\mathrm{em}}}{\pi}\frac{dy}{y}\frac{dQ^{2}}{Q^{2}}\displaybreak[0]\nonumber\\
\times&\left[(1-y)\frac{Q^{2}-Q^{2}_{\mathrm{min}}}{Q^{2}}F_{1}(Q^{2})+\frac{y^{2}}{2}F_{2}(Q^{2})\right],\displaybreak[0]
\end{align}
this form is derived from Eq.~(\ref{fgamma.Gen.}) by assuming that $Q^{2}_{\mathrm{min}}=y^{2}m_{\alpha}^{2}/(1-y)$, which is the LO term of the following complete expression in the expansion of $\mathcal{O}(m_{\alpha}^{2})$,
\begin{align}\label{Q2lim.}
Q^{2}_{\mathrm{min}}=&-2m_{\alpha}^{2}+\frac{1}{2s_{\alpha\beta}}\bigg[(s_{\alpha\beta}+m_{\alpha}^{2})(s_{\alpha\beta}-\hat{s}+m_{p}^{2})\displaybreak[0]\nonumber\\
&\left.-(s_{\alpha\beta}-m_{\alpha}^{2})\sqrt{(s_{\alpha\beta}-\hat{s}+m_{\alpha}^{2})^{2}-4s_{\alpha\beta}m_{\alpha}^{2}}\right].\displaybreak[0]\nonumber\\
\end{align}
This approximation is only valuable when $m_{\alpha}^{2}\ll1~\mathrm{GeV}^2$, however $m_{p}^{2}\approx 0.88~\mathrm{GeV}^{2}$ does not satisfy this condition, and causes about $10\%$ errors in various spectra.

In the case of coherent-photon emission of proton, Drees and Zeppenfeld (DZ) provided a approximate form of Eq.~(\ref{fgamma.Gen.V})~\cite{Drees:1988pp}, this photon flux is widely employed in the literature~\cite{Zhu:2015qoz,Yu:2015kva,Yu:2017rfi,Yu:2017pot,Fu:2011zzf}.
By taking $Q^{2}_{\mathrm{max}}\rightarrow\infty$, setting $Q^{2}-Q^{2}_{\mathrm{min}}\approx Q^{2}$, and $F_{1}(Q^{2})=F_{2}(Q^{2})=G_{\mathrm{E}}^{2}(Q^{2})$ (which equals to neglect the effect of magnetic form factor), they obtained
\begin{align}\label{fgamma.DZ.}
&f_{\gamma}^{\mathrm{DZ}}(y)\displaybreak[0]\nonumber\\
=&\frac{\alpha_{\mathrm{em}}}{2\pi}\frac{1+(1-y)^{2}}{y}\left(\ln A-\frac{11}{6}+\frac{3}{A}-\frac{3}{2A^{2}}+\frac{1}{3A^{2}}\right),\displaybreak[0]\nonumber\\
\end{align}
where $A=(1+0.71~\mathrm{GeV^{2}}/Q^{2}_{\mathrm{min}})$.

In the case of incoherent-photon emission, a widely employed photon spectrum was derived from Eq.~(\ref{fgamma.Gen.V}) in Ref.~\cite{Drees:1994zx}, by neglecting the WF in Eqs.~(\ref{F12el}) and (\ref{F12inel}), and setting $Q_{\mathrm{min}}^{2}=1~\mathrm{GeV}^{2}$ and $Q^{2}_{\mathrm{max}}=\hat{s}/4$ ~\cite{Zhu:2015qoz,Yu:2015kva,Yu:2017rfi,Yu:2017pot,Fu:2011zzf,Fu:2012xm,Fu:2011zzm},
\begin{align}\label{fgamma.incohI.}
f_{\gamma}^{\mathrm{incoh}}(y)=e_{\alpha}^{2}\frac{\alpha_{\mathrm{em}}}{2\pi}\frac{1+(1-y)^{2}}{y}\ln\frac{Q_{\mathrm{max}}^{2}}{Q_{\mathrm{min}}^{2}}.\displaybreak[0]
\end{align}
Another important form of incoherent photon spectrum is given by Brodsky, Kinoshita and Terazawa in Ref.~\cite{Brodsky:1971ud}, which is originally derived for $ep$ scattering, and is directly expanded to describe the probability of finding a photon in any relativistic fermion,
\begin{align}\label{fgamma.incohBKT}
&f_{\gamma}^{\mathrm{BKT}}(y)\displaybreak[0]\nonumber\\
=&\frac{e_{\alpha}^{2}\alpha_{\mathrm{em}}}{\pi}\Bigg\{\frac{1+(1-y)^{2}}{y}\left(\ln\frac{E}{m}-\frac{1}{2}\right)\displaybreak[0]\nonumber\\
&+\frac{y}{2}\left[\ln(\frac{2}{y}-2)+1\right]+\frac{(2-y)^{2}}{2y}\ln(\frac{2-2y}{2-y})\Bigg\}.\displaybreak[0]
\end{align}

\section{Numerical results}
\label{Numerical results}

\begin{figure*}[htbp]
\setlength{\abovecaptionskip}{1mm}
  \centering
  \includegraphics[width=0.36\textwidth]{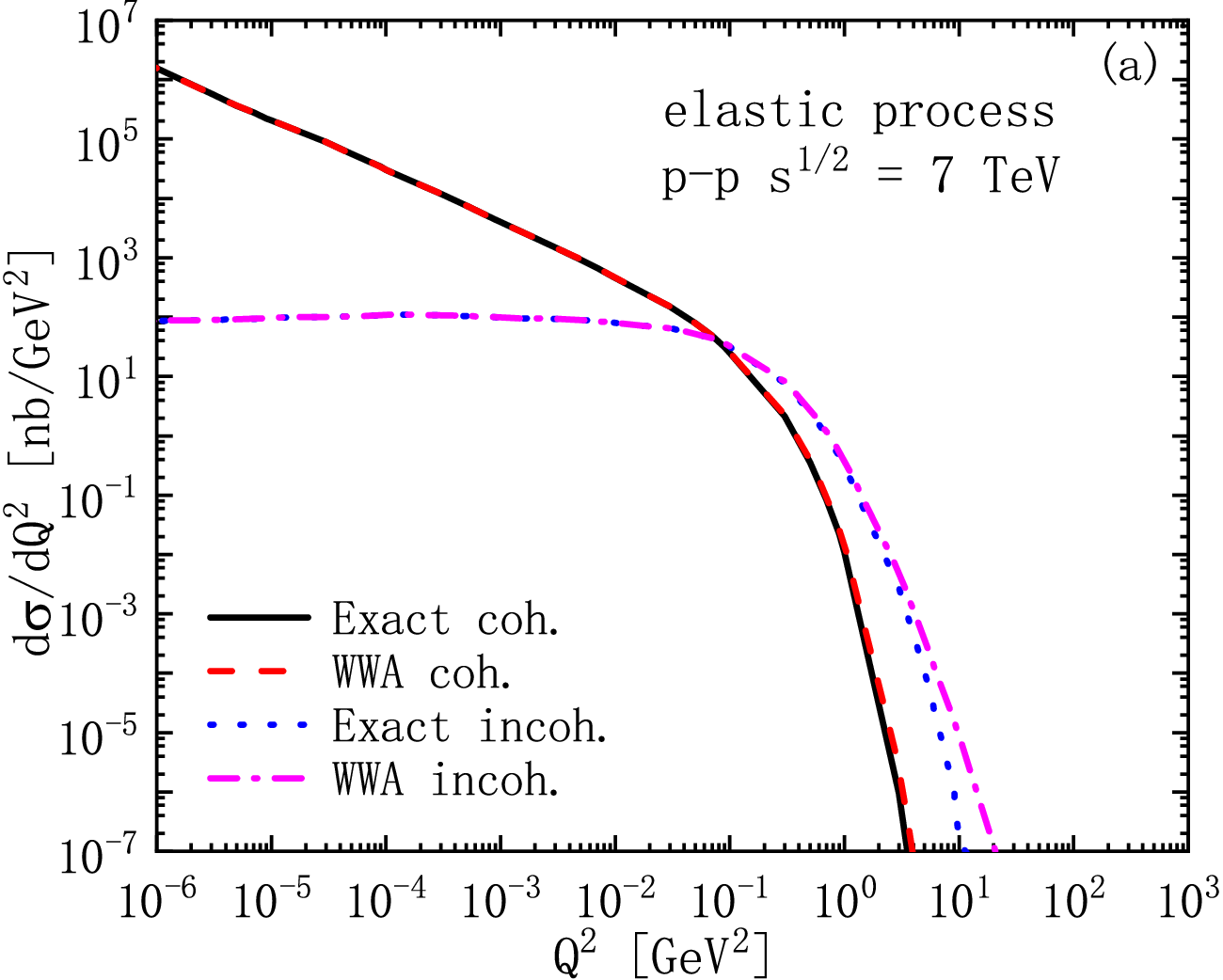}
  \includegraphics[width=0.36\textwidth]{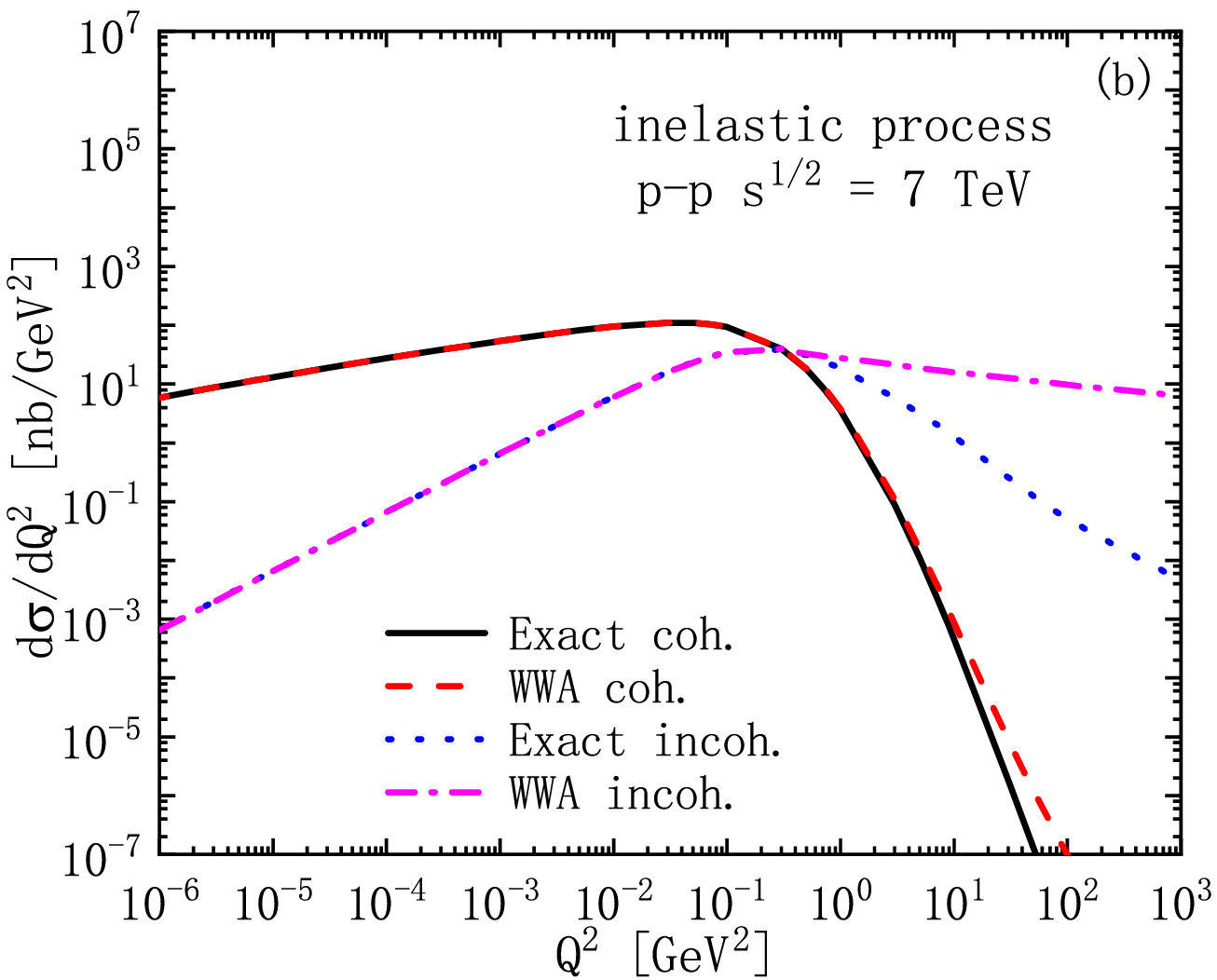}\\
    \includegraphics[width=0.36\textwidth]{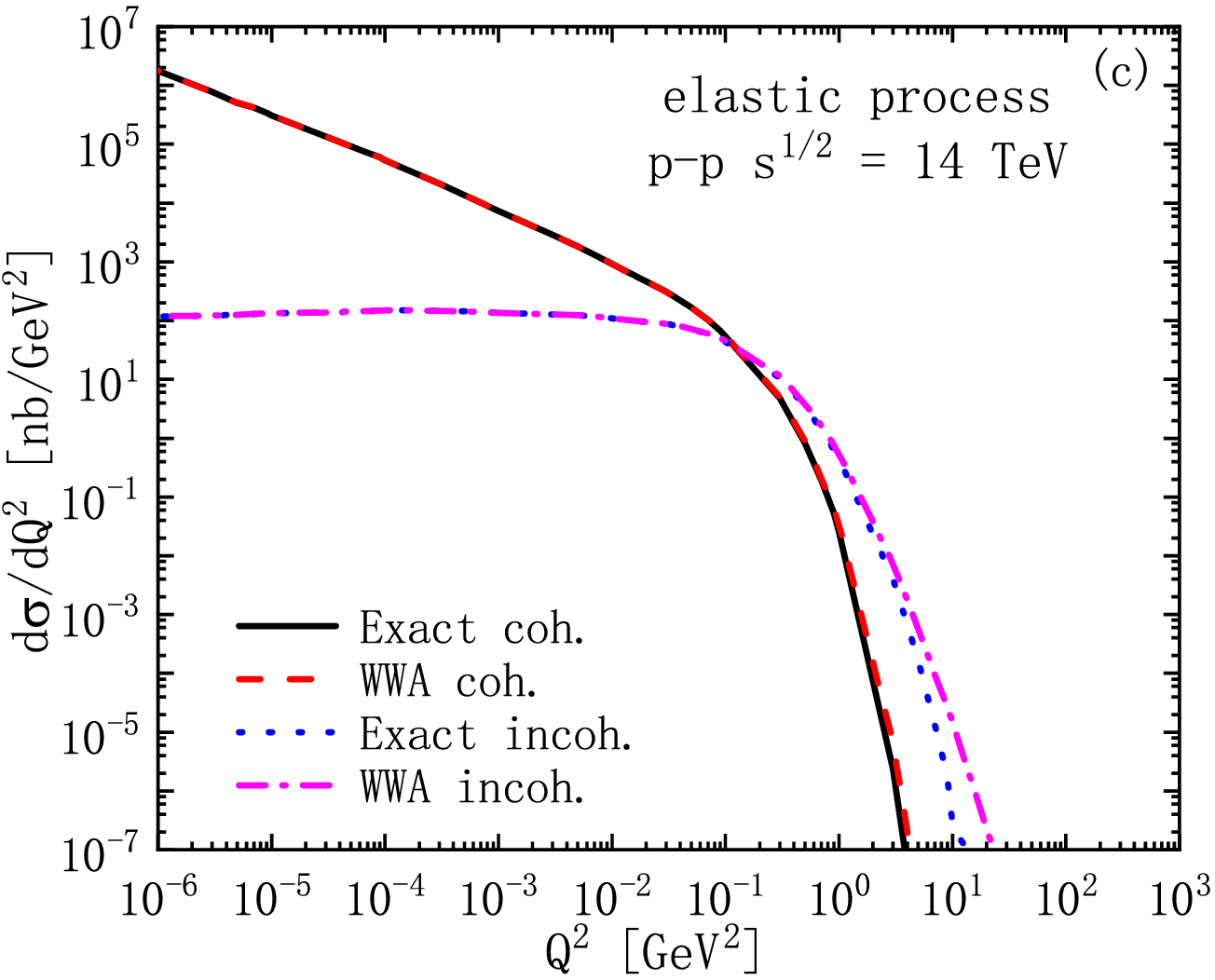}
    \includegraphics[width=0.36\textwidth]{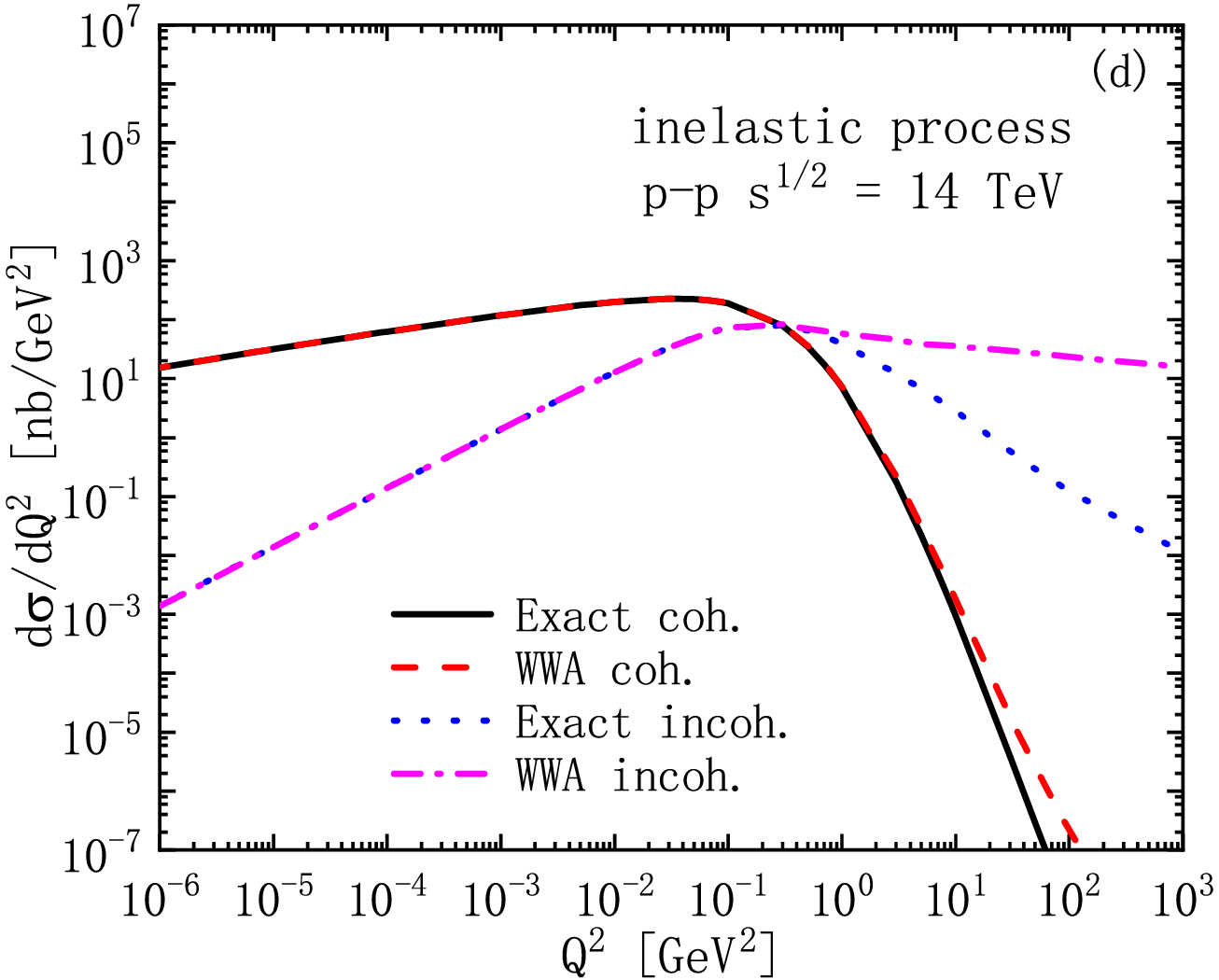}
  \caption{The $Q^{2}$ distribution of $J/\psi$ photoproduction in p-p collisions at LHC energies.
  The left panels show the results of $Q^{2}$ dependent differential cross sections for elastic photoproduction processes at different energies.
  The right panels show the corresponding results for inelastic photoproduction processes.
  The black solid and red dashed lines represent the exact results and the WWA ones for coherent-photon emission [coh.(dir.+res.)], respectively.
  The blue dotted and magenta dot-dashed lines represent the exact results and the WWA ones for incoherent-photon emission [incoh.(dir.+res.)], respectively.}
  \label{fig:Q2}
\end{figure*}

In this section, we provide the numerical results for the $J/\psi$ photoproduction in $p$-$p$ collisions at LHC energies.
There are several theoretical inputs, and the kinematic conditions need to be provided.
The mass of proton and $J/\psi$ are $m_{p}=0.938~\mathrm{GeV}$, $m_{J/\psi}=3.097~\mathrm{GeV}$ \cite{Agashe:2014kda}.
The LDMEs of NRQCD used in this paper are~\cite{Sharma:2012dy},
\begin{align}\label{LDMEs.S.}
&\langle\mathcal{O}^{J/\psi}[^{3}S_{1}^{(1)}]\rangle=1.2 \ \mathrm{GeV}^{3},\displaybreak[0]\nonumber\\
&\langle\mathcal{O}^{J/\psi}[^{1}S_{0}^{(8)}]\rangle=1.8 \times 10^{-2} \ \mathrm{GeV}^{3},\displaybreak[0]\nonumber\\
&\langle\mathcal{O}^{J/\psi}[^{3}S_{1}^{(8)}]\rangle=1.3 \times 10^{-3} \ \mathrm{GeV}^{3},\displaybreak[0]
\end{align}
and the multiplicity relations
\begin{align}\label{LDMEs.P.}
&&\langle\mathcal{O}^{J/\psi}[^{3}P_{0}^{(8)}]\rangle=m_{\mathrm{charm}}^{2} \langle\mathcal{O}^{J/\psi}[^{1}S_{0}^{(8)}]\rangle,\displaybreak[0]\nonumber\\
&&\langle\mathcal{O}^{J/\psi}[^{3}P_{J}^{(8)}]\rangle=(2J+1)\langle\mathcal{O}^{J/\psi}[^{3}P_{0}^{(8)}]\rangle,\displaybreak[0]
\end{align}
are adopted.
The strong coupling constant is taken as the one-loop form~\cite{Ma:2015ykd}
\begin{align}\label{alfas}
&&\alpha_{s}=\frac{12 \pi}{(33-2n_{f})\ln(\mu^{2}/\Lambda^{2})},\displaybreak[0]
\end{align}
with $n_{f}=3$ and $\Lambda=0.2~\mathrm{GeV}$.
Furthermore, we choose the MSHT20 set for the parton distribution function of proton with $n_{f}=3$~\cite{Bailey:2020ooq}.
The complete kinematical relations and the boundaries of involved variables are summarized in Appendix~\ref{FKR}.
Finally, the differential cross section for the LO initial parton hard scattering (had.scat. and had.scat.-frag.) satisfies the following forms
\begin{align}\label{hard.scat.}
&d\sigma_{\mathrm{had.scat.}}\displaybreak[0]\nonumber\\
=&\sum_{a,b}\int dx_{a}dx_{b}f_{a/p}(x_{a},\mu_{a}^{2})f_{b/p}(x_{b},\mu_{b}^{2})\displaybreak[0]\nonumber\\
&\times\sum_{n}\langle\mathcal{O}^{J/\psi}[n]\rangle d\sigma_{ab\rightarrow c\bar{c}[n]d},\displaybreak[0]\\
\displaybreak[0]\nonumber\\
&d\sigma_{\mathrm{had.scat.-frag.}}\displaybreak[0]\nonumber\\
=&\sum_{a,b,c}\int dx_{a}dx_{b}dz_{c}f_{a/p}(x_{a},\mu_{a}^{2})f_{b/p}(x_{b},\mu_{b}^{2})\displaybreak[0]\nonumber\\
&\times\sum_{n}\langle\mathcal{O}^{J/\psi}[n]\rangle \frac{D_{c\rightarrow Q\bar{Q}[n]}(z_{c},Q^{2})}{z_{c}}d\sigma_{ab\rightarrow cd},\displaybreak[0]
\end{align}
where the partonic cross sections are given in Ref.~\cite{Owens:1986mp}.

\begin{figure*}[htbp]
\setlength{\abovecaptionskip}{1mm}
  \centering
  \includegraphics[width=0.36\textwidth]{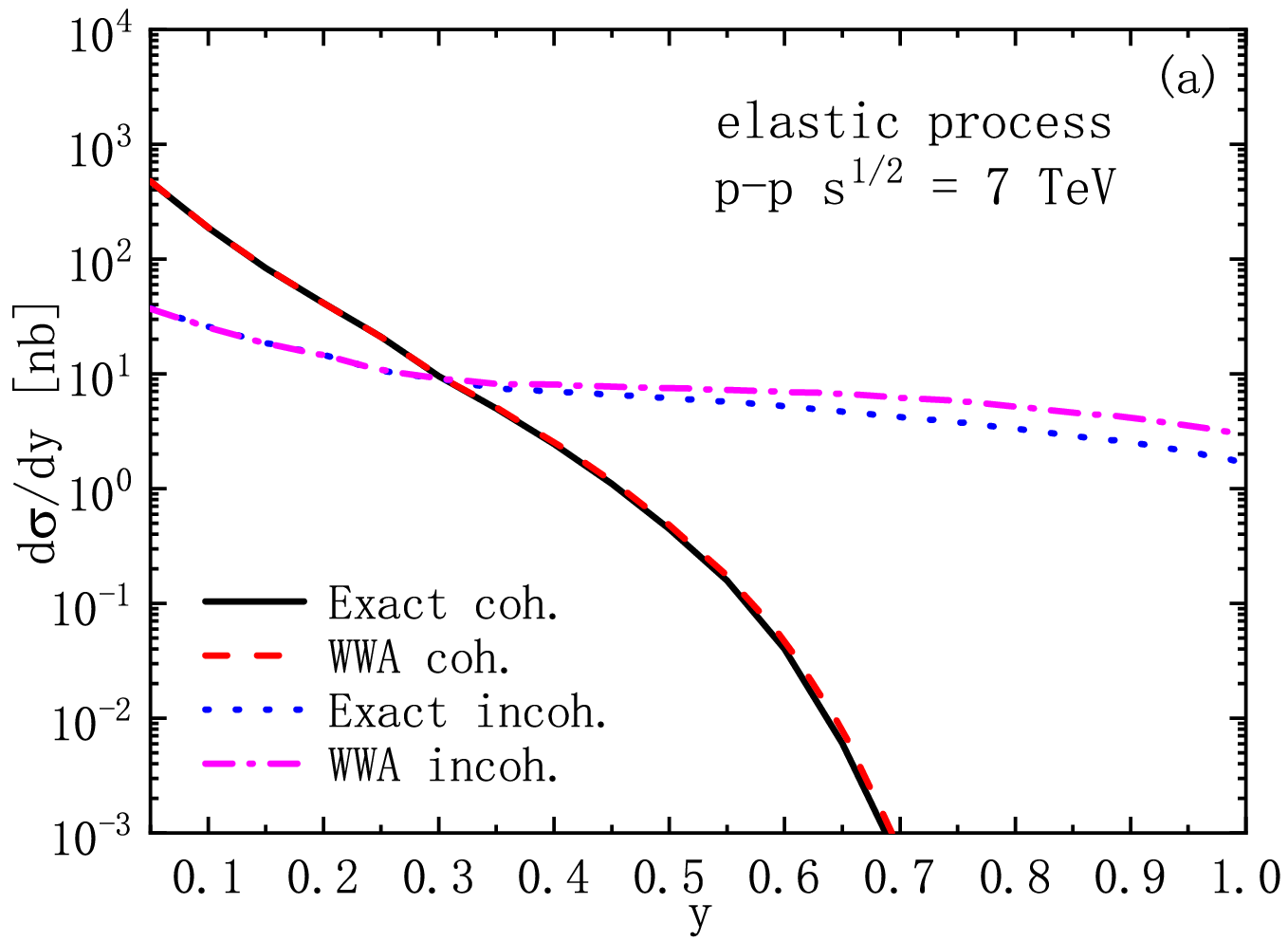}
  \includegraphics[width=0.36\textwidth]{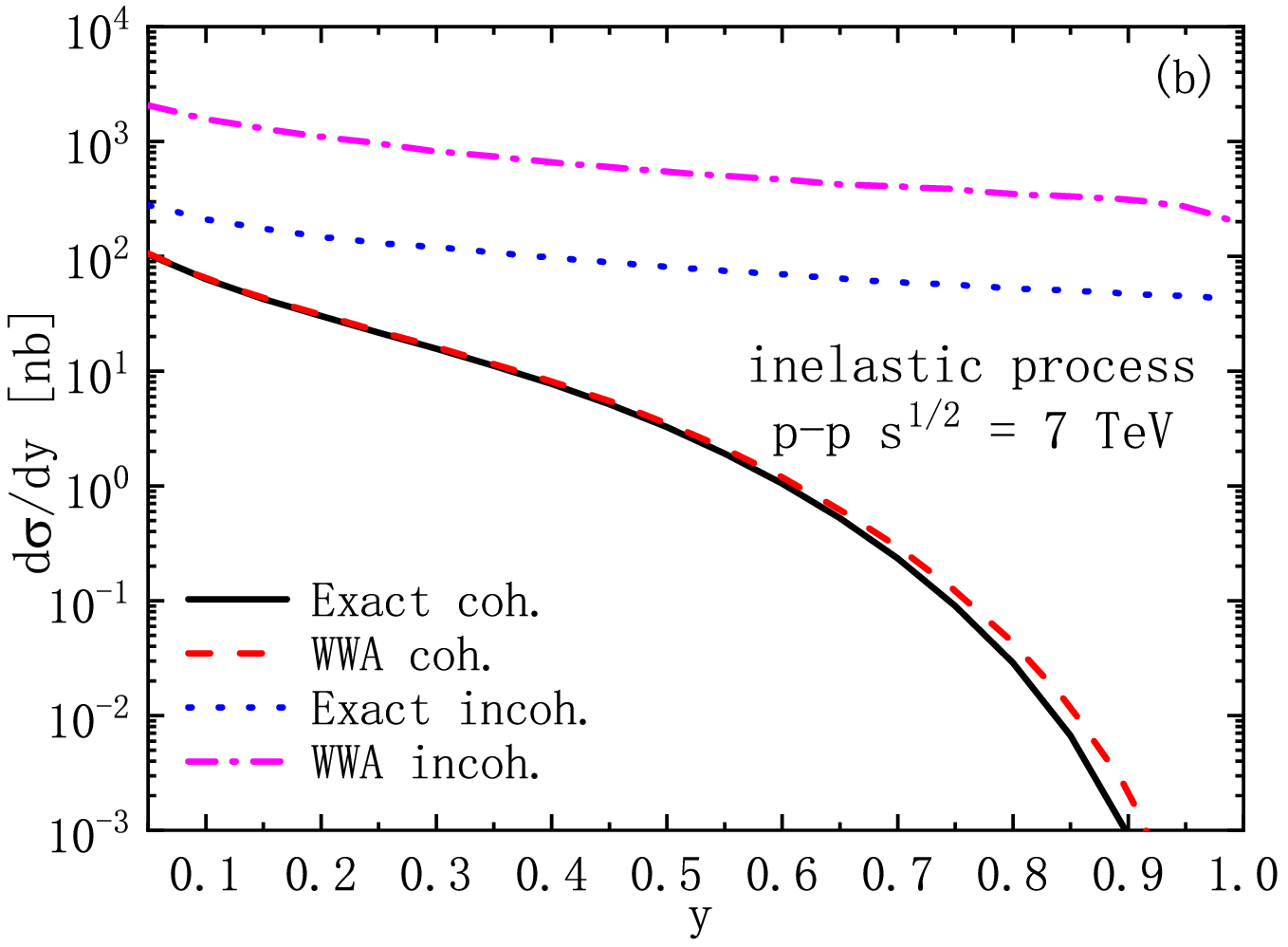}\\
    \includegraphics[width=0.36\textwidth]{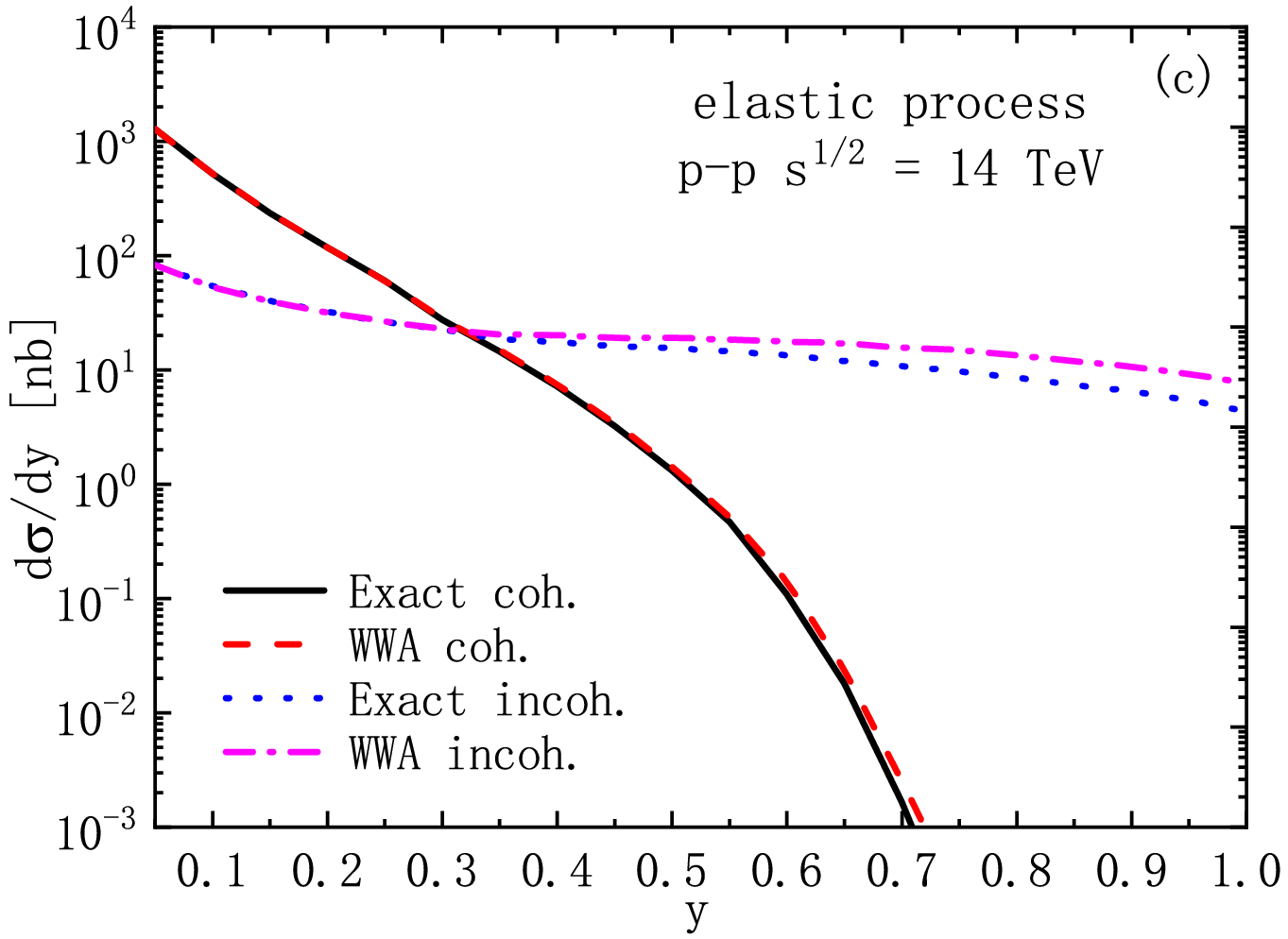}
    \includegraphics[width=0.36\textwidth]{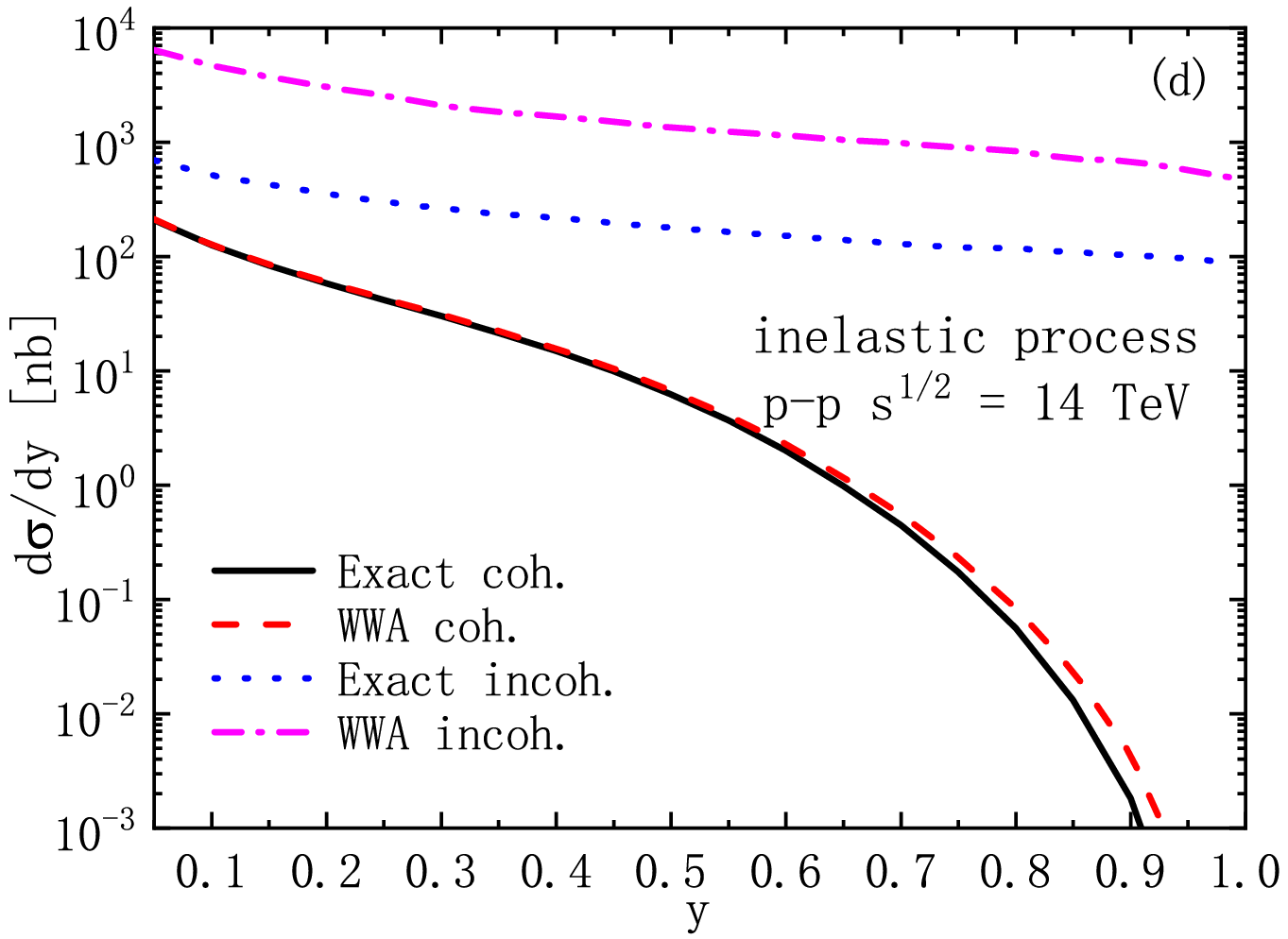}
  \caption{Same as Fig. \ref{fig:Q2} but for $y$ distribution.}
  \label{fig:y}
\end{figure*}

In Fig.~\ref{fig:Q2}, the $Q^{2}$ distribution of $J/\psi$ photoproduction in $p$-$p$ collisions at LHC energies is plotted.
The left panels show the results of $Q^{2}$ dependent differential cross sections for elastic photoproduction processes at different energies;
while the right panels show the corresponding results for inelastic processes.
The curves of WWA and exact results share the same trend, and are consistent with each other in the small $Q^{2}$ region,
this reflects one of the advantages of exact treatment that it recovers WWA in the limit $Q^{2}\rightarrow 0$.
The differences between WWA and exact results appear when $Q^{2}>0.3~\mathrm{GeV}^{2}$ and become evident at large values of $Q^{2}$.
These differences become more and more obvious from elastic to inelastic processes, and also from coherent to incoherent-photon emissions.
Therefore, WWA is only applicable in the small $Q^{2}$ domain, and its error appears when $Q^{2}>0.3~\mathrm{GeV}^{2}$.

We find that the contributions of elastic and inelastic photoproduction processes dominate the small and large $Q^{2}$ regions, respectively.
Furthermore, the contributions of coherent and incoherent-photon emissions also concentrate on the small and large $Q^{2}$ regions, respectively.
They become comparable around $Q^{2}=0.3~\mathrm{GeV}^{2}$.
Comparing with the WWA feature derived from early discussion, we can deduce that WWA can be employed for elastic photoproduction processes, and also for coherent-photon emission.
Especially in the case of elastic-coherent processes [Fig.~\ref{fig:Q2} (a), (c)], the exact results almost have no difference compared to the WWA ones.
Since the elastic-coherent processes rapidly decreased with increasing $Q^{2}$, which effectively avoid WWA errors.
However, WWA is not a good approximation for inelastic photoproduction processes, and also for incoherent-photon emission.
Especially in the case of inelastic-incoherent processes [Fig.~\ref{fig:Q2} (b), (d)], the WWA is inapplicable, in which the deviations between the WWA results and the exact ones become largest.
That is, the WWA results are about $100$ times larger than the exact ones at $Q^{2}=10^{2}~\mathrm{GeV}^{2}$.
This is because the inelastic-incoherent processes concentrate on the large $Q^{2}$ domain, where the WWA errors are largest.

\begin{table}[htbp]\footnotesize
\begin{threeparttable}
\renewcommand\arraystretch{1.2}
\centering
\caption{\label{Total.CS.coh.7TeV}Total cross sections of the $J/\psi$ photoproduction in the coherent-photon emission [coh.(dir.+res.)] at $\sqrt{s}=7~\mathrm{TeV}$.}
\begin{tabular}{L{2.6cm}C{1.2cm}C{1.2cm}m{0cm}C{1.3cm}C{1.3cm}}
\hline
\hline
\multirow{2}{*}{\makecell*[c]{Coherent \\ $s^{\frac{1}{2}}=7~\mathrm{TeV}$}} &
\multicolumn{2}{c}{elastic} & &
\multicolumn{2}{c}{inelastic}\\
\cline{2-3}\cline{5-6}
    & $\sigma~[\mathrm{nb}]$ & $\delta\tnote{a}~[$\%$]$ & & $\sigma~[\mathrm{nb}]$ & $\delta~[$\%$]$\\
    \hline
    Exact                                         & 39.47      & 0.0        && 3.64    & 0.0    \\
    WWA ($Q^{2}_{\mathrm{max}}\backsim \hat{s}$)  & 39.75      & 0.7        && 9.74    & 167.4  \\
    WWA $(y_{\mathrm{max}}=1)$                    & 98.42      & 149.4      && 14.35   & 294.2  \\
    WWA No WF                                     & 50.28      & 27.4       && 43.81   & 1103.3 \\
\hline
\hline
\end{tabular}
\begin{tablenotes}
\scriptsize
\item[a] Relative error with respect to the exact result: $\delta=\sigma/\sigma_{\mathrm{Exact}}-1$.
\end{tablenotes}
\end{threeparttable}
\end{table}

\begin{table}[htbp]\footnotesize
\begin{threeparttable}
\renewcommand\arraystretch{1.2}
\centering
\caption{\label{Total.CS.coh.14TeV}Same as Table~\ref{Total.CS.coh.7TeV}, but at $\sqrt{s}=14~\mathrm{TeV}$.}
\begin{tabular}{L{2.6cm}C{1.2cm}C{1.2cm}m{0cm}C{1.3cm}C{1.3cm}}
\hline
\hline
\multirow{2}{*}{\makecell*[c]{Coherent \\ $s^{\frac{1}{2}}=14~\mathrm{TeV}$}} &
\multicolumn{2}{c}{elastic} & &
\multicolumn{2}{c}{inelastic}\\
\cline{2-3}\cline{5-6}
    & $\sigma~[\mathrm{nb}]$ & $\delta~[$\%$]$ & & $\sigma~[\mathrm{nb}]$ & $\delta~[$\%$]$\\
    \hline
    Exact                                         & 84.26      & 0.0        && 6.01    & 0.0    \\
    WWA ($Q^{2}_{\mathrm{max}}\backsim \hat{s}$)  & 84.84      & 0.7        && 19.54   & 225.0  \\
    WWA $(y_{\mathrm{max}}=1)$                    & 187.63     & 122.7      && 28.16   & 368.3  \\
    WWA No WF                                     & 110.36     & 31.0       && 91.81   & 1427.1 \\
\hline
\hline
\end{tabular}
\end{threeparttable}
\end{table}

In Fig.~\ref{fig:y}, the results are expressed vs $y$.
The coherent-photon emission are important when $y<0.5$ and rapidly deceased with $y$ increasing.
Inversely, the contribution of incoherent-photon emission is important in the whole $y$ regions and much higher than those of coherent ones in right panels.
The WWA results nicely agree with the exact ones when $y<0.35$, but the differences appear with increasing $y$.
Especially, when $y>0.7$ the differences become evident (at $y=0.35$, the WWA results deviate from the exact ones by about $2.6\%$ and $9.7\%$ for coherent and incoherent-photon emissions, respectively; at $y=0.7$ the deviations are about $29.1\%$ and $45.6\%$, respectively).
Therefore, WWA can be a good approximation in the small $y$ region, although its error is evident at large values of $y$.
One exception is the case of inelastic-incoherent process, where the WWA results are about an order of magnitude (OOM) larger than the exact ones in the whole $y$ regions.
This verifies again the inapplicability of WWA in inelastic-incoherent process.

\begin{table}[htbp]\footnotesize
\begin{threeparttable}
\renewcommand\arraystretch{1.2}
\centering
\caption{\label{Total.CS.incoh.7TeV}Total cross sections of the $J/\psi$ photoproduction in the incoherent-photon emission [incoh.(dir.+res.)] at $\sqrt{s}=7~\mathrm{TeV}$.}
\begin{tabular}{L{2.6cm}C{1.2cm}C{1.2cm}m{0cm}C{1.3cm}C{1.3cm}}
\hline
\hline
\multirow{2}{*}{\makecell*[c]{Incoherent \\ $s^{\frac{1}{2}}=7~\mathrm{TeV}$}} &
\multicolumn{2}{c}{elastic} & &
\multicolumn{2}{c}{inelastic}\\
\cline{2-3}\cline{5-6}
    & $\sigma~[\mathrm{nb}]$ & $\delta ~[$\%$]$ & & $\sigma~[\mathrm{nb}]$ & $\delta~[$\%$]$\\
    \hline
    Exact       & 8.51      & 0.0           && 33.83     & 0.0    \\
    WWA         & 8.63      & 1.3           && 235.20    & 595.2  \\
    WWA No WF   & 211.45    & 2383.8        && 389.98    & 1052.6 \\
    WWA No WF $(Q^{2}_{\mathrm{min}}\backsim 1~\mathrm{GeV^{2}})$
                & 121.28    & 1324.6        && 220.82    & 552.7  \\
\hline
\hline
\end{tabular}
\end{threeparttable}
\end{table}

\begin{table}[htbp]\footnotesize
\begin{threeparttable}
\renewcommand\arraystretch{1.2}
\centering
\caption{\label{Total.CS.incoh.14TeV}Same as Table~\ref{Total.CS.incoh.7TeV}, but at $\sqrt{s}=14~\mathrm{TeV}$.}
\begin{tabular}{L{2.6cm}C{1.2cm}C{1.2cm}m{0cm}C{1.3cm}C{1.3cm}}
\hline
\hline
\multirow{2}{*}{\makecell*[c]{Incoherent \\ $s^{\frac{1}{2}}=14~\mathrm{TeV}$}} &
\multicolumn{2}{c}{elastic} & &
\multicolumn{2}{c}{inelastic}\\
\cline{2-3}\cline{5-6}
    & $\sigma~[\mathrm{nb}]$ & $\delta ~[$\%$]$ & & $\sigma~[\mathrm{nb}]$ & $\delta~[$\%$]$\\
    \hline
    Exact       & 19.01     & 0.0           && 67.96    & 0.0     \\
    WWA         & 19.47     & 2.4           && 479.09   & 605.5   \\
    WWA No WF   & 492.46    & 2490.5        && 860.33   & 1166.1  \\
    WWA No WF $(Q^{2}_{\mathrm{min}}\backsim 1~\mathrm{GeV^{2}})$
                & 296.86    & 1461.6        && 517.83   & 762.5   \\
\hline
\hline
\end{tabular}
\end{threeparttable}
\end{table}

\begin{figure*}[htbp]
\setlength{\abovecaptionskip}{1mm}
  \centering
  \includegraphics[width=0.3\textwidth]{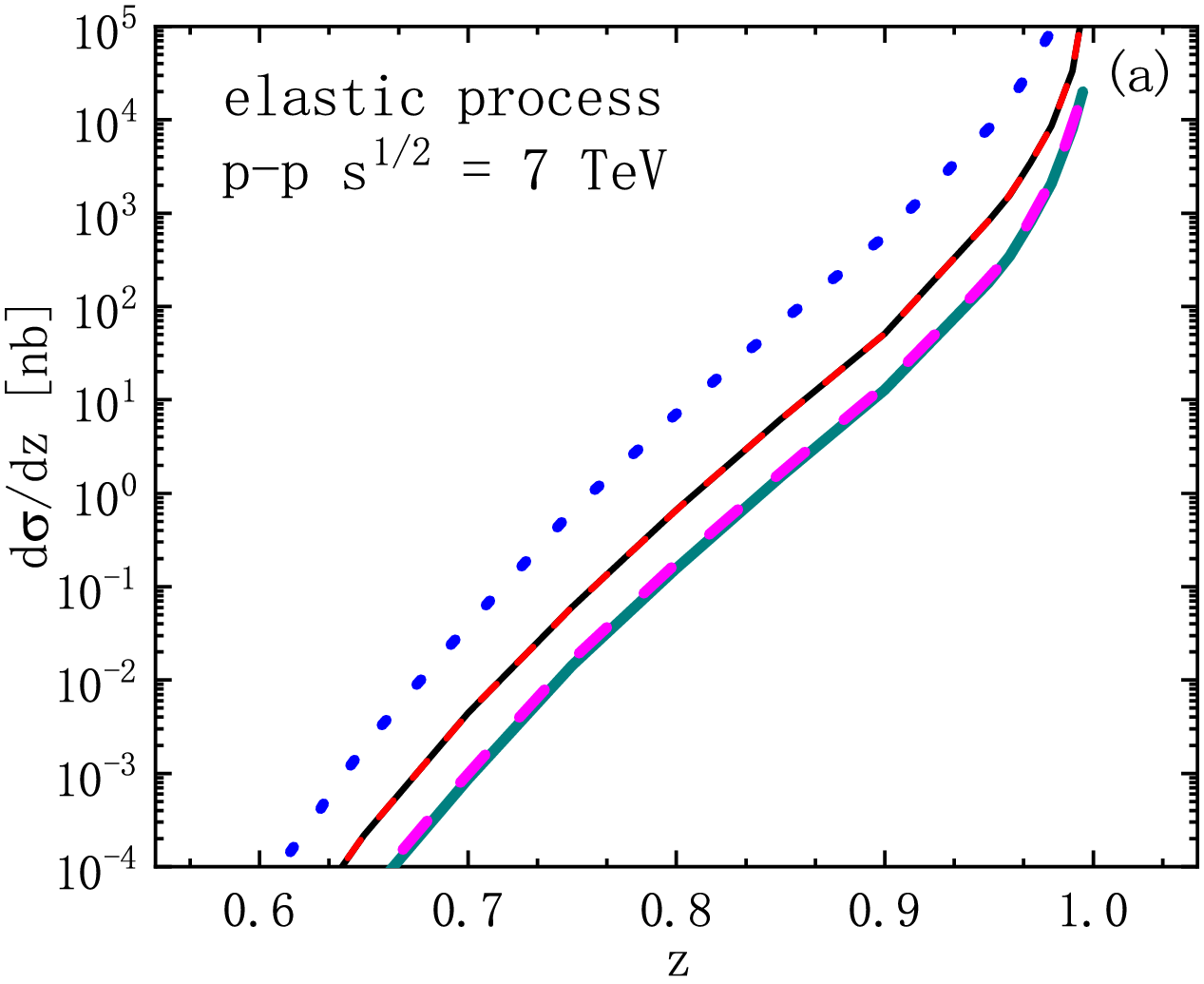}
  \includegraphics[width=0.3\textwidth]{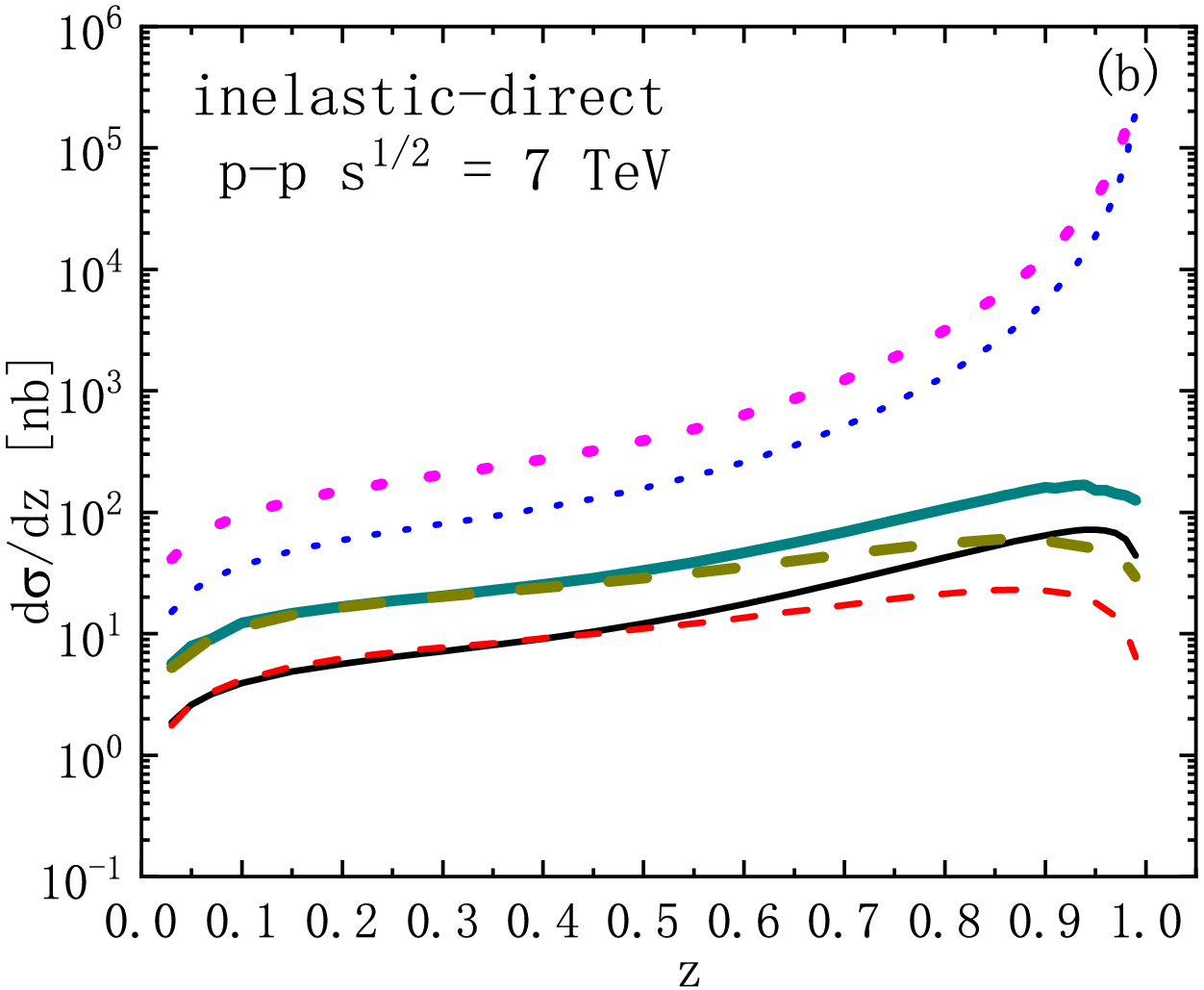}
  \includegraphics[width=0.3\textwidth]{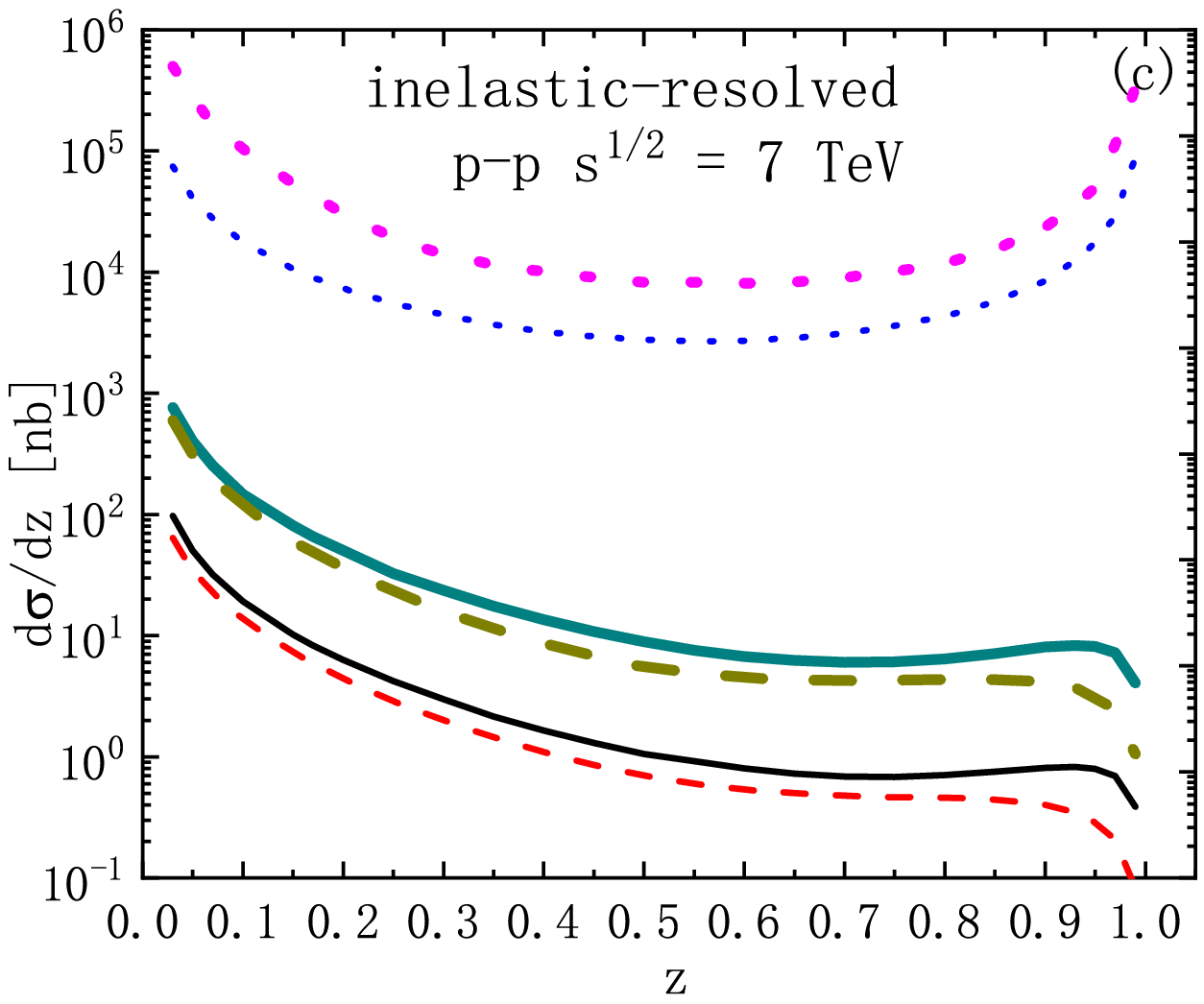}\\
    \includegraphics[width=0.3\textwidth]{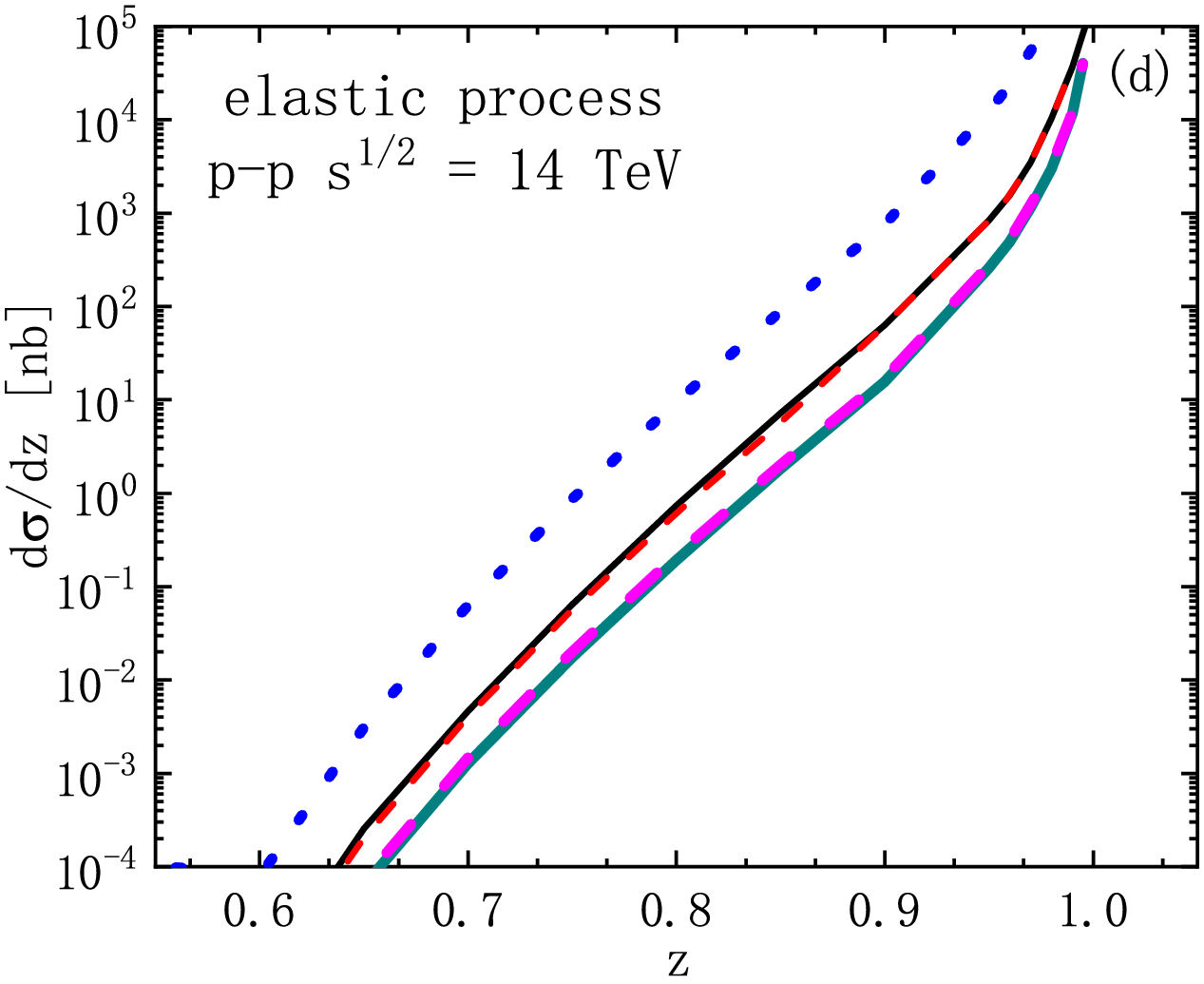}
    \includegraphics[width=0.3\textwidth]{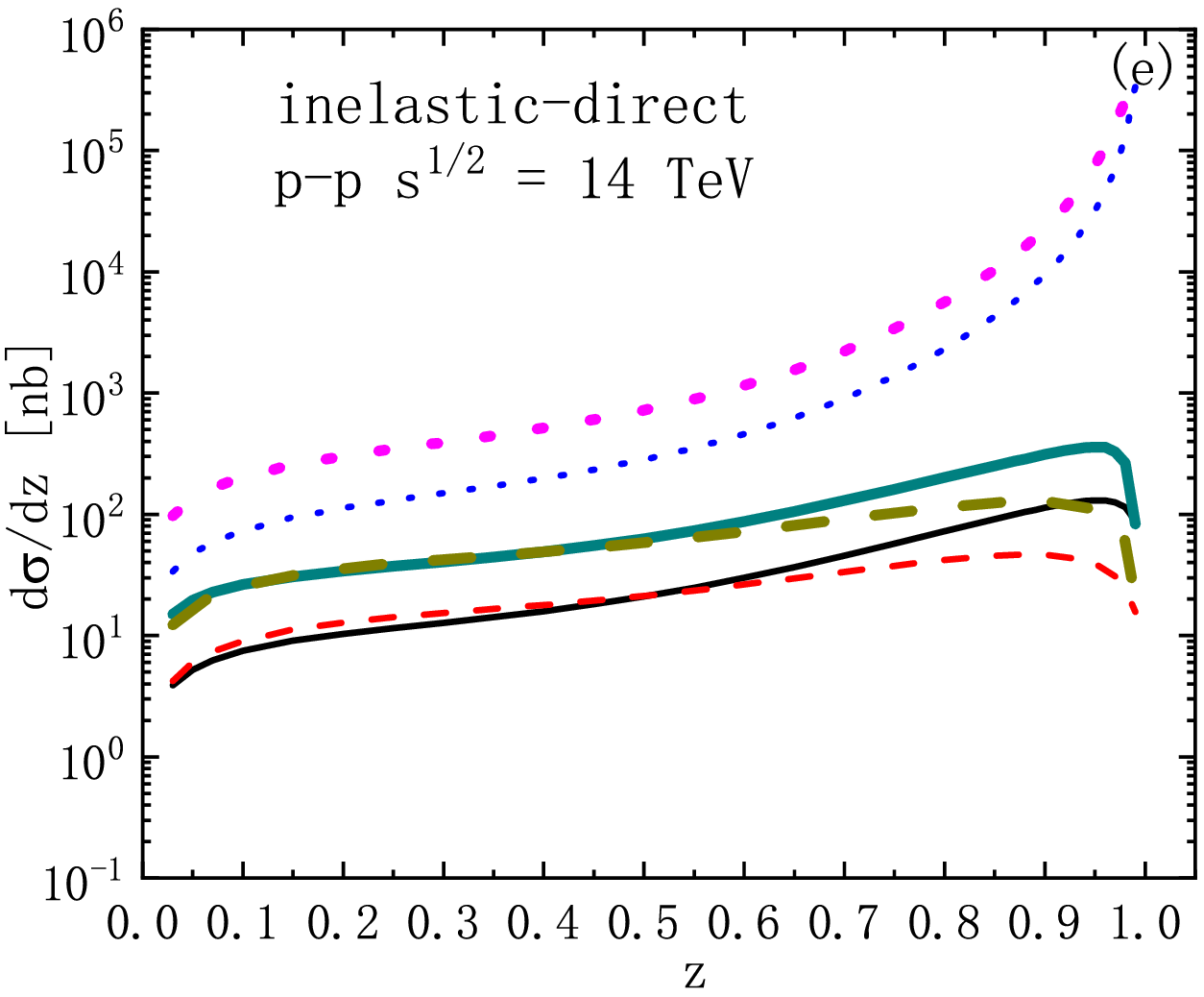}
    \includegraphics[width=0.3\textwidth]{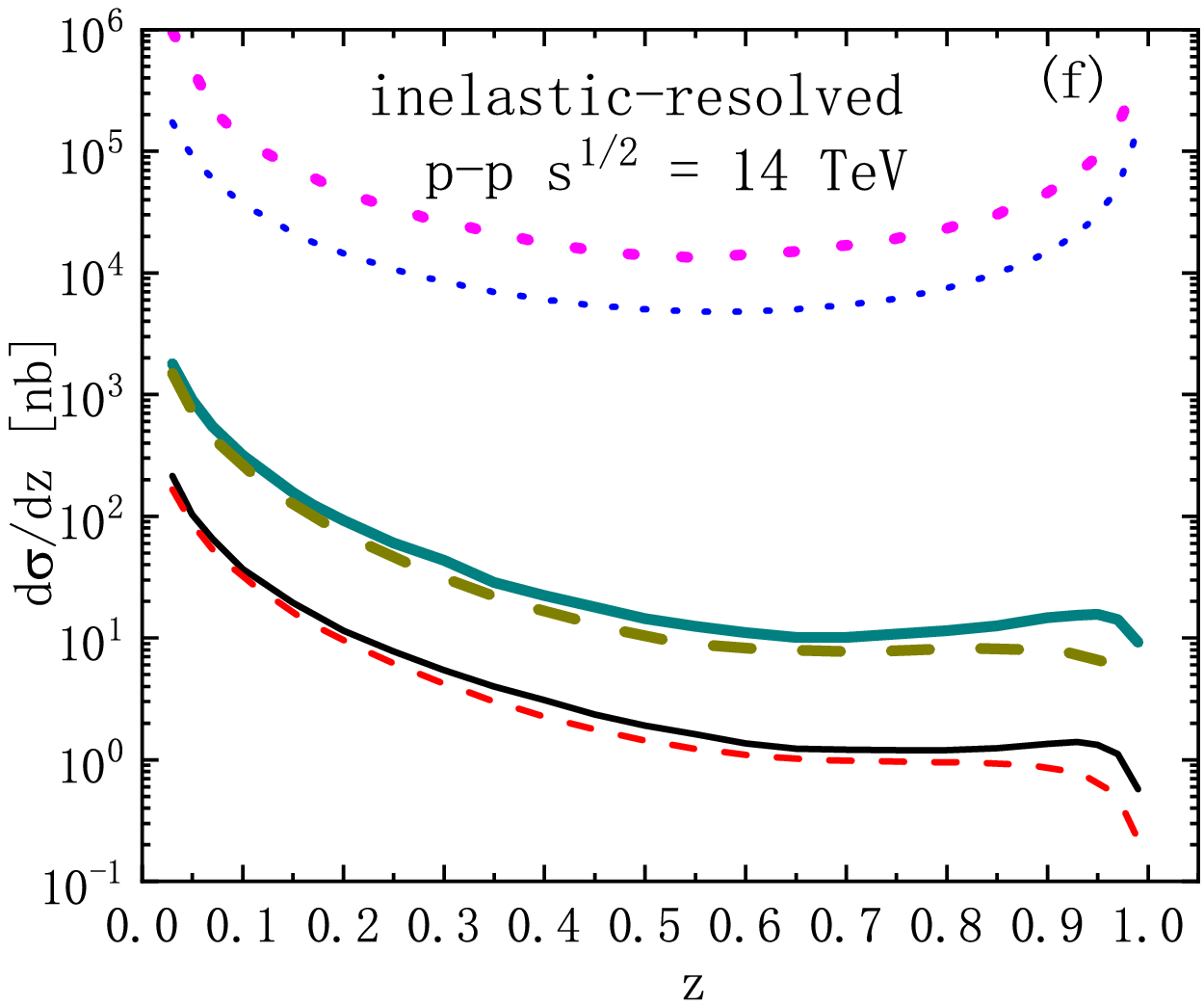}
  \caption{The $z$ distribution of $J/\psi$ photoproduction in p-p collisions at LHC energies.
  The left panels show the results of $z$ dependent differential cross sections for elastic photoproduction processes at different energies.
  The central and right panels show the corresponding results for inelastic-direct and inelastic-resolved photoproduction processes, respectively.
  (a), (d): Black solid and red dashed lines denote the exact results and the WWA ones for coherent-photon emissions, respectively;
  bolded dark cyan solid line is for the exact results, magenta dashed (blue dotted) line is for the WWA results (with no WF), for incoherent-photon emission.
  (b), (c), (e), and (f): Black solid line denotes the exact results, red dashed (blue dotted) line depicts the exact results with no WF but (no) adopts $p_{T \mathrm{min}}=M_{J/\psi}$ for coherent-photon emission; those for incoherent-photon emission are depicted by the same type lines but are bolded and with different colors.}
  \label{fig:z}
\end{figure*}

\begin{figure*}[htbp]
\setlength{\abovecaptionskip}{1mm}
  \centering
  \includegraphics[width=0.38\textwidth]{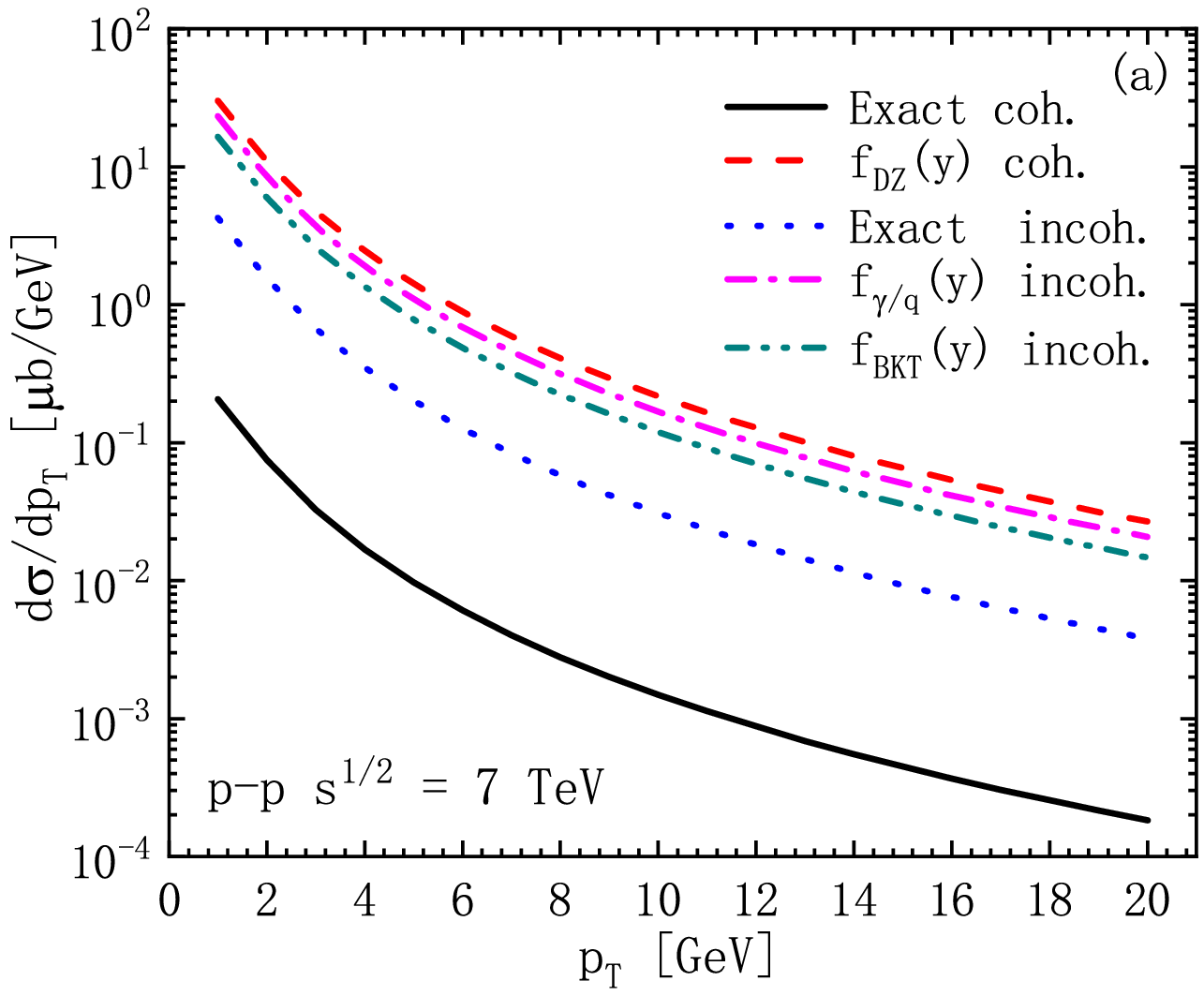}
  \includegraphics[width=0.38\textwidth]{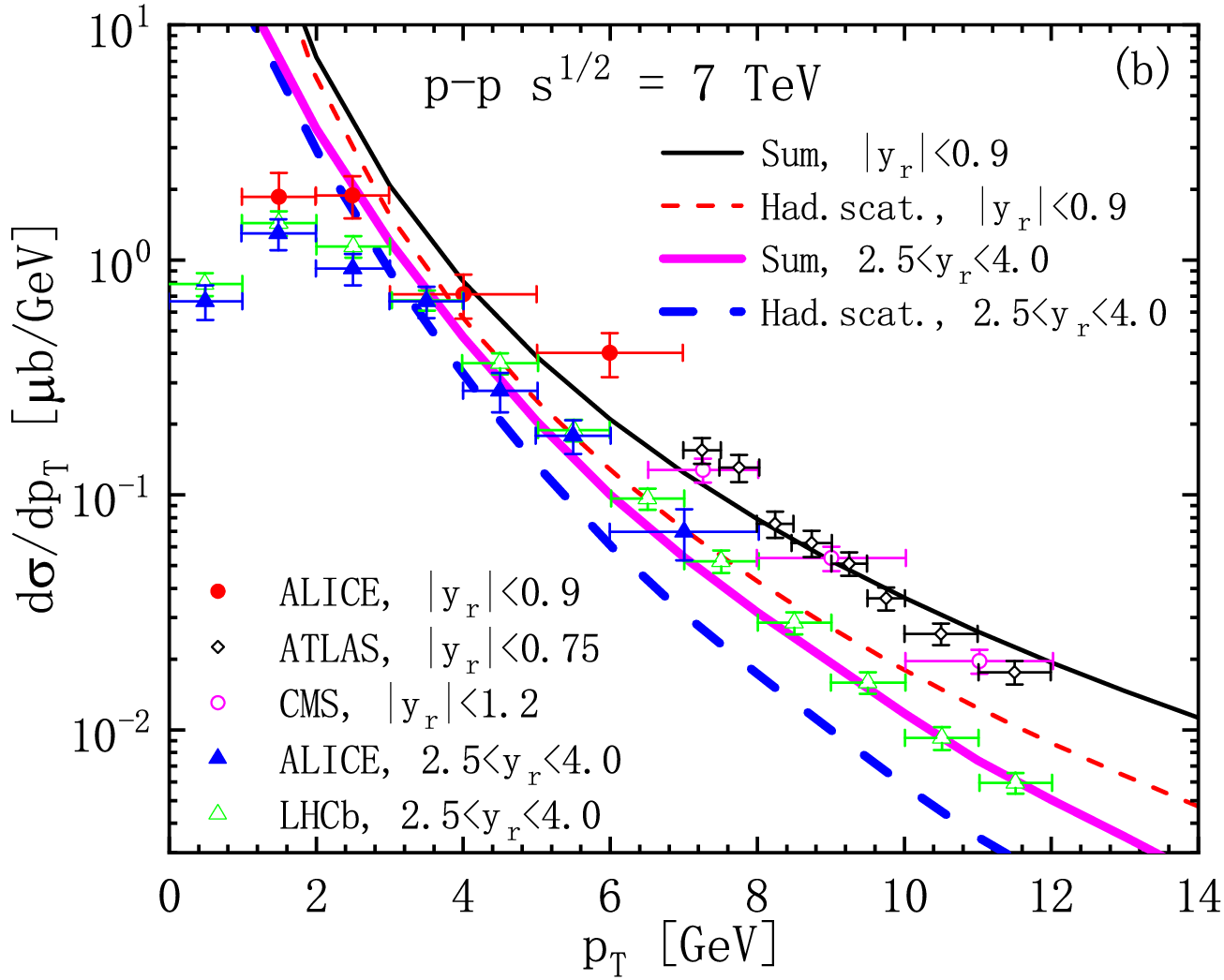}\\
    \includegraphics[width=0.38\textwidth]{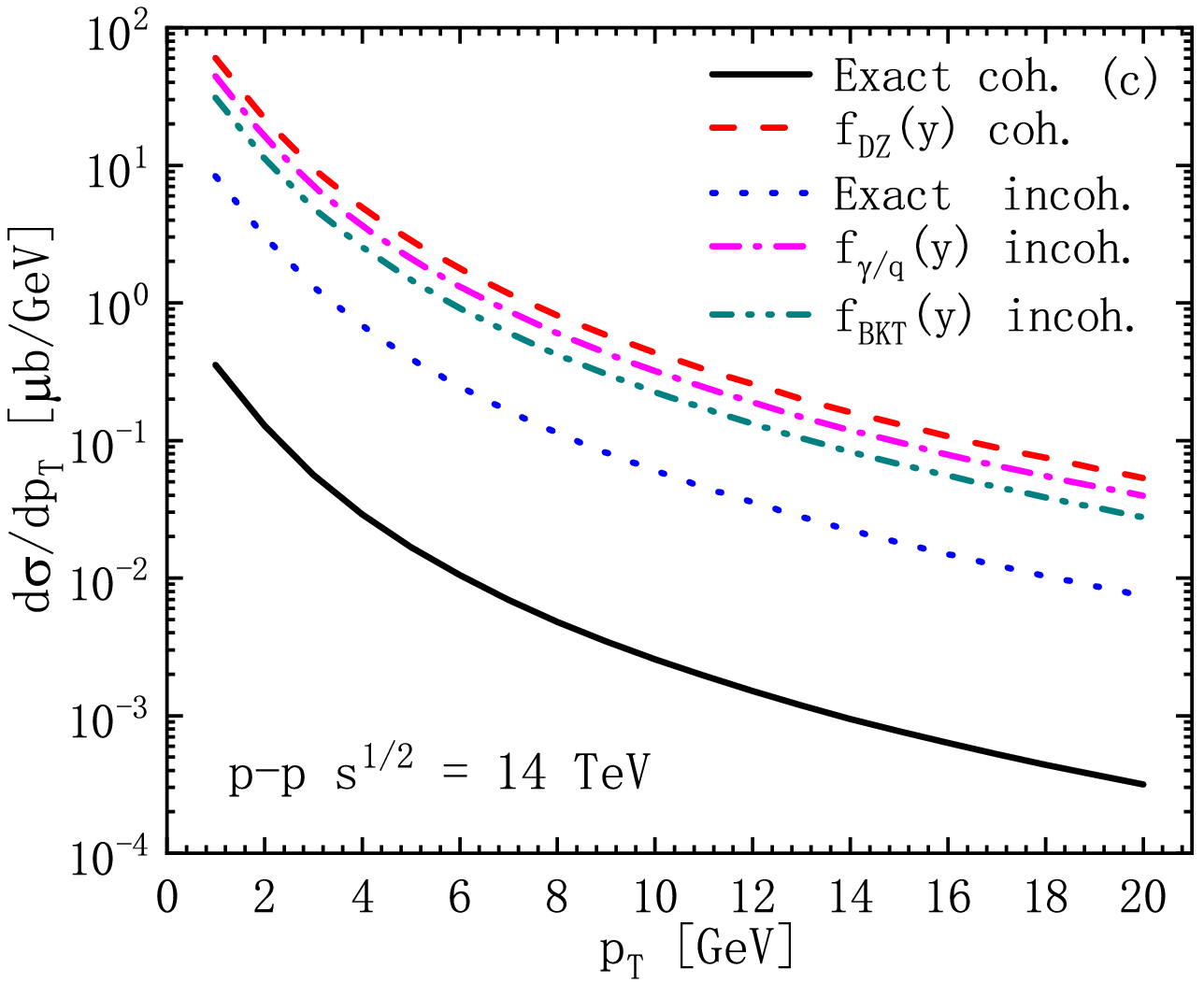}
    \includegraphics[width=0.38\textwidth]{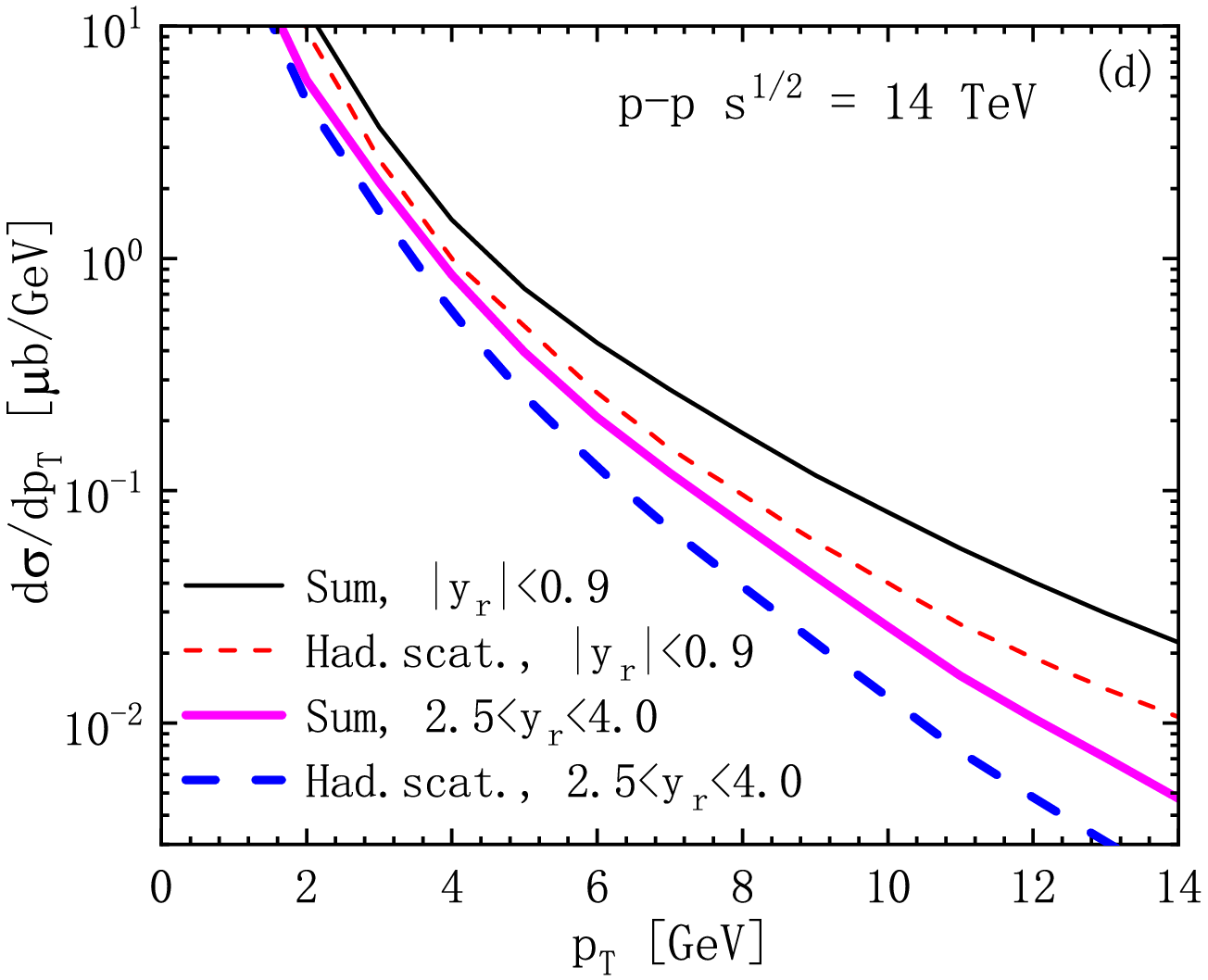}
  \caption{The $p_{T}$ distribution of inelastic $J/\psi$ photoproduction in p-p collisions at LHC energies.
  The left panels show the results of $p_{T}$ dependent differential cross sections in different approximations, while the right panels show the comparison between the photoproduction processes and the hard scattering of initial partons (had.scat.).
  (a), (c): Black solid and red dashed lines denote the exact results and the WWA ones based on the spectrum Eq.~(\ref{fgamma.DZ.}) for coherent-photon emission [coh.(dir.+res.)+coh.(dir.+res.)-frag.], respectively.
  Blue dotted, magenta dot-dashed, and dark cyan dot-dot dashed lines represent the exact results, the WWA ones based on the spectra Eqs.~(\ref{fgamma.incohI.}) and (\ref{fgamma.incohBKT}), respectively.
  (b), (d): Red dashed line is for had.scat., black solid line is for the sum of had.scat. and photoproduction processes;
  those thick curves with different colors are the same results but with different rapidity range.
  The $J/\psi$ data are from Ref.~\cite{ALICE:2011zqe}.
  }
  \label{fig:PT}
\end{figure*}

\begin{figure*}[htbp]
\setlength{\abovecaptionskip}{1mm}
  \centering
  \includegraphics[width=0.36\textwidth]{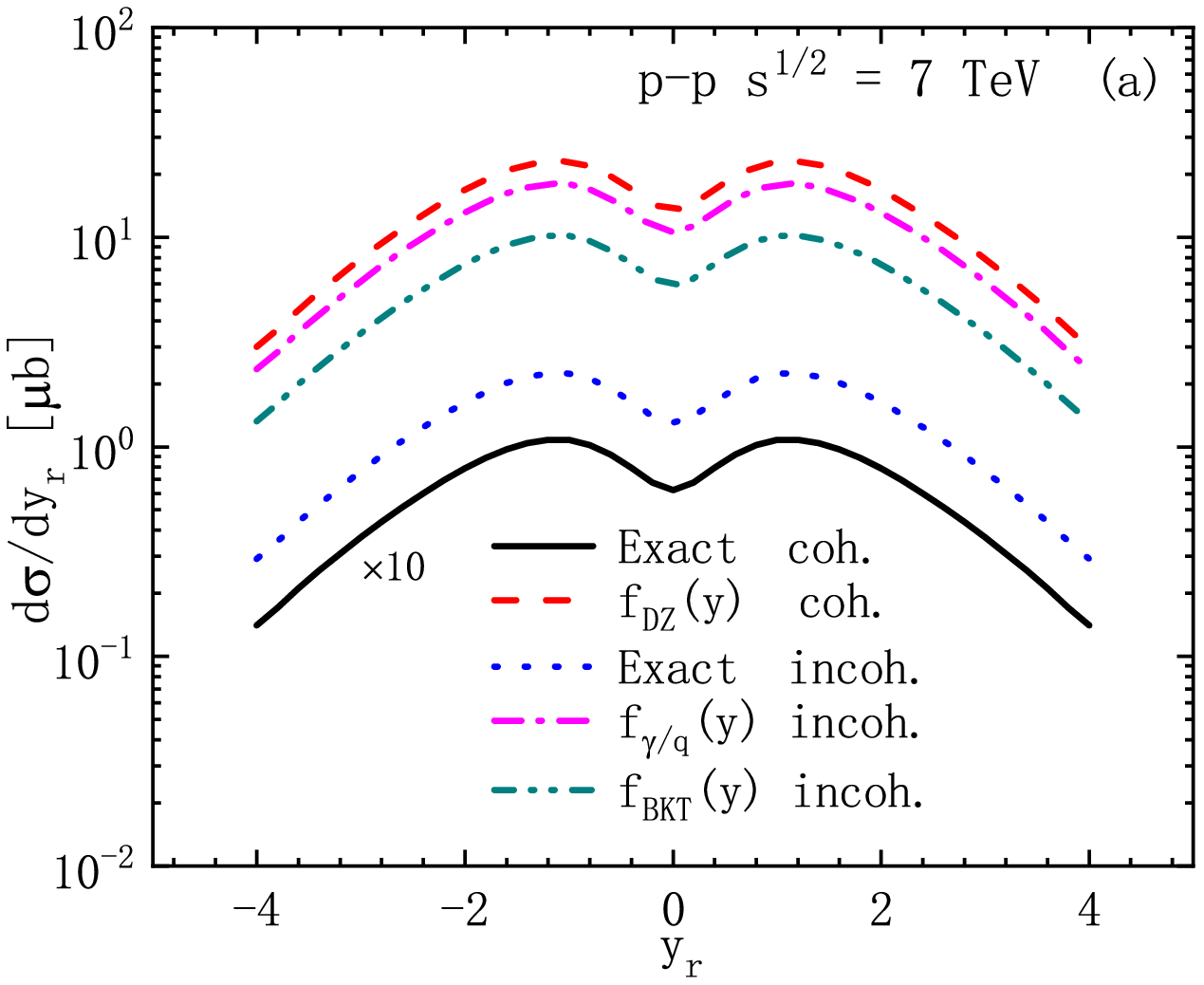}
  \includegraphics[width=0.36\textwidth]{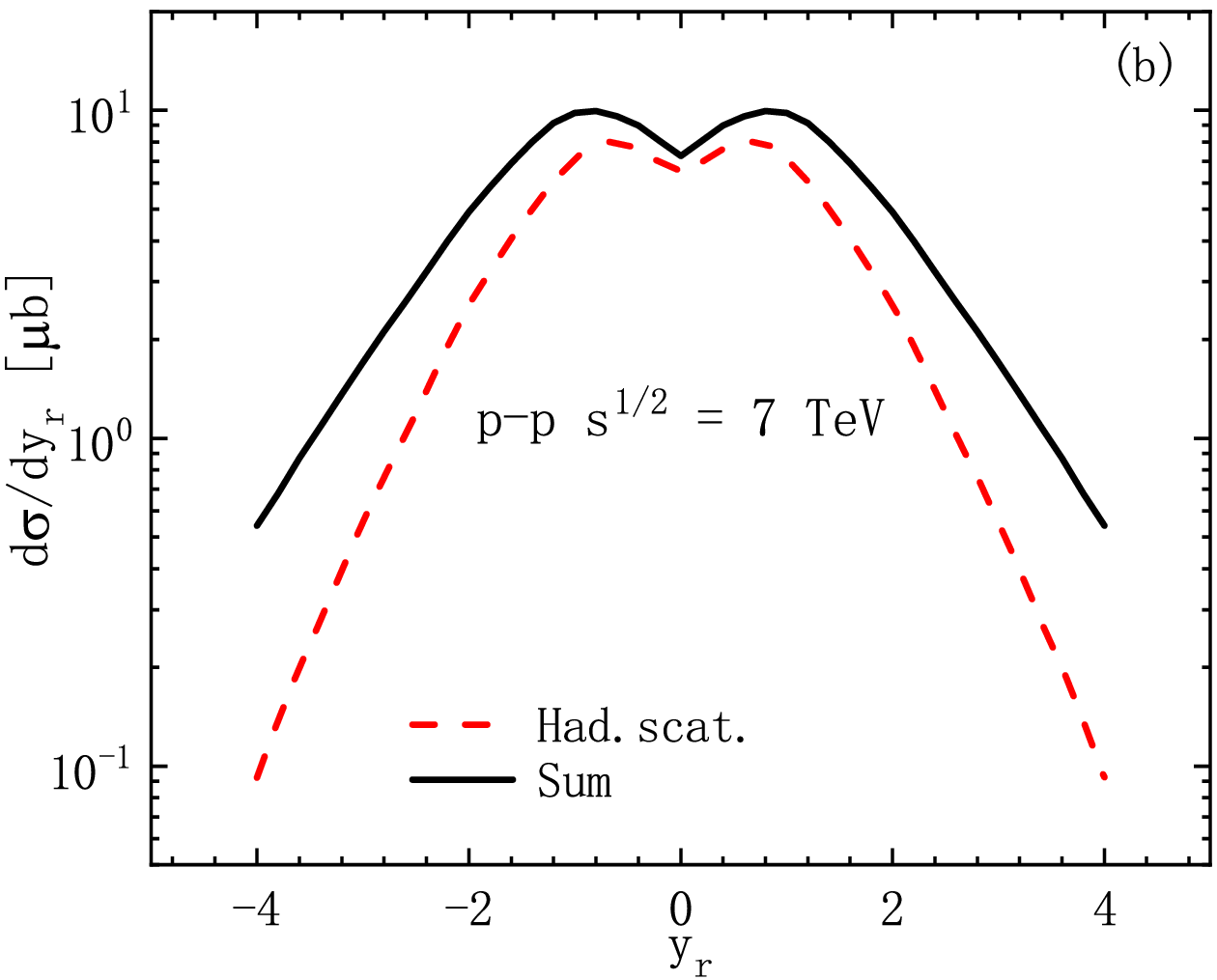}\\
    \includegraphics[width=0.36\textwidth]{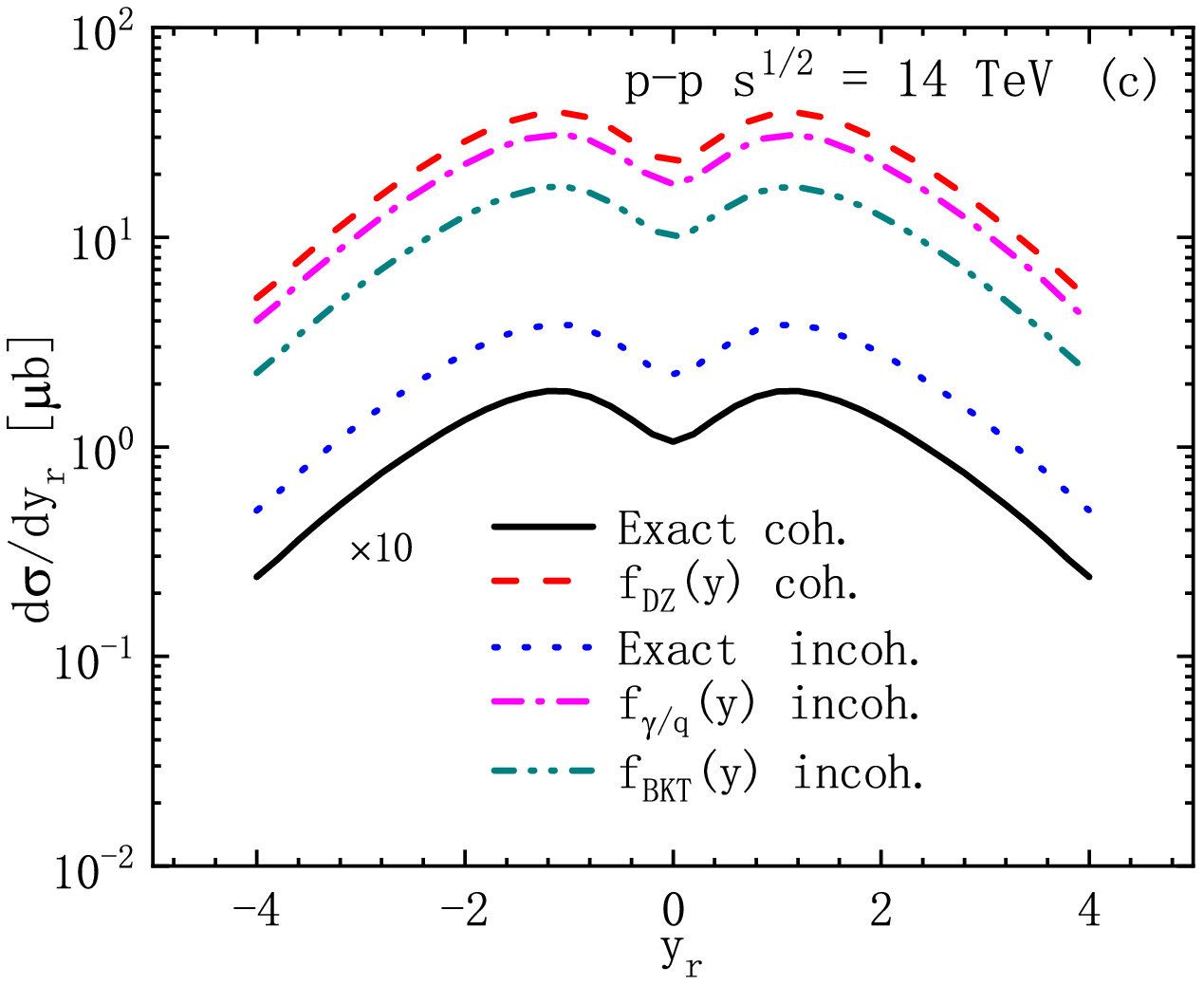}
    \includegraphics[width=0.36\textwidth]{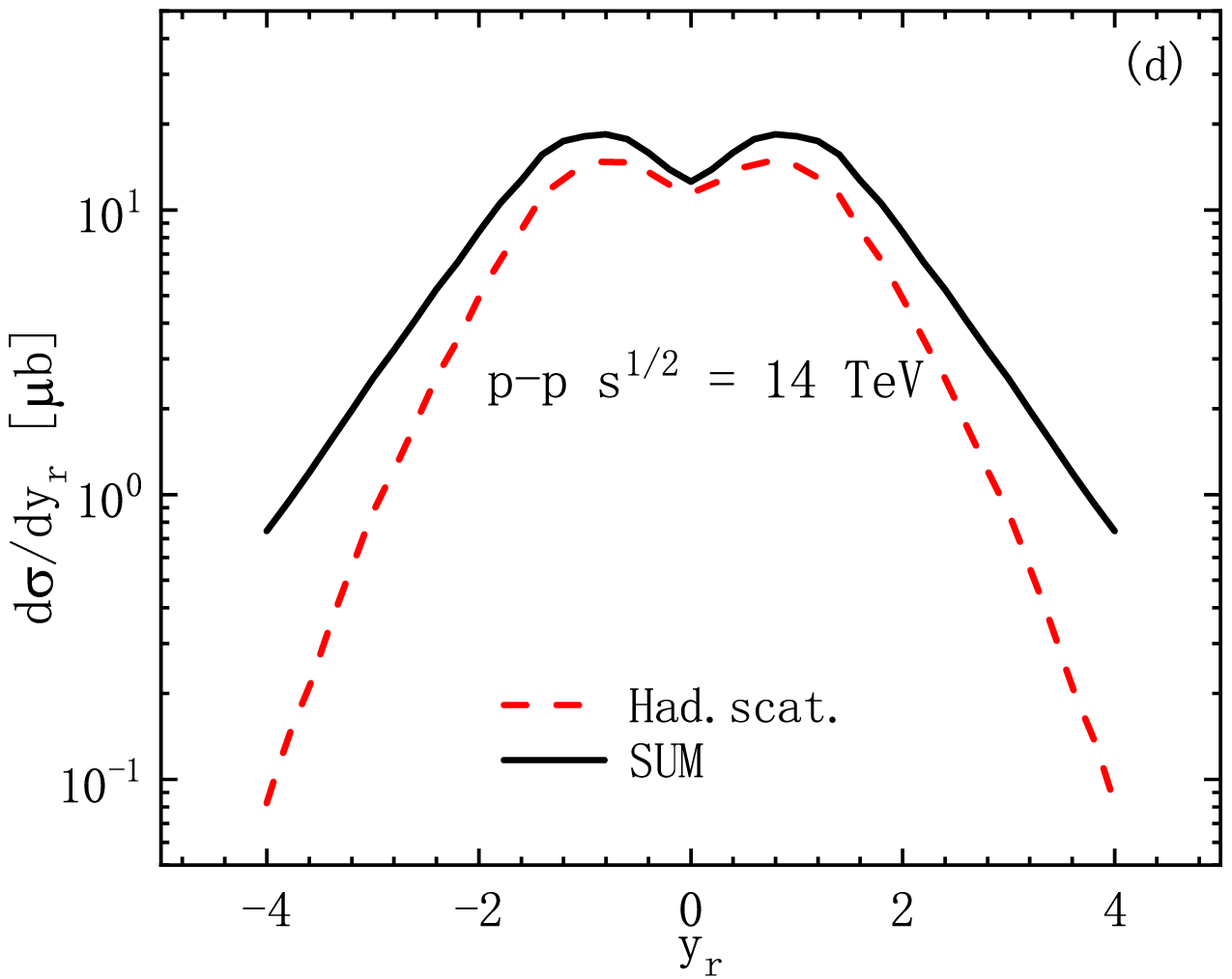}
  \caption{Same as Fig.~\ref{fig:PT} but for $y_{r}$ distribution.
}
  \label{fig:yr}
\end{figure*}

\begin{figure*}[htbp]
\setlength{\abovecaptionskip}{1mm}
  \centering
  \includegraphics[width=0.38\textwidth]{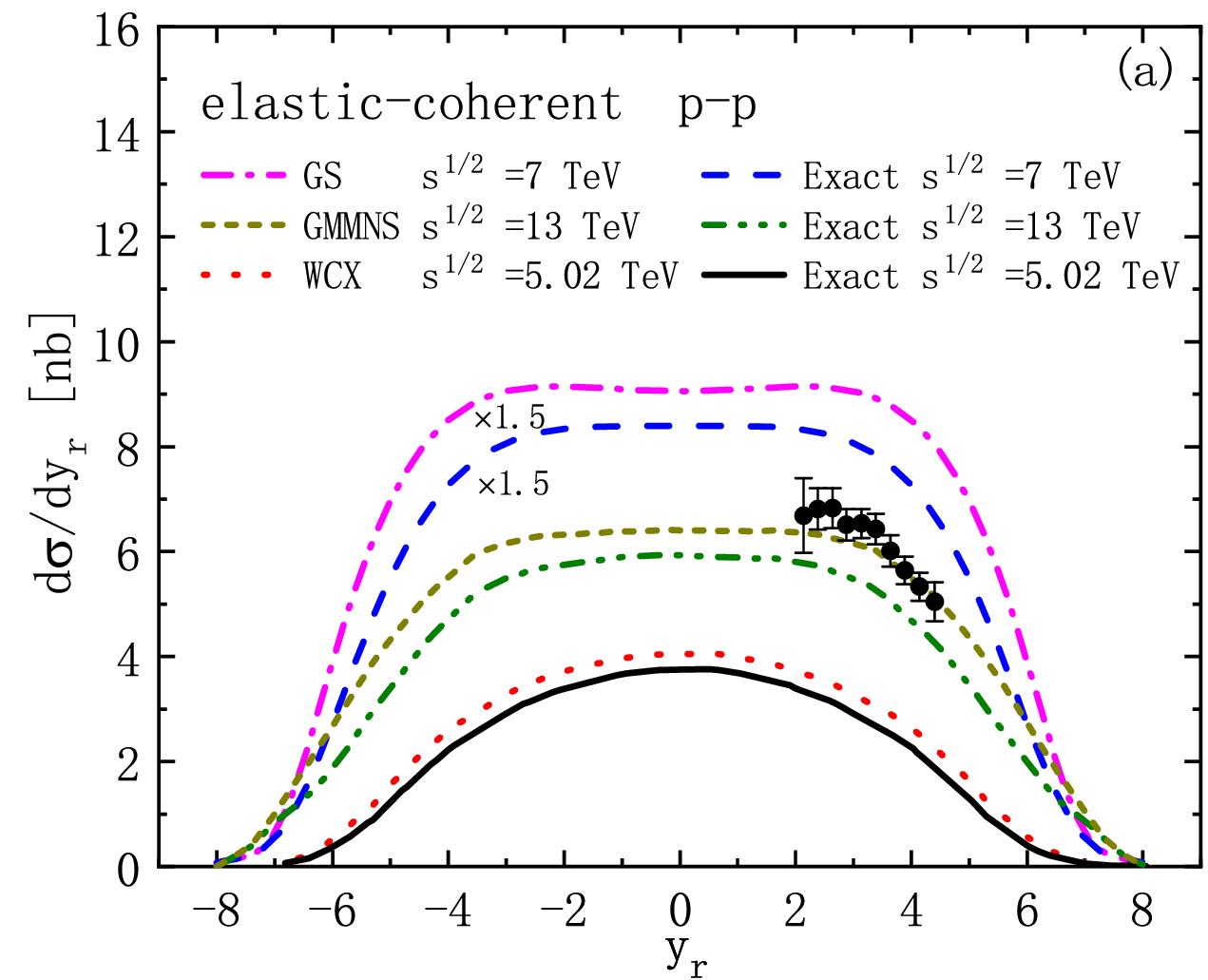}
  \includegraphics[width=0.38\textwidth]{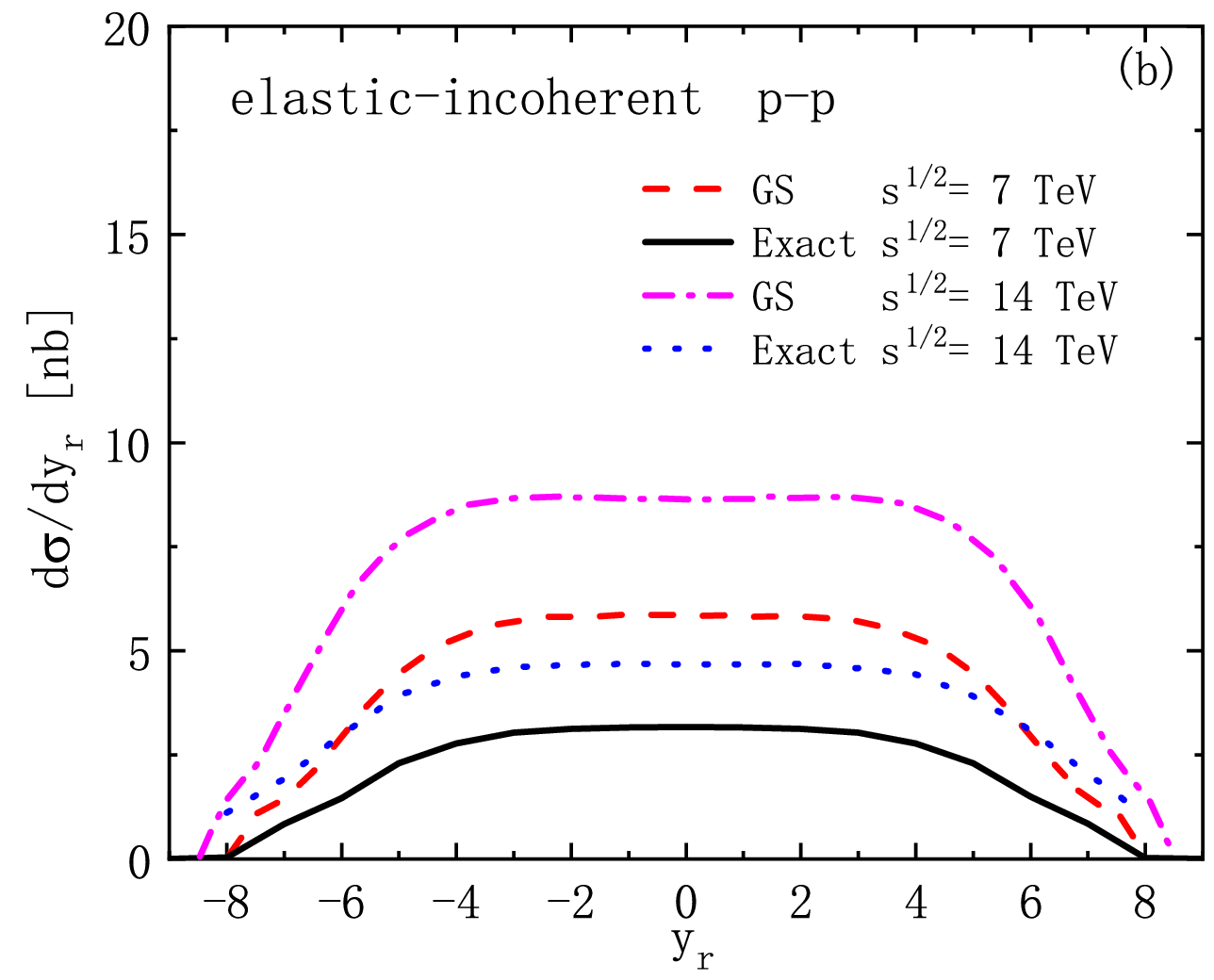}\\
    \includegraphics[width=0.38\textwidth]{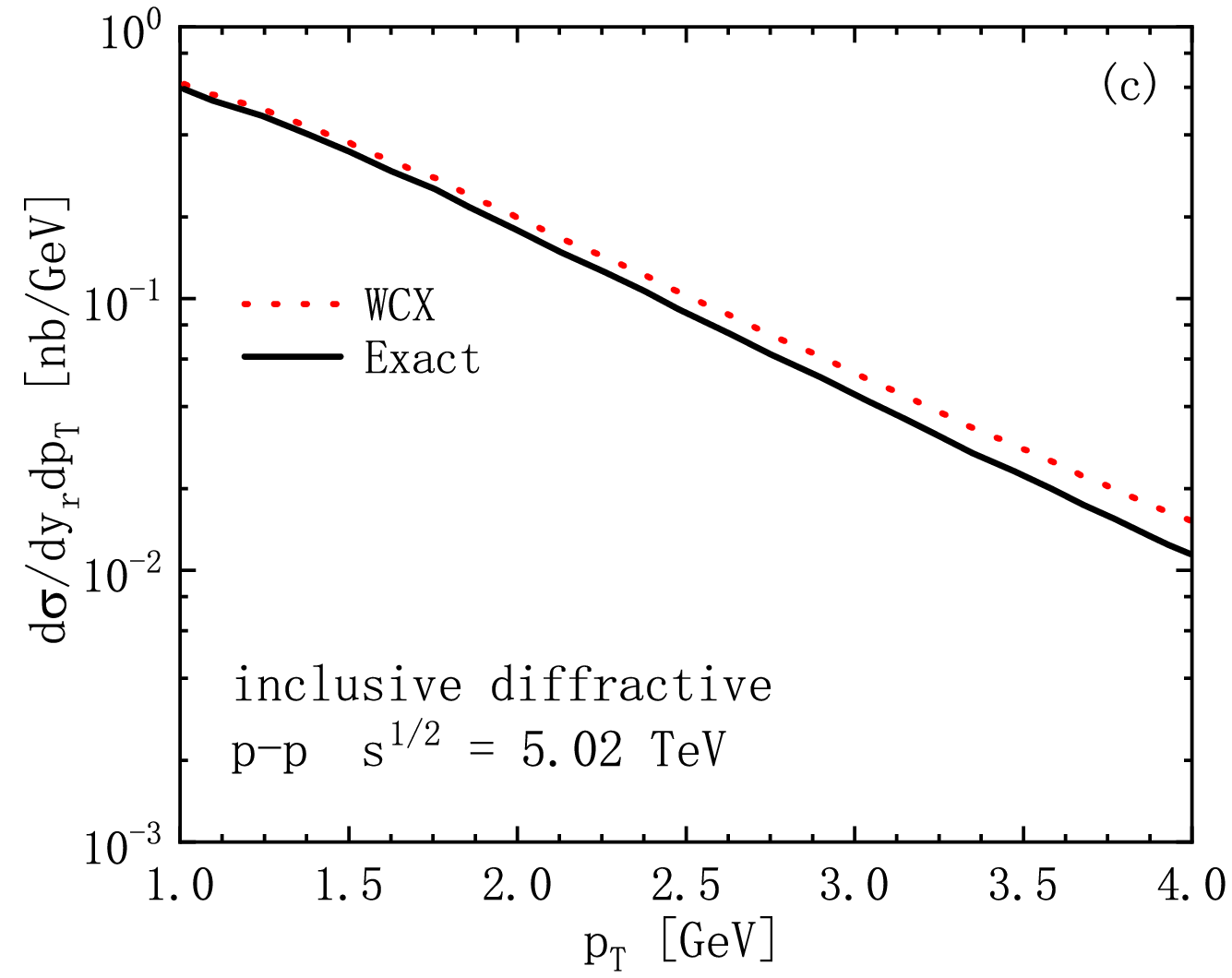}
    \includegraphics[width=0.38\textwidth]{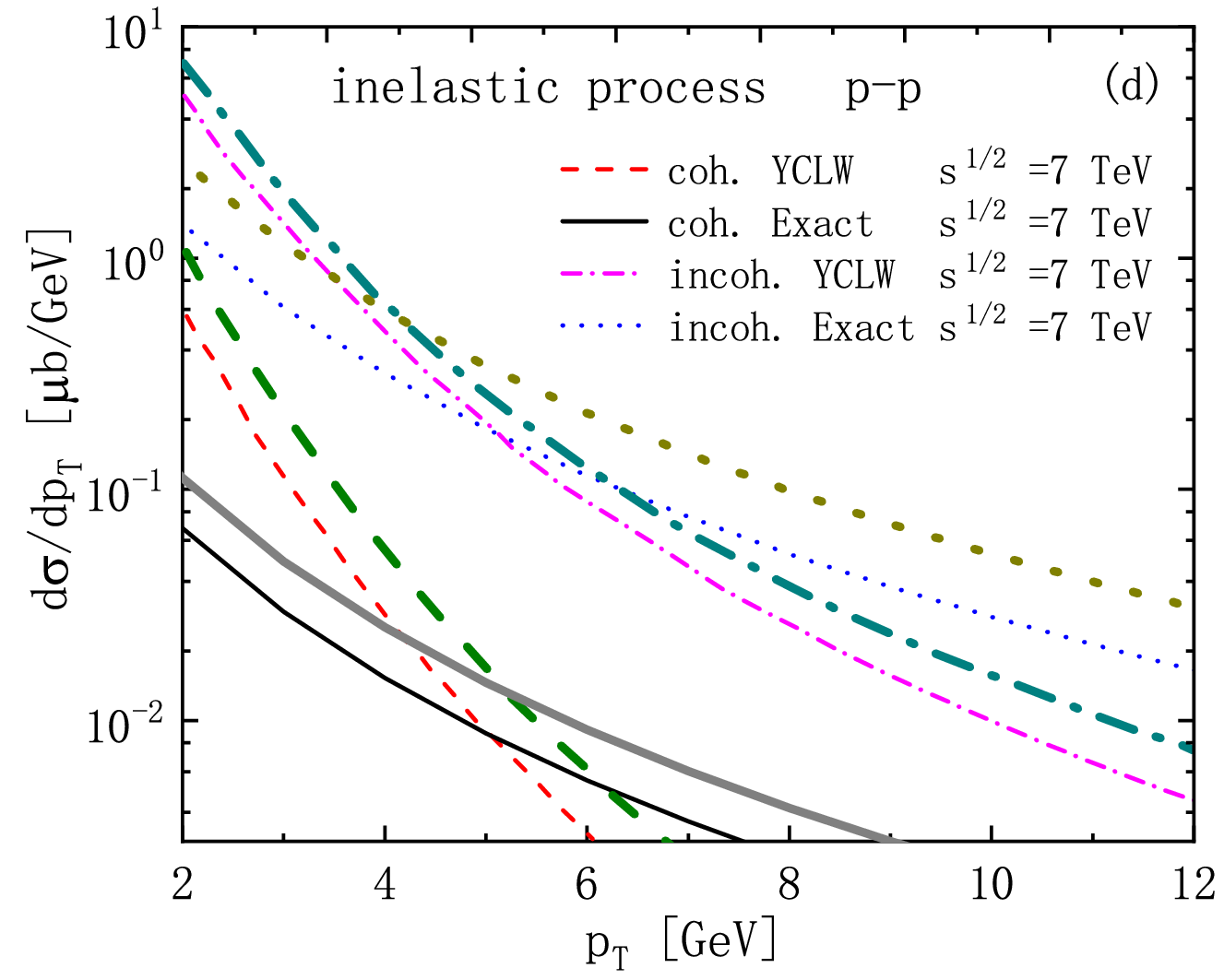}
  \caption{Comparison of the exact results with the WWA ones in the literature for $J/\psi$ photoproduction.
  (a), (c): Magenta dot-dashed line---the elastic result of Fig.~4 in Ref.~\cite{Goncalves:2015dia}.
      Dark yellow short dashed line---the result based on the bCGC model in Fig.~4 of Ref.~\cite{Goncalves:2017wgg}.
      Red dotted line---the total results in Figs. 4 and 7 of Ref.~\cite{Wu:2020ujf}.
  (b): Red dashed and magenta dot-dashed lines are for the inelastic results based on the naive approach in Fig. 4 of Ref.~\cite{Goncalves:2015dia}.
  (d): Red dashed and magenta dot-dashed lines are for the coherent and incoherent results in Fig.~1 (a), (b) of Ref.~\cite{Yu:2017pot}.
  Those bolded same type curves with different colors are for the same results but at $s^{1/2}=14~\mathrm{TeV}$.
  Data from LHCb Collaboration~\cite{LHCb:2013nqs, LHCb:2014acg, LHCb:2016oce}, which was adopted in Ref.~\cite{Goncalves:2017wgg}.
  }
  \label{fig:DC}
\end{figure*}

We calculate the total cross sections in Tables~\ref{Total.CS.coh.7TeV}-\ref{Total.CS.incoh.14TeV}, to quantitatively estimate the influence from different choices of kinematical boundaries, which are widely employed in various photon spectra.
In the case of coherent-photon emission [Tables~\ref{Total.CS.coh.7TeV}, \ref{Total.CS.coh.14TeV}], we can see that the usual choices of $Q^{2}_{\mathrm{max}}\sim \hat{s}$ and $y_{\mathrm{max}}=1$ bring the large relative errors.
These relative errors are much larger in the inelastic photoproduction processes, and become more obvious with increasing $\sqrt{s}$. 
One exception is the elastic-coherent process, where the errors from $Q^{2}_{\mathrm{max}}\sim \hat{s}$ can be neglected. 
In the case of incoherent-photon emission [Table~\ref{Total.CS.incoh.7TeV}, \ref{Total.CS.incoh.14TeV}], the WWA errors are non-negligible.
Especially in the inelastic-incoherent process, the WWA errors are prominent and become larger with increasing $\sqrt{s}$;
this quantitatively verifies the inapplicability of WWA in this case.
Therefore, the values of $Q^{2}_{\mathrm{max}}$ and $y_{\mathrm{max}}$ is crucial to the precision of WWA, the common choices will cause large errors.

We also check the relative contribution of each process by comparing their exact results in the Tables.
We observe that in the case of elastic photoproduction processes, the main channel is the coherent-photon emission; this agrees with the traditional perspective.
However, the incoherent-photon emission can also provide the meaningful contribution (about $18\%$).
In the case of inelastic photoproduction processes, the main channel is the incoherent-photon emission, its contribution reaches up to $91\%$.
Therefore, the contribution of incoherent-photon emission can not be neglected in elastic photoproduction, and starts to play a fundamental role in inelastic photoproduction.

On the other hand, it is necessary to discuss the double counting encountered in most of works.
In the Tables, all of the WWA results without weighting factor (WWA No WF) are the unrealistic large values. 
One exception is the elastic-coherent process, where the errors are relatively small. 
For coherent-photon emissions in Tables~\ref{Total.CS.coh.7TeV} and \ref{Total.CS.coh.14TeV}, this will cause the large fictitious contributions.
And for incoherent-photon emissions, a traditional way for avoiding these unphysical results is to adopt an artificial cutoff $Q^{2}>1\ \mathrm{GeV}^{2}$, but we can see that in Tables~\ref{Total.CS.incoh.7TeV} and \ref{Total.CS.incoh.14TeV} the corresponding results are still not accurate.
Thus, the weighting factor adopted in exact treatment can effectively avoid double counting.

In Fig.~\ref{fig:z}, the $z$ distribution of $J/\psi$ photoproduction is plotted.
The left panels show the results of $z$ dependent differential cross sections for elastic photoproduction processes at different energies;
while the central and right panels show the corresponding results for inelastic-direct and inelastic-resolved photoproduction processes, respectively.
It can be seen that the results without weighting factor are much larger than the exact ones; this reflects the serious double counting we discussed above.
In the left panels, the WWA results are consistent with exact ones in the whole $z$ regions; this verifies again that WWA is a good approximation for elastic photoproduction processes.
And the curves are negligible in most of $z$ regions and become important when $z>0.9$, especially near the endpoint $z=1$, the curves show a pronounced rising.

In the case of inelastic photoproduction processes [central and right panels], the contributions are important in most of $z$ regions.
We can see that the results without WF are divergent near $z=1$, since the NRQCD prediction breaks down and the color-octet channels exhibit collinear singularities in the region of $z\lesssim1$, where diffractive production takes place.
In order to screen the collinear singularities and suppress the elastic production, the traditional way is to imposes the following cuts in $z$ and $p_{T}$: $z<0.9$ and $p_{T}>1~\mathrm{GeV}$ or $M_{J/\psi}$ (actually, if $p_{T}$ has a nonzero minimum value, the maximum value of $z$ will naturally less than one).
Another possibility to suppress the elastic production at $z\lesssim1$ would be to require that $Q^{2}$ be sufficiently large~\cite{Kniehl:2001tk}.
However, then also the bulk of the inelastic contribution would be also sacrificed.
Comparing with the traditional way, it can be seen that these diffractive contributions are effectively suppressed in the exact results (the solid lines fall down rapidly with increasing $z$ at the endpoint region).
In addition, these exact results also agree well with the calculations which adopt the cut $p_{T~\mathrm{min}}=M_{J/\psi}$ (the dashed lines).
The reason is that the weighting factor is employed in the exact treatment.
Therefore, the exact treatment can naturally suppress the elastic production at $z\lesssim1$, and thus effectively avoid double counting.

Finally, we find that the curves of inelastic-direct processes [central panels] is comparable with elastic photoproduction processes [left panels] near $z=0.9$, but rapidly deceased near $z=1$;
these features are in agreement with the traditional perspective that $z=0.9$ is the boundary to distinguish the elastic and inelastic photoproductions~\cite{Berger:1981, Ko:1996xw}.
And the exact results of inelastic-resolved processes [right panels] dominate the lower $z$ region and are smaller than those of inelastic-direct processes when $z>0.2$; these features also agree with the traditional perspective that the resolved contribution is only important in the lower $z$ region ($z<0.2$)~\cite{Ko:1996xw, Yuan:1999eb}.

In order to estimate the contribution of $J/\psi$ photoproduction to LO hard scattering of initial partons (had.scat.), and discuss the feature of the photon spectra which are widely employed in most works.
We plot the $p_{T}$ distribution of inelastic $J/\psi$ photoproduction in Fig.~\ref{fig:PT}, where all the curves are the sum of direct and fragmentation $J/\psi$ contributions.
The left panels show the results of $p_{T}$ dependent differential cross sections in different approximations, while the right panels show the comparison between the photoproduction processes and the had.scat.
Since the intrinsic motion of incident partons inside colliding hadrons renders the differential cross section uncertain for $p_{T}<2~\mathrm{GeV}$, we have not attempt to remove the divergences from the small $p_{T}$ domain.
Instead, we simply regard the portion of the plot which runs below $p_{T}=2~\mathrm{GeV}$ as untrustworthy.

In the left panels, we observe that the results based on the photon spectra mentioned in Sec.~\ref{WWA} generally have the prominent deviations from the exact ones.
In the case of coherent-photon emission, the curves of $f_{\mathrm{DZ}}$ are larger than the exact ones by about two OOMs;
while in the case of incoherent-photon emission, the curves of $f_{\mathrm{\gamma/q}}$ and $f_{\mathrm{BKT}}$ are larger than the exact ones by about one OOM.
There are two common reasons.
Firstly, the integrations of these spectra are performed in the entire kinematical allowed regions: $Q^{2}_{\mathrm{max}}\sim\hat{s}$ or $\infty$, and $y_{\mathrm{max}}=1$, which include the large WWA errors.
Actually, in most of the physically interesting cases such a dynamical cut off $\Lambda_{\gamma}$ exists such that, the WWA errors can be effectively avoided and the photo-absorption cross sections differ only slightly from their values on the mass shell.
Thus, for the practical use of WWA, besides considering the kinematically allowed regions, one should also elucidate whether there is a dynamical cut off $\Lambda^{2}_{\gamma}$, and estimate it.
Secondly, the term of weighting factor is neglected in these WWA parameterizations, which causes the double counting problem.
In addition, the reason for error of $f_{\mathrm{BKT}}$ is that, $f_{\mathrm{BKT}}$ is originally derived from $ep$ scattering, but is directly expanded to describe the probability of finding a photon in any relativistic fermion and to deal with hadronic collisions in Ref.~\cite{Kniehl:1990iv}, this will overestimate the cross sections.
Therefore, the referred photon spectra will provide the large fictitious contributions to the $J/\psi$ production, and the results in Ref.~\cite{Drees:1989vq, Drees:1988pp, Frixione:1993yw, Zhu:2015via, Zhu:2015qoz, Fu:2011zzm, Fu:2011zzf, Chin.Phys.C_36_721, Yu:2015kva, Yu:2017rfi, Yu:2017pot, Nystrand:2004vn, Kniehl:2001tk, Kniehl:1990iv, Fu:2012xm, sp} are not accurate enough, where the mentioned spectra are adopted and the serious double counting exists.

In the right panels, we compare the exact results with data derived from relevant collaborations.
We observe that the contribution of $J/\psi$ produced by photoproduction processes is non-negligible.
Especially in the large $p_{T}$ domain, the corrections of photoproduction processes to LO had.scat. are evident.

\begin{table}[htbp]\footnotesize
\begin{threeparttable}
\renewcommand\arraystretch{1.2}
\centering
\caption{\label{Total.CS.DF}Total cross sections of the direct and fragmentation $J/\psi$ photoproductions.}
\begin{tabular}{L{2.7cm}C{1.2cm}C{2.8cm}C{1.3cm}}
\hline
\hline
    &$\sigma_{\textrm{dir.}}~[\mathrm{nb}]$ & $\sigma_{\textrm{frag.}}~[\mathrm{nb}]$ ($\sigma_{\textrm{frag.}}/\sigma_{\textrm{total}}$)& $\sigma_{\textrm{total}}~[\mathrm{nb}]$ \\
    \hline
    coherent ($7~\mathrm{TeV}$)     & 2.35      & 1.70 (0.42)    & 4.05     \\
    incoherent ($7~\mathrm{TeV}$)   & 77.91      & 51.94 (0.40)   & 129.85  \\
    coherent ($14~\mathrm{TeV}$)    & 4.30       & 3.66 (0.46)    & 7.96    \\
    incoherent ($14~\mathrm{TeV}$)  & 140.18      & 97.41 (0.41)   & 237.59 \\
\hline
\hline
\end{tabular}
\end{threeparttable}
\end{table}

In Fig.~\ref{fig:yr}, the $y_{r}$ distribution is plotted.
In the left panels, the WWA results based on the referred photon spectra generally have prominent errors compared to the exact ones in the whole $y_{r}$ region.
In the right panels, the photoproduction processes give the evident corrections to had.scat., especially in the large $y_{r}$ domain.
In order to estimate the relative contribution of fragmentation $J/\psi$, we calculate the total cross sections in Table~\ref{Total.CS.DF}.
We observe that the contributions of fragmentation processes are important.
Especially in the coherent-photon emission, its contribution reaches up to $46\%$.
Furthermore, the incoherent-photon emissions are about $30$ times larger than the coherent ones;
this verifies again that the incoherent-photon emission plays a fundamental role in the inelastic $J/\psi$ photoproduction.

Finally, in Fig.~\ref{fig:DC} we present the direct comparison of the current results with works in literature based on WWA approach.
In Ref.~\cite{Goncalves:2015dia}, Gon\c{c}alves and Silveira (GS) studied the diffractive quarkonium photoproduction in $pp$ collisions at LHC energies to probe the photon flux associated with a proton.
In Ref.~\cite{Goncalves:2017wgg}, a comprehensive analysis of exclusive vector meson photoproduction using the Color Dipole formalism was performed by Gon\c{c}alves, Machado, Moreira, Navarra and Santos (GMMNS).
Wu, Cai, and Xiang (WCX) studied the inclusive diffractive heavy quarkonium photoproduction using the resolved pomeron model in Ref.~\cite{Wu:2020ujf}.
All of these works are associated with the equivalent photon spectrum $f^{\mathrm{DZ}}_{\gamma}(Q^{2})$ [Eq.~(\ref{fgamma.DZ.})].
In Ref.~\cite{Yu:2017pot}, Yu, Cai, Li and Wang (YCLW) studied the heavy quarkonium photoproduction in ultrarelativistic heavy ion collisions by using the incoherent photon spectrum $f_{\gamma}^{\mathrm{BKT}}(Q^{2})$ [Eq.~(\ref{fgamma.incohBKT})].
We take into account above works in Fig.~\ref{fig:DC}.

In panels (a) and (c), the exact results agree well with GS and WCX predictions, this is consistent with the views derived in Fig.~\ref{fig:Q2} that the WWA can reach the high precision in elastic-coherent process.
The errors are mainly from the spectrum $f^{\mathrm{DZ}}_{\gamma}(Q^{2})$, which is performed in the whole kinematical allowed regions.
In panel (b), the exact results qualitatively agree with the GS predictions that the incoherent-photon emission has the meaningful contribution in elastic photoproduction processes.
The deviations are originating from the spectrum $f^{\mathrm{incoh}}_{\gamma}(Q^{2})$, which neglects the WF factor and set $Q^{2}_{\mathrm{max}}=\hat{s}/4-m_{q}^{2}$.
In panel (d), the exact results have the visible deviations from the YCLW predictions, these errors are caused by the inapplicability of WWA in inelastic-incoherent processes.

\section{Summary and conclusions}
\label{Summary and conclusions}

We have investigated the production of $J/\psi$ induced by photoproduction and fragmentation processes in $p$-$p$ collisions at LHC energies.
The elastic and inelastic photoproduction processes, and coherent and incoherent-photon emissions are considered simultaneously.
By performing a consistent analysis of the terms neglected in going from the accurate expression to the WWA one, the exact treatment which reduces to the WWA in the region $Q^{2}\rightarrow0$ is achieved, where the density of virtual photon is expanded by using the linear combinations, and the square of electric form factor $F^{2}_{1}(Q^{2})$ is applied as weighting factor to weight the different channels.
And the full partonic kinematics matched with exact treatment is also obtained.
In order to systematically study the properties of WWA in $J/\psi$ photoproduction in $p$-$p$ collisions at LHC energies, and to estimate the relative contributions of different channels to the $J/\psi$ production, we presented the comparison between the exact results and the WWA ones as the $Q^{2}$, $y$, $z$, $p_{T}$, and $y_{r}$ distributions.
And the total cross sections are also calculated.

The numerical results indicate that the coherent-photon emission is the main part of elastic photoproduction processes, but its contribution can be neglected in inelastic processes.
The contribution of incoherent-photon emission should not be neglected in the $J/\psi$ production, since it can provide the meaningful contributions in elastic photoproduction processes, and even starts to play a fundamental role in the inelastic processes.
In addition, the contributions of photoproduction and fragmentation processes are evident in the $J/\psi$ production, especially in the large $y_{r}$ and large $p_{T}$ regions.

On the other hand, the WWA is only effective in restricted domains (small $Q^{2}$ and $y$ domains), and the values of $y_{\mathrm{max}}$ and $Q^{2}_{\mathrm{max}}$ is very crucial to its precision.
The mentioned photon spectra integrating over the entire allowed kinematical regions, this will lead large errors.
These feature permits one to employ the WWA for elastic photoproduction processes, and also for coherent-photon emission.
Especially in the case of elastic-coherent process, the precision of WWA is highest.
However, WWA is not a good approximation for inelastic photoproduction processes, and also for incoherent-photon emission.
Especially in the case of inelastic-incoherent process, the WWA is inapplicable.
Moreover, the double counting exists when the different channels are considered simultaneously.
Therefore, the exact treatment can effectively suppress the WWA errors and can naturally avoid double counting, which needs to be adopted in $J/\psi$ photoproduction in $p$-$p$ collisions at LHC energies.

\section*{ACKNOWLEDGMENTS}

This work is supported in part by National Key R \& D Program of China under grant No. 2018YFA0404204, the NSFC (China) grant Nos. 11747086 and 12150013, and by the Young Backbone Teacher Training Program of Yunnan University.
Z. M. is funded by China Postdoctoral Science Foundation under grant No. 2021M692729.


\appendix
\section{Full kinematical relations}
\label{FKR}

\begin{table}[htbp]
\renewcommand\arraystretch{1.7}
\centering
\caption{\label{Kine.el.Q2.y} Kinematical boundaries of $Q^{2}$ and $y$ distributions for elastic processes, where $\hat{s}_{\mathrm{min}}=(M_{J/\psi}+m_{p})^{2}$.
}
\begin{tabular}{C{1.2cm}C{2.2cm}C{4.8cm}}
\hline
\hline
Variables & Coherent & Incoherent \\
    \hline
    $\hat{t}_{\mathrm{min}}$  &  \multicolumn{2}{c}{$M_{J/\psi}^{2}-Q^{2}-2(E_{q}E_{J/\psi}+p_{\mathrm{CM}}p_{\mathrm{CM}}')$}  \\
    $\hat{t}_{\mathrm{max}}$  &  \multicolumn{2}{c}{$M_{J/\psi}^{2}-Q^{2}-2(E_{q}E_{J/\psi}-p_{\mathrm{CM}}p_{\mathrm{CM}}')$}  \\
    $x_{a \mathrm{min}}$      &  $\backslash$   &   $(\hat{s}_{\mathrm{min}}+Q^{2}-m_{\beta}^{2})/y(s_{\alpha\beta}-m_{\alpha}^{2}-m_{\beta}^{2})$  \\
    $x_{a \mathrm{max}}$      &  $\backslash$   &   1   \\
    $y_{\mathrm{min}}$        &  \multicolumn{2}{c}{$(\hat{s}_{\mathrm{min}}+Q^{2}-m_{\beta}^{2})/(s_{\alpha\beta}-m_{\alpha}^{2}-m_{\beta}^{2})$}  \\
    $y_{\mathrm{max}}$        & \multicolumn{2}{c}{$\frac{1}{2m_{\alpha}^{2}}\left(\frac{\sqrt{Q^{2}(4m_{\alpha}^{2}+Q^{2})
    [(s-m_{\alpha}^{2}-m_{\beta}^{2})^{2}-4m_{\alpha}^{2}m_{\beta}^{2}]}}{s_{\alpha\beta}-m_{\alpha}^{2}-m_{\beta}^{2}}-Q^{2}\right)$}  \\
    $Q^{2}_{\mathrm{min}}$   &  \multicolumn{2}{c}{$y^{2}m^{2}_{\alpha}/(1-y)$} \\
    $Q^{2}_{\mathrm{max}}$   &  \multicolumn{2}{c}{$(1-y)(s-2m_{p}^{2})$} \\
\hline
\hline
\end{tabular}
\end{table}

\begin{table*}[htbp]
\renewcommand\arraystretch{1.5}
\centering
\caption{\label{Kine.inel.Q2.y} Same as Table~\ref{Kine.el.Q2.y} but for inelastic photoproduction processes, where $\hat{s}_{\mathrm{min}}=\hat{s}_{\gamma \mathrm{min}}$, $p_{T}^{2}=\hat{t}(\hat{s}\hat{u}+Q^{2}M_{J/\psi}^{2})/(\hat{s}+Q^{2})^{2}$.
}
\begin{tabular}{L{1cm}C{4cm}C{4cm}C{4cm}C{4cm}}
\hline
\hline
Variables & Coherent direct & Incoherent direct & Coherent resolved & Incoherent resolved \\
    \hline
    $z_{\mathrm{min}}$  &  \multicolumn{4}{c}{$\left[(M_{J/\psi}^{2}+\hat{s})-\sqrt{(\hat{s}-M_{J/\psi}^{2})^{2}-4p_{T \mathrm{min}}^{2}\hat{s}}\right]/2\hat{s}$} \\
    $z_{\mathrm{max}}$  &  \multicolumn{4}{c}{$\left[(M_{J/\psi}^{2}+\hat{s})+\sqrt{(\hat{s}-M_{J/\psi}^{2})^{2}-4p_{T \mathrm{min}}^{2}\hat{s}}\right]/2\hat{s}$} \\
    $\hat{t}_{\mathrm{min}}$  &  \multicolumn{2}{c}{$-(1-z_{\mathrm{min}})(\hat{s}+Q^{2})$}  & \multicolumn{2}{c}{$-(1-z_{\mathrm{min}})\hat{s}_{\gamma}$} \\
    $\hat{t}_{\mathrm{max}}$  &  \multicolumn{2}{c}{$-(1-z_{\mathrm{max}})(\hat{s}+Q^{2})$}  & \multicolumn{2}{c}{$-(1-z_{\mathrm{max}})\hat{s}_{\gamma}$} \\
    $z_{a' \mathrm{min}}$    &   $\backslash$   &   $\backslash$   &   $\hat{s}_{\gamma \mathrm{min}}/yx_{b}(s-2m_{p}^{2})$  &  $\hat{s}_{\gamma \mathrm{min}}/yx_{a}x_{b}(s-2m_{p}^{2})$ \\
    $z_{a' \mathrm{max}}$    &   $\backslash$   &   $\backslash$   &   1   &   1 \\
    $x_{b \mathrm{min}}$    & $(\hat{s}_{\mathrm{min}}+Q^{2})/y(s-2m_{p}^{2})$  &  $(\hat{s}_{\mathrm{min}}+Q^{2})/yx_{a}(s-2m_{p}^{2})$  &  $(\hat{s}_{\gamma \mathrm{min}})/z_{a' \mathrm{max}}y(s-2m_{p}^{2})$  &  $(\hat{s}_{\gamma \mathrm{min}})/z_{a' \mathrm{max}}yx_{a}(s-2m_{p}^{2})$ \\
    $x_{b \mathrm{max}}$    &  \multicolumn{4}{c}{1}  \\
    $x_{a \mathrm{min}}$    &   $\backslash$   &   $(\hat{s}_{\mathrm{min}}+Q^{2})/y(s-2m_{p}^{2})$   &   $\backslash$   &   $(\hat{s}_{\gamma \mathrm{min}})/z_{a' \mathrm{max}}y(s-2m_{p}^{2})$ \\
    $x_{a \mathrm{max}}$    &   $\backslash$   &   1   &   $\backslash$   &   1 \\
    $y_{\mathrm{min}}$      & \multicolumn{2}{c}{$(\hat{s}_{\mathrm{min}}+Q^{2})/(s-2m_{p}^{2})$} &  \multicolumn{2}{c}{$\hat{s}_{\gamma \mathrm{min}}/z_{a'
    \mathrm{max}}(s-2m_{p}^{2})$} \\
    $y_{\mathrm{max}}$      & \multicolumn{4}{c}{$\left[\sqrt{Q^{2}(4m_{\alpha}^{2}+Q^{2})}-Q^{2}\right]/2m_{\alpha}^{2}$} \\
\hline
\hline
\end{tabular}
\end{table*}

\begin{table*}[htbp]
\renewcommand\arraystretch{1.8}
\centering
\caption{\label{Kine.inel.pT.yr} Same as Table~\ref{Kine.inel.Q2.y} but for $p_{T}$ and $y_{r}$ distributions.
$x_{1}=\hat{s}/s$, $\tau=M_{J/\psi}^{2}/\hat{s}_{\mathrm{max}}$, $\tau'=m_{d}^{2}/\hat{s}_{\mathrm{max}}$, and $z_{a\mathrm{max}}=1/(1+Q^{2}/4p_{T}^{2})$~\cite{Rossi:1983xz}.
The bounds of $y$ are the same as Table~\ref{Kine.inel.Q2.y}, we are not list it here.
}
\begin{tabular}{L{1cm}C{4cm}C{4cm}C{4cm}C{4cm}}
\hline
\hline
Variables & Coherent direct & Incoherent direct & Coherent resolved & Incoherent resolved \\
    \hline
    $x_{b \mathrm{min}}$    &  $\backslash$  & $\backslash$  & $\hat{s}_{\gamma}/z_{a \mathrm{max}}y(s-2m_{p}^{2})$  & $\hat{s}_{\gamma}/z_{a \mathrm{max}}yx_{a}(s-2m_{p}^{2})$ \\
    $x_{b \mathrm{max}}$    &  $\backslash$  & $\backslash$  & 1  & 1 \\
    $x_{a \mathrm{min}}$    &  $\backslash$  & $(\hat{s}+Q^{2})/y(s-2m_{p}^{2})$  & $\backslash$  & $\hat{s}_{\gamma}/z_{a \mathrm{max}}y(s-2m_{p}^{2})$ \\
    $x_{a \mathrm{max}}$    &  $\backslash$  & 1  & $\backslash$  & 1 \\
    $Q^{2}_{\mathrm{min}}$   &  \multicolumn{4}{c}{$x^{2}_{1}m^{2}_{\alpha}/(1-x_{1})$} \\
    $Q^{2}_{\mathrm{max}}$   &  $\hat{s}/4$  & $(1-x_{1})(s-2m_{p}^{2})$  & $\hat{s}/4$ & $(1-x_{1})(s-2m_{p}^{2})$ \\
    $|y_{r \mathrm{max}}|$  & \multicolumn{4}{c}{$\frac{1}{2}\ln\frac{\hat{s}_{\mathrm{max}}+M_{J/\psi}^{2}-m_{\beta}^{2}+\sqrt{(\hat{s}_{\mathrm{max}}-M_{J/\psi}^{2}-m_{\beta}^{2})^{2}
-4(M_{J/\psi}^{2}m_{\beta}^{2}+p^{2}_{T}\hat{s}_{\mathrm{max}})}}{\hat{s}_{\mathrm{max}}+M_{J/\psi}^{2}-m_{\beta}^{2}-\sqrt{(\hat{s}_{\mathrm{max}}-M_{J/\psi}^{2}-m_{\beta}^{2})^{2}
-4(M_{J/\psi}^{2}m_{\beta}^{2}+p^{2}_{T}\hat{s}_{\mathrm{max}})}}$} \\
    $p_{T\mathrm{min}}$ & \multicolumn{4}{c}{$M_{J/\psi}$} \\
    $p_{T\mathrm{max}}$ &\multicolumn{4}{c}{$\sqrt{\hat{s}_{\mathrm{max}}}\left[(1-\tau)^{2}+(1-\tau')^{2}-2\tau\tau'
    -4\tau\sinh^{2}y_{r}-1\right]^{\frac{1}{2}}/2\cosh y_{r}$} \\
\hline
\hline
\end{tabular}
\end{table*}

We give here the detailed treatment of partonic kinematics matched with the exact treatment in Section~\ref{Exac}.

The energy and momentum in $\alpha\beta$ CM frame read
\begin{align}\label{Mant.coh.dir}
E_{\alpha}&=\frac{1}{2\sqrt{s_{\alpha\beta}}}(s_{\alpha\beta}+m_{\alpha}^{2}-m_{\beta}^{2}),\displaybreak[0]\nonumber\\
E_{\beta}&=\frac{1}{2\sqrt{s_{\alpha\beta}}}(s_{\alpha\beta}-m_{\alpha}^{2}+m_{\beta}^{2}),\displaybreak[0]\nonumber\\
p_{\mathrm{CM}}&=\frac{1}{2\sqrt{s_{\alpha\beta}}}\sqrt{(s_{\alpha\beta}-m_{\alpha}^{2}-m_{\beta}^{2})^{2}-4m_{\alpha}^{2}m_{\beta}^{2}},\displaybreak[0]
\end{align}
while those in $\gamma^{*}\beta$ CM frame are
\begin{align}\label{Epab.}
\hat{E}_{\gamma}&=\frac{1}{2\sqrt{\hat{s}}}(\hat{s}-Q^{2}-m_{\beta}^{2}),\displaybreak[0]\nonumber\\
\hat{E}_{J/\psi}&=\frac{1}{2\sqrt{\hat{s}}}(\hat{s}+M_{J/\psi}^{2}-m_{\beta}^{2}),\displaybreak[0]\nonumber\\
\hat{p}_{\mathrm{CM}}&=\frac{1}{2\sqrt{\hat{s}}}\sqrt{(\hat{s}+Q^{2}-m_{\beta}^{2})^{2}+4Q^{2}m_{\beta}^{2}},\displaybreak[0]\nonumber\\
\hat{p}_{\mathrm{CM}}'&=\frac{1}{2\sqrt{\hat{s}}}\sqrt{(\hat{s}-M_{J/\psi}^{2}-m_{\beta}^{2})^{2}-4M_{J/\psi}^{2}m_{\beta}^{2}}.\displaybreak[0]
\end{align}

The Mandelstam variables involved in the case of direct photoproduction processes are
\begin{align}\label{Mant.}
\hat{s}&=(q+p_{\beta})^{2}=y(s_{\alpha\beta}-m_{\alpha}^{2}-m_{\beta}^{2})+m_{\beta}^{2}-Q^{2},\displaybreak[0]\nonumber\\
\hat{t}&=(q-p_{J/\psi})^{2}=-(1-z)(\hat{s}+Q^{2}),\displaybreak[0]\nonumber\\
\hat{u}&=(p_{\beta}-p_{J/\psi})^{2}=M_{J/\psi}^{2}-z(\hat{s}+Q^{2}),\displaybreak[0]
\end{align}
while those for resolved processes are
\begin{align}\label{Mant.}
\hat{s}_{\gamma}&=(p_{a'}+p_{\beta})^{2}=yz_{a'}(s_{\alpha\beta}-m_{\alpha}^{2}-m_{\beta}^{2})+m_{a'}^{2}+m_{\beta}^{2},\displaybreak[0]\nonumber\\
\hat{t}_{\gamma}&=(p_{a'}-p_{J/\psi})^{2}=-(1-z)\hat{s}_{\gamma},\displaybreak[0]\nonumber\\
\hat{u}_{\gamma}&=(p_{\beta}-p_{J/\psi})^{2}=M_{J/\psi}^{2}-z\hat{s}_{\gamma}.\displaybreak[0]
\end{align}

We summarize the kinematical boundaries in Table~\ref{Kine.el.Q2.y}-\ref{Kine.inel.pT.yr}.
The kinematical boundaries of $z$ distribution are the same as Table~\ref{Kine.el.Q2.y}, \ref{Kine.inel.Q2.y}, but instead of $\hat{t}$, $Q^{2}$ should be integrated out.
Finally, we give here the complete expressions of the Jacobian determinant $\mathcal{J}$ for each distribution.
In the case of the $Q^{2}$ and $y$ distributions,
\begin{align}\label{Jac.Q2.y}
\mathcal{J}=\frac{2r^{2}\left|\textbf{p}_{\beta}\right|}{E_{\alpha'}E_{\beta}}=2r^{2}\sqrt{\frac{(s_{\alpha\beta}-m_{\alpha}^{2}-m_{\beta}^{2})^{2}-4m_{\alpha}^{2}m_{\beta}^{2}}
{(s_{\alpha\beta}-m_{\alpha}^{2}-m_{\beta}^{2})^{2}(r^{2}+m_{\alpha}^{2})}}.\displaybreak[0]
\end{align}
In the case of the $p_{T}$ and $y_{r}$ distributions,
\begin{align}\label{Jac.pT.yr}
\mathcal{J}=&\frac{2\hat{p}_{\mathrm{CM}}\hat{s}}{(s_{\alpha\beta}-m_{\alpha}^{2}
-m_{\beta}^{2})(\sqrt{\hat{s}}-\cosh y_{r}m_{T})},\displaybreak[0]
\end{align}
for elastic-coherent processes.
The relations between Eq.~({\ref{Jac.pT.yr}}) and the rest cases are: $\mathcal{J}^{\mathrm{elastic}}_{\mathrm{incoherent}}=\mathcal{J}^{\mathrm{inelastic}}_{\mathrm{coh.dir.}}=\mathcal{J}/y$, $\mathcal{J}^{\mathrm{inelastic}}_{\mathrm{incoh.dir.}}=\mathcal{J}/yx_{a}$, $\mathcal{J}^{\mathrm{inelastic}}_{\mathrm{coh.res.}}=\mathcal{J}/yx_{b}$, and
$\mathcal{J}^{\mathrm{inelastic}}_{\mathrm{incoh.res.}}=\mathcal{J}/yx_{a}x_{b}$.
In the case of fragmentation processes, $\mathcal{J}=(\hat{s}+Q^{2})/\cosh y_{r}\sqrt{\hat{s}}$.

\end{document}